\RequirePackage{fix-cm}
\documentclass[twocolumn]{svjour3}          
\smartqed  
\usepackage{graphicx}
\usepackage{multirow}
\usepackage{subfigure}
\usepackage{url}
\usepackage{hyperref}
\usepackage{color}
\usepackage{xspace}
\usepackage{amsmath}
\usepackage{amssymb}

\usepackage{lipsum}

\newcommand{\tabincell}[2]{\begin{tabular}{@{}#1@{}}#2\end{tabular}}

\newcommand{\kw}[1]{{\ensuremath {\mathsf{#1}}}\xspace}
\newcommand{\truss}{${k}$-\kw{truss}}

\newcommand{\ktruss}{$k$-truss\xspace}

\newcommand{\ctcp}{\kw{CTC}-\kw{Problem}}

\newcommand{\keyw}{\kw{f }}

\newcommand{\score}{\kw{score }}

\newcommand{\ati}{\kw{AT index}}

\newcommand{\kdtruss}{\kw{(k,d)}-\kw{truss}}

\newcommand{\xin}[1]{{\color{black}{#1}}}

\begin{document}

\title{A Survey of Community Search Over Big Graphs}

\author{Yixiang Fang    \and
        Xin Huang       \and
        Lu Qin          \and
        Ying Zhang      \and\\
        Wenjie Zhang    \and
        Reynold Cheng   \and
        Xuemin Lin
}

\institute{Yixiang Fang, Xuemin Lin \at
              University of New South Wales, Australia\\
              Zhejiang Lab, Hangzhou, China\\
              \email{yixiang.fang@unsw.edu.au, lxue@cse.unsw.edu.au}
           \and
           Xin Huang \at
           Hong Kong Baptist University, Hong Kong\\
           \email{xinhuang@comp.hkbu.edu.hk}
           \and
           Lu Qin, Ying Zhang \at
           The University of Technology Sydney, Australia\\
           \email{lu.qin@uts.edu.au, ying.zhang@uts.eud.au}
           \and
           Wenjie Zhang \at
           University of New South Wales, Australia\\
           \email{zhangw@cse.unsw.edu.au}
           \and
           Reynold Cheng \at
           The University of Hong Kong, Hong Kong\\
           \email{ckcheng@cs.hku.hk}
}

\date{Received: date / Accepted: date}

\maketitle

\newcommand\blfootnote[1]{%
\begingroup
\renewcommand\thefootnote{}\footnote{#1}%
\addtocounter{footnote}{-1}%
\endgroup
}

\begin{abstract}

With the rapid development of information technologies, various big graphs are prevalent in many real applications (e.g., social media and knowledge bases). An important component of these graphs is the network community. Essentially, a community is a group of vertices which are densely connected internally. Community retrieval can be used in many real applications, such as event organization, friend recommendation, and so on. Consequently, how to efficiently find high-quality communities from big graphs is an important research topic in the era of big data. Recently a large group of research works, called community search, have been proposed. They aim to provide efficient solutions for searching high-quality communities from large networks in real-time. Nevertheless, these works focus on different types of graphs and formulate communities in different manners, and thus it is desirable to have a comprehensive review of these works.

In this survey, we conduct a thorough review of existing community search works. Moreover, we analyze and compare the quality of communities under their models, and the performance of different solutions. Furthermore, we point out new research directions. This survey does not only help researchers to have better understanding of existing community search solutions, but also provides practitioners a better judgement on choosing the proper solutions.

\end{abstract}

\section{Introduction}
\label{sec:intro}

With the rapid development of information technologies, various big graphs are prevalent in many real applications (e.g., social media and knowledge bases). An important component of these graphs is the network community. Essentially, a community is a group of vertices which are densely connected internally. For example, in Facebook, communities consist of users that are with strong friendship \cite{acquisti2006imagined}; on the World Wide Web, communities contain web sites which share similar topics \cite{broder2000graph}; in protein-protein interaction networks \cite{article05clique} and metabolic networks \cite{guimera2005functional}, communities correspond to functionality modules.
Retrieving communities from a network is a fundamental problem in network science, and it can be applied to many real-life applications. Here are some typical applications, to name a few:
\begin{itemize}
  \item {\it Event organization.} A social event (e.g., a party or a conference) often involves a group of users and its organization can benefit from communities. For example, to hold a cocktail part, a user can find his community, i.e., a group of researchers, each of which is well acquainted.
  \item {\it Friend recommendation.} Many social media platforms (e.g., Facebook) often maintain a friendship network. To suggest candidate friends to a specific user $u$, intuitively we can recommend $u$ those who are in $u$'s community but are not yet $u$'s friends.
  \item {\it Protein complex identification.} In biology, proteins interact with each other and a gene is often regulated by a set of proteins. To study a gene, a biologist may focus on a set of proteins that highly interact with each other, which is a community of proteins.
  \item {\it Advertisement in e-commence.} Users of the same community often share similar interests. To push advertisements for a user $u$, we may find her community first and then select advertisements that are checked by members of her community.
\end{itemize}

Owing to the importance of communities, how to effectively and efficiently find communities from large graphs is an important research topic in the era of big data. With a careful observation on these applications, we identify a list of factors that the community retrieval solutions should satisfy:
\begin{itemize}
  \item {\it High efficiency.} For many real applications (e.g., event organization), the communities often need to be retrieved in real-time, based on query requests. Thus, the community retrieval solutions should be able to respond in real-time.
  \item {\it High scalability.} Nowadays, many real networks contain millions or billions of vertices. As a result, the solutions should be scalable to real big graphs.
  \item {\it High personalization.} In practice, for large networks, people usually are interested in communities of some specific users, rather than all the users. Thus, the solutions should allow users to specify query vertices. Moreover, some personalized requirements on structures (and attributes) could be imposed.
  \item {\it High quality.} The vertices in the communities retrieved should be cohesively linked. Moreover, the communities should be easy for interpretation.
  \item {\it Support for dynamic graphs.} Since real networks often involve as the time goes on, the solutions should be able to adapt for the dynamic changes easily.
\end{itemize}

Towards the goals above, recently a large group of research works, called community search (CS),
have been proposed \cite{Huang:ICDE:2017}. Generally, the goal of CS is to search high-quality communities in an online manner, based on a query request. Specifically, given a vertex $q$ of a graph $G$, it aims to find a community, or a dense subgraph, which contains $q$ and satisfies the properties: (1) {\it connectivity}, i.e., vertices in the community are connected; and (2) {\it cohesiveness}, i.e., vertices in the community are intensively linked to each other w.r.t. a particular goodness metric \cite{KDD2010,KDD2010,local2014,barbieri2015efficient,online-sigmod2013}. The metric is often defined by using some classical subgraph cohesiveness metrics such as:
\begin{itemize}
  \item $k$-core. The $k$-core~\cite{md1983,kcore2003} is the largest subgraph of $G$, in which each vertex's degree is at least $k$ within the subgraph.
  \item $k$-truss. The $k$-truss~\cite{cohen2008trusses,k-truss2014} is the largest subgraph of $G$ in which every edge is contained in at least ($k-2$) triangles within the subgraph.
  \item $k$-clique. A $k$-clique~\cite{kclique} is a set of $k$ vertices of $G$ such that each pair of vertices has an edge.
  \item $k$-ECC. A $k$-ECC ($k$-edge connected component)~\cite{gibbons1985algorithmic} is a subgraph of $G$ such that after removing any $k$--1 edges, it is still connected.
\end{itemize}

\begin{figure}[]
\centering
	\includegraphics[width=0.6\columnwidth]{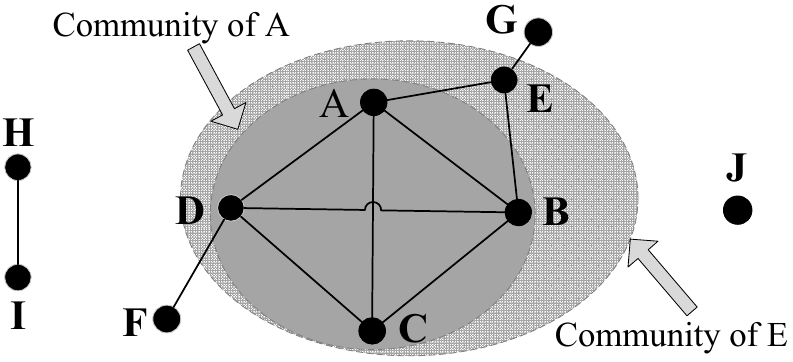}
\caption{An example of community search.}
\label{fig:intro}
\end{figure}

Let us illustrate CS by an example. Consider the graph with ten vertices in Fig. \ref{fig:intro}, and CS solutions \cite{KDD2010,local2014,barbieri2015efficient}, which are based on the $k$-core model. Let $q$=$A$. Then, the induced subgraph of vertices \{$A$, $B$, $C$, $D$\} will be returned as the community. Note that the subgraph forms a $k$-core with $k$=3, since each vertex's degree is 3 within the subgraph, and it is also the core attaining the maximum value of $k$.

In the literature, there is a highly related group of research works, called community detection (CD) \cite{CD:Survey:2009,CD:Survey:2011,CD:Survey:2011a,CD:Survey:2013,CD:Survey:2017}.
Generally, it has similar goals with CS, but there are three key differences:
(1)	The problem definitions are different. CS aims to search communities regarding a set of query vertices and some query parameters, while CD often detects all communities in the graph.
(2)	The criteria of defining communities are different. In CS, the criteria of defining communities are based on query parameters given by the users. In other words, communities are retrieved depending on user-defined parameters. In contrast, CD methods often use the same global criterion to detect communities by partitioning the entire graph.
For example, in Fig.~\ref{fig:intro}, if $q$=$A$, CS solutions \cite{KDD2010,local2014} will find the community \{$A$, $B$, $C$, $D$\}, and if $q$=$E$, they will find the community \{$A$, $B$, $C$, $D$, $E$\}. In contrast, if using a CD method (e.g., the spectral clustering \cite{von2007tutorial}) with setting the number of communities to 3, we will obtain three communities, each of which forms a connectivity component, where $B$ and $E$ are in the same community.
(3)	The algorithms are different. As shown in existing studies, CS solutions can search communities efficiently in an online manner, while CD solutions are often time consuming and unscalable to big graphs. Moreover, CS queries can often be supported by indexes and handle dynamic graphs easily.
Thus, compared to CD solutions, CS solutions can better satisfy factors aforementioned.

\begin{table*}[ht]
  \centering
  \caption {Classification of works of community search (``P." means Problem).}
  \label{tab:methods}
  \begin{tabular}{c|c|c|c|c|c|c}
     \hline
          \multirow{2}*{\textbf{Metric}}
          & \multirow{2}*{\textbf{Simple graphs}}
          & \multicolumn{5}{|c}{\textbf{Attributed graphs}}
          \\
     \cline{3-7}
        & & \textbf{Keyword} & \textbf{Location} & \textbf{Temporal} & \textbf{Influence (weight)} & \textbf{Profile}\\
     \hline\hline
        $k$-core
        & \tabincell{c}{\cite{KDD2010,local2014,barbieri2015efficient,Fang:TKDE:CSD}\\
           (P. \ref{prob:kcoreGlobal}, \ref{prob:kcoreLocal}, \ref{prob:kcoreGlobalSize}, \ref{prob:kcoreMinSize}, \ref{prob:CSD})}
        & \tabincell{c}{\cite{Fang:VLDB:2016,Fang:VLDBJ:2017}\\
           (P. \ref{prob:acq})}
        & \tabincell{c}{\cite{Fang:VLDB:2017,Fang:TKDE:SAC,wangkai2018,zhu2017geo}\\
           (P. \ref{prob:sac}, \ref{prob:rbkcore}, \ref{prob:gsgq})}
        & \tabincell{c}{\cite{Li:ICDE:2018}\\
           (P. \ref{prob:timecs})}
        & \tabincell{c}{\cite{Li:vldb:2015,Li:VLDBJ:2017,chen2016efficient,zheng2017querying,Bi:2018,Li:SIGMOD:2018} \\
           (P. \ref{prob:nic}, \ref{prob:skyCS})}
        & \tabincell{c}{\cite{Yankai18}\\
           (P. \ref{prob:PCS})}
        \\
     \hline
        $k$-truss
        & \tabincell{c}{\cite{k-truss2014,Akbas:VLDB:2017,huang2015approximate}\\
          (P. \ref{problem:TTC}, \ref{prob:ctc})}
        & \tabincell{c}{\cite{Huang:2017:ATC}\\
          (P. \ref{prob:atc})}
        & --
        & --
        & \tabincell{c}{\cite{Zheng:IS:2017}\\
          (P. \ref{problem:WTC})}
        & --
        \\
     \hline
        $k$-clique
        & \tabincell{c}{\cite{online-sigmod2013,kclique2018,yang2011social,wang2017query}\\
          (P. \ref{prob:kcliquesearch}, \ref{prob:densest}, \ref{prob:sgq}, \ref{prob:mckpq})}
        & --
        & --
        & \tabincell{c}{\cite{kclique2017}\\
          (P. \ref{prob:infCS})}
        & --
        & --
        \\
     \hline
        $k$-ECC
        & \tabincell{c}{\cite{Chang:SIGMOD:2015,hu2016querying,hu2017querying}\\
          (P. \ref{prob:keccMax}, \ref{prob:keccMinimum}, \ref{prob:keccMinimal})}
        & --
        & --
        & --
        & --
        & --
        \\
     \hline
                 Others & \multicolumn{6}{l}{
                        local modularity: \cite{clauset2005finding,Luo2006ELC}
                        query biased density: \cite{Wu:VLDB:2015}
                        pagerank: \cite{andersen2006communities,Kloumann2014} (P. \ref{prob:PPR})
                        neighbors: \cite{Mehler2009}
                       }\\
    \hline
  \end{tabular}
  \vspace{-0.1in}
\end{table*}

Although there are many CS solutions, they deal with different types of graphs and formulate communities in different manners. Meanwhile, there is a lack of systematic survey of CS solutions. Thus, it is desirable to organize these works and understand how well they perform in terms of efficiency and quality. To this end, in this paper we will provide a thorough review of these works. We will also compare different CS solutions so that readers can better understand the state-of-the-art, and point out directions for future study.

As shown in Table~\ref{tab:methods}, we classify CS solutions into five categories such that solutions in each category (except the last category) adopt the same structure cohesiveness metric. Moreover, for works in each category, we further partition them into two groups, where the first group focuses on simple graphs while the second group targets attributed graphs. Note that the IDs of CS problems are also included in the brackets of Table~\ref{tab:methods}.
For simple graphs, CS solutions search communities purely based on link information, while for attributed graphs, CS solutions often consider both links and attributes.
We remark that these cohesiveness metrics are orthogonal to graph types. This implies that if a metric has not been studied for a particular type of graphs, then it is a possible future research direction to study CS by applying the metric on this type of graphs.

In summary, our main contributions are as follows:
\begin{itemize}
  \item First, we provide a systematic classification of studies on CS. Specifically, we classify these studies according to the community cohesiveness metrics. For each class of works, we review the representative studies on different types of graphs.
  \item Second, we perform a thorough analysis and comparison of different community cohesiveness metrics. Moreover, we analyze and compare CS solutions on simple graphs and attributed graphs.
  \item Third, we offer insightful suggestions for future study on CS. This may give researchers new to CS an understanding of the recent development of CS, as well as a good starting point to work in this field.
\end{itemize}

The rest of this paper is organized as follows. In Section~\ref{sec:pre}, we introduce and discuss community cohesiveness metrics. In Sections~\ref{sec:kcore}, \ref{sec:ktruss}, \ref{sec:kclique}, \ref{sec:kecc}, and \ref{sec:other}, we extensively discuss CS solutions in each category. We also present two CS systems in Section \ref{sec:demo}. We review the related work in Section~\ref{sec:related}. Finally, we present a list of future topics in Section~\ref{sec:future} and conclude in Section~\ref{sec:conclude}.

\section{Preliminaries}
\label{sec:pre}

In this section, we first formally introduce the commonly-used community cohesiveness metrics, and then compare their cohesiveness and computational efficiency.

\subsection{Cohesiveness Metrics}
\label{sec:definitions}

For ease of exposition, we consider a simple undirected graph $G(V,E)$, with vertex set $V$ and edge set $E$. Let $n$ and $m$ be the corresponding sizes of $V$ and $E$. The degree of a vertex $v$ of $G$ is denoted by $deg_G(v)$.

\noindent\underline{$\bullet$ \textbf{$k$-core.}} We introduce its formal definition as follows.

\begin{definition}[$k$-core~\cite{md1983,kcore2003}]
\label{def:kcore}
Given an integer $k$ ($k\geq 0$), the $k$-core of $G$,
denoted by $H_{k}$, is the largest subgraph of $G$, such that $\forall v \in H_k$, $deg_{H_k}(v) \geq k$.
\end{definition}

We say that $H_k$ has an order of $k$.  Notice that $H_k$ may not be a connected graph~\cite{kcore2003}. Observe that $k$-$core$s are ``nested''~\cite{kcore2003}: given two positive integers $i$ and $j$, if $i<j$, then $H_j \subseteq H_i$.

\begin{figure}[h]
\centering
\begin{tabular}{c c}
  \begin{minipage}{4.0cm}
	\includegraphics[width=4.0cm]{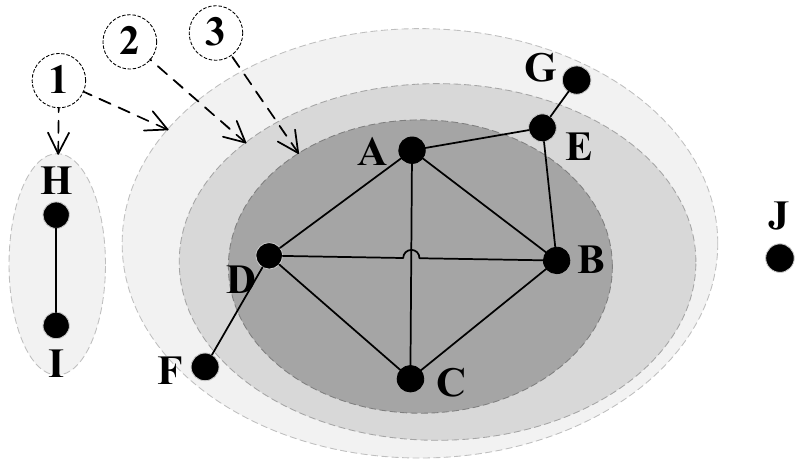}
  \end{minipage}
  &
  \begin{minipage}{3.8cm}
	\includegraphics[width=3.379cm]{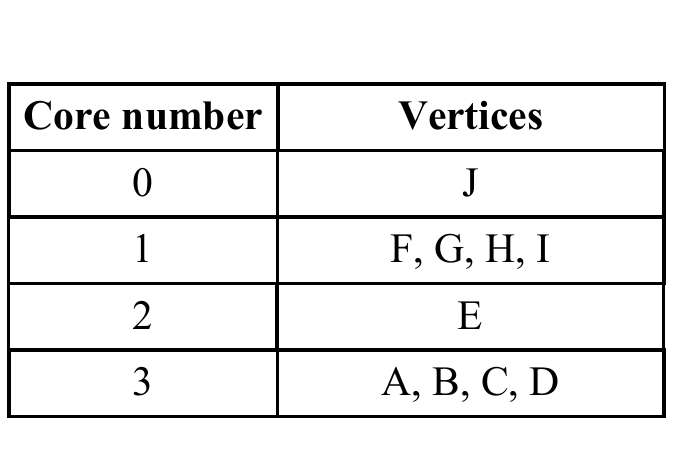}
  \end{minipage}
  \\
  \small (a) $k$-cores
  &
  \small (b) Core numbers
\end{tabular}
\caption{Illustrating $k$-core.}
\label{fig:kcoreEg}
\end{figure}

\begin{example}
\label{eg:kcore}
In Fig. \ref{fig:kcoreEg}(a), the subgraph of $\{A,B,C,D\}$ is the 3-core. The 1-core has vertices $\{A,B,C, D,E,F$, $G,H,I\}$, and is composed of two connected components: $\{A,B,C$, $D,E,F,G\}$ and $\{H,I\}$. The number $k$ in each circle represents the $k$-core contained in that ellipse. Clearly, $H_3\subset H_2 \subset H_1$.
\end{example}

\begin{definition}[core number]
\label{def:coreNum}
Given a vertex $v \in V$, its core number, denoted by $core_G[v]$, is the highest order of a $k$-core that contains $v$.
\end{definition}

A list of core numbers and their respective vertices for Example~\ref{eg:kcore} are shown in Fig. \ref{fig:kcoreEg}(b).
Equivalently, the $k$-core is the induced subgraph of vertices, whose core numbers are at least $k$.

\noindent\underline{$\bullet$ \textbf{$k$-truss.}} The $k$-truss is defined based on triangles. Specifically, a triangle in $G$ is a cycle of length 3. Let $u$, $v$, $w\in V$ be the three vertices on the cycle. Then, we denote this triangle by $\triangle_{uvw}$.

\begin{definition}[support]
\label{def:edgeSupport}
Given a graph $G(V,E)$, the support of an edge ($u$, $v$)$\in E$, denoted by $sup(e, G)$, is defined as
$|\{\triangle_{uvw}: u,v,w\in V\}|$.
\end{definition}

\begin{definition}[$k$-truss~\cite{saito2008extracting,cohen2008trusses,zhang2012extracting}]
\label{def:ktruss}
Given a graph $G$, the $k$-truss of $G$, denoted by $J_k$, is the largest subgraph of $G$, such that $\forall e\in J_k$, $sup(e, J_k)\geq (k-2)$.
\end{definition}

\begin{figure}[h]
\hspace*{-.2cm}
\centering
\begin{tabular}{c c}
  \begin{minipage}{4.0cm}
	\includegraphics[width=4.0cm]{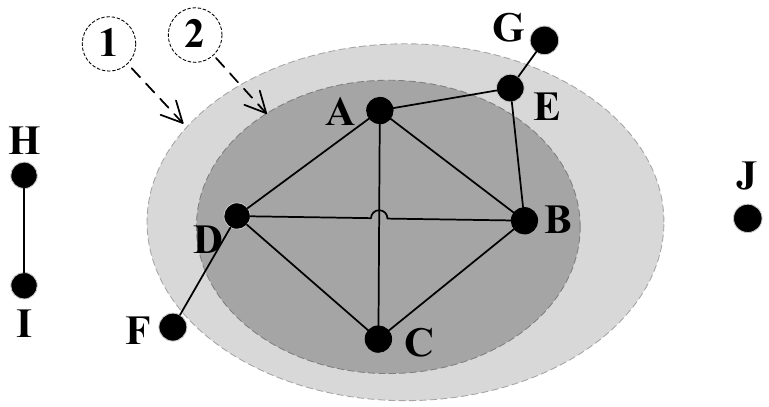}
  \end{minipage}
  &
  \begin{minipage}{3.9cm}
	\includegraphics[width=3.9cm]{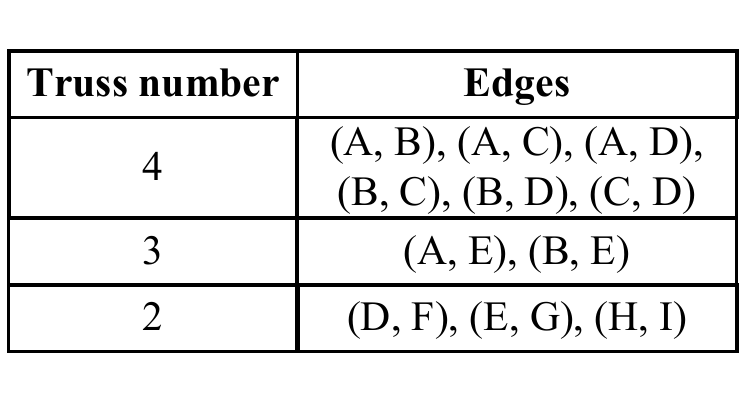}
  \end{minipage}
  \\
  \small (a) $k$-trusses
  &
  \small (b) Trusses
\end{tabular}
\caption{Illustrating $k$-truss.}
\label{fig:ktrussEg}
\end{figure}

\begin{example}
\label{eg:ktruss}
Let us reconsider the graph $G$ in Fig. \ref{fig:kcoreEg}(a). The induced subgraph of $G$ by vertex set $\{A,B,C,D\}$ is the 4-truss. The 3-truss has vertices $\{A,B,C, D,E\}$. The number $k$ in each circle represents the $k$-truss contained in that ellipse.
\end{example}



\xin{
\begin{definition}[truss number~\cite{wang2012truss}]
\label{def:trussNum}
Given a graph $G$, the truss number (trussness) of an edge $e\in G$, denoted by $\tau(e)$, is the largest $k$ such that there is a $k$-truss containing $e$.
\end{definition}

A list of truss numbers and their respective edges for Example~\ref{eg:ktruss} are shown in Fig.~\ref{fig:ktrussEg}(b).
}
Equivalently, the $k$-truss is the induced subgraph of edges, whose truss numbers are at least $k$. Similar to $k$-core, a $k$-truss may contain multiple connected components.

\noindent\underline{$\bullet$ \textbf{$k$-clique.}} It is defined as follows.

\begin{definition}[$k$-clique~\cite{kclique,article05clique}]
\label{def:kclique}
A $k$-clique is a complete graph with $k$ vertices where there is an edge between every pair of vertices.
\end{definition}

\begin{example}
\label{eg:kclique}
In the graph in Fig. \ref{fig:kcoreEg}(a). The subgraph of $\{A,B,C,D\}$ is a 4-clique and any three vertices of them form a 3-clique (i.e., triangle). The subgraph of $\{A,B,E\}$ is also a 3-clique. Any edge is a 2-clique.
\end{example}

\noindent\underline{$\bullet$ \textbf{$k$-ECC.}} We first introduce some related concepts.

\begin{definition}[edge connectivity~\cite{gibbons1985algorithmic,hu2016querying}]
\label{def:connectivity}
Given a graph $G(V,E)$ and two vertices $u,v\in V$, the connectivity $\lambda(u,v)$ between $u$ and $v$ is the minimum number of edges whose removal disconnects $u$ and $v$.
\end{definition}

\begin{definition}[graph connectivity~\cite{gibbons1985algorithmic,hu2016querying}]
\label{def:connectivity}
Given a graph $G(V,E)$, the connectivity of the graph $G$, $\lambda(G)$= $\min_{u,v\in V}\lambda(u,v)$, is the minimum connectivity between any two distinct vertices in $G$, i.e., the minimum number of edges whose removal disconnects $G$.
\end{definition}

\begin{definition}[$k$-ECC~\cite{gibbons1985algorithmic,hu2016querying}]
\label{def:connectivity}
Given a graph $G(V,E)$, a subgraph $G'$ of $G$ is a $k$-edge connected component, or $k$-ECC, if $\lambda(G')\geq k$ and the connectivity of any super-graph of $G'$ in $G$ is less than $k$.
\end{definition}

\begin{example}
\label{eg:kECC}
In the graph in Fig. \ref{fig:kcoreEg}(a). The subgraph of $\{A,B,C,D\}$ is the 3-ECC, because for any pair of vertices in it, to disconnect them, we need to remove at least 3 edges. The 2-ECC has vertices $\{A,B,C, D,E\}$. There are two 1-ECCs, which contain vertices $\{H, I\}$ and $\{A,\cdots$, $G\}$ respectively.
\end{example}

\subsection{Cohesiveness and Computational Efficiency}
\label{sec:analysis}

Generally, in terms of structure cohesiveness, $k$-clique is the most cohesive one, since each vertex of a $k$-clique is linked to all the other ($k-1$) vertices. For each connected component of the $k$-truss, it is more cohesive than a $k$-ECC. This is because $k$-truss is more restrictive as it is defined based on triangle, which is a local concept, whereas $k$-ECC is more global~\cite{akiba2013linear}.

Obviously, the $k$-truss is more cohesive than the $k$-core, since in a $k$-truss, each pair of vertices within an edge must have ($k-2$) common neighbors, while in a $k$-core, any pair of vertices within an edge may have no common neighbors. Also, the $k$-ECC is more cohesive than $k$-core, since it is a connected subgraph and requires that each vertex has at least $k$ neighbors, while a $k$-core may contain multiple connected components.
We further analyze their inclusion-ship as follows. Let $G(V,E)$ be a graph and $k$ be an integer ($k$$\geq$0). We have:
\begin{enumerate}
  \item a $k$-clique must be a subgraph of the $k$-truss;
  \item each connected component of the $k$-truss must be a subgraph of a particular $k$-ECC;
  \item the $k$-truss must be a subgraph of the ($k$--1)-core;
  \item a $k$-ECC must be a subgraph of the $k$-core;
\end{enumerate}

In summary, in terms of structure cohesiveness, the four metrics above can be roughly ranked as:
$k$-core $\preceq$ $k$-ECC $\preceq$ $k$-truss $\preceq$ $k$-clique.

Next, we discuss their computational efficiency~\footnote{Here, we only consider algorithms that assume the graph can be kept in the memory of a single machine.}. Note that for each metric, there may exist multiple algorithms for enumerating its subgraphs, but here we only discuss complexities of the most efficient ones.

In~\cite{kcore2003}, a linear $k$-core decomposition algorithm, which computes all the $k$-cores in the graph $G$, takes $O(m+n)$ time and $O(m+n)$ space.
In~\cite{Chang:SIGMOD:2013}, Chang et al. proposed an algorithm, which computes all the $k$-ECCs for a specific $k$, and it takes $O(h\cdot l\cdot m)$ time and $O(m+n)$ space, where $h$ and $l$ are usually bounded by smaller constants for real graphs~\cite{Chang:SIGMOD:2013}.
In~\cite{wang2012truss}, an efficient algorithm for computing the $k$-truss, for all $k\geq 3$, takes $O(m^{1.5})$ time and $O(m+n)$ space.
In~\cite{mauro2018}, an algorithm, which enumerates all the $k$-cliques for a specific $k$, completes in
$O(c(G)\cdot\Sigma_{l=2}^{k-1}N^l+k\cdot N^k)$ time and $O(m+n)$ space, where $c(G)$ denotes the maximum core number of vertices in $G$ and $N^l$ is the number of $l$-cliques. Notice that $N^l$ could be exponentially large.
As a result, considering their computational efficiency, we can rank these metrics as: $k$-core $\succeq$ $k$-ECC $\succeq$ $k$-truss $\succeq$ $k$-clique.

\begin{figure}[t]
\centering
\includegraphics[width=0.88\linewidth]{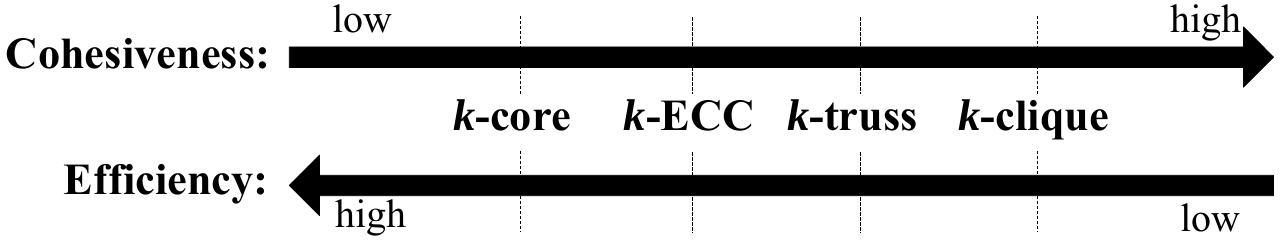}
\caption{Comparison of cohesiveness models.}
\label{fig:comparison}
\end{figure}

In summary, there is a trade-off between the structure cohesiveness and computational efficiency, as shown in Fig. \ref{fig:comparison}. That is, a more cohesive metric often takes more computational cost.
In addition, we have performed  a comparison study of the efficiency for these metrics on four real graphs \footnote{Email-Enron, Google, Livejournal are downloaded from \url{https://snap.stanford.edu/data/index.html}, and Wise is downloaded from \url{http://www.wise2012.cs.ucy.ac.cy/challenge.html}.}.
namely Email-Enron ($|V|$= 36.7\textbf{K}, $|E|$=183.8\textbf{K}), Google ($|V|$=876\textbf{K}, $|E|$=5.1\textbf{M}), Livejournal ($|V|$=4.8\textbf{M}, $|E|$=69\textbf{M}), and Wise ($|V|$=58.6 \textbf{M}, $|E|$=265.1\textbf{M}), where \textbf{K}$=10^3$ and \textbf{M}$=10^6$. Clearly, the efficiency results well confirm the analysis above.

\begin{table}
  \centering
  \caption {Efficiency comparison for different metrics.
  }
  \label{tab:metricExp}
  \begin{tabular}{c|c|c|c|c}
    \hline \textbf{Datasets}
                      & \textbf{$k$-core}
                      & \textbf{$k$-ECC}
                      & \textbf{$k$-truss}
                      & \textbf{$k$-clique}\\
    \hline\hline
        email-Enron &
         0.2s  &
         0.8s  &
        5s   &
        201s  \\
   \hline
        Google &
        8.9s   &
        40.8s   &
        65s   &
        $>$24 hours   \\
    \hline
        Livejournal &
        85s   &
        854s  &
        1726s   &
        $>$24 hours   \\
    \hline
        Wise &
        553s   &
        5764s   &
        32221s   &
        $>$24 hours   \\
    \hline
  \end{tabular}
\end{table}

Based on the comparison analysis above, we would like to make some suggestions:
(1) For small or moderate-size graphs, $k$-clique and $k$-truss not only achieve higher cohesiveness but also reasonable efficiency.
(2) For large graphs, $k$-core and $k$-ECC should be better choices since they can be computed more efficiently.
(3) For graphs with higher clustering coefficient which can be decomposed into more triangles, $k$-truss is preferable.
(4) For some special graphs (e.g., bipartite graphs), there may not exist any triangles and thus the $k$-truss model may not work.

\section{$K$-Core-Based Community Search}
\label{sec:kcore}

In this section, we review CS works that use the $k$-core as structure cohesiveness metric. We classify these works into several groups according to the types of graphs, namely undirected graphs, directed graphs, and attributed graphs including keyword-based, location-based, temporal, influence value-based, and profile-based graphs, and then discuss them respectively.

\subsection{Undirected Graphs}
\label{sec:kcoreSimple}

An undirected graph, denoted by $G(V,E)$, contains a set $V$ of vertices and a set $E$ of edges. Existing CS works on simple undirected graphs can be classified as {\it size-unbounded} and {\it size-bounded} CS, where the former one has no constraint on the size of the community and the latter one imposes constraint on the community size.

\subsubsection{Size-Unbounded Community Search}
\label{sec:sizeUnbounded}

In~\cite{KDD2010}, Sozio et al. proposed and studied the problem of community search, defined as follows:

\begin{problem}
\label{prob:kcoreGlobal}
Given an undirected simple graph $G(V,E)$, a set of query vertices $Q\subseteq V$, and a goodness function $f$, return a subgraph $H(V_H,E_H)$ of $G$, such that
\begin{enumerate}
  \item $V_H$ contains $Q$;
  \item $H$ is connected;
  \item $f(H)$ is maximized among all feasible choices for $H$.
\end{enumerate}
\end{problem}

Here, $f(H)$ is a general goodness function for measuring cohesiveness of the community $H$. Intuitively, the value of $f(H)$ should be larger, if $H$ is densely connected. There are many possible choices for $f$, and an outstanding one is defined based on the {\it minimum degree}, i.e., $f(H)$=$\min_{\forall v\in H} deg_H(v)$. The reasons why the minimum degree is a good metric for the community are three-fold:
First, minimum degree is one of the most fundamental characteristics of a graph. For instance, it is adopted for describing the evolution of random graphs and graph visualization~\cite{local2014}.
Second, it is often used to measure the cohesiveness of user groups in social media. In~\cite{md1983}, Seidman et al. compared the minimum degree with many other metrics of cohesiveness (e.g., connectedness and diameter) and found that the minimum degree is indeed a good metric for social network analysis.
Third, for community search tasks, Sozio et al.~\cite{KDD2010} also showed that it is better than some other metrics, including the average degree and density.
In the following, we assume that the minimum degree metric is adopted in $f$.

To solve Problem~\ref{prob:kcoreGlobal}, there are two online algorithms, which are based on global and local search~\cite{KDD2010,local2014} respectively, and one index-based algorithm~\cite{barbieri2015efficient}.

\noindent\textbf{$\bullet$ A global search algorithm.}
Sozio et al.~\cite{KDD2010} proposed a greedy algorithm, which follows the peeling framework \cite{charikar2000greedy} of computing the densest subgraphs~\cite{goldberg1984finding} and removes vertices iteratively. Specifically, let $G_0$=$G$ and $G_t$ be the graph in $t$-th iteration ($1\leq t\textless n$). At the $t$-th ($1\leq t \textless n$) step, it removes the vertex which has the minimum degree in $G_{t-1}$ and obtain an updated graph $G_t$. The above operation iterates and stops at the $T$-th step, if either (1) at least one of the query vertices $Q$ has minimum degree in the graph $G_{T-1}$, or (2) the query vertices $Q$ are no longer connected. Let $G_t'$ be the connected component containing $Q$ in $G_t$. Then, the subgraph $G_O$=$\arg\max\{f(G_t')\}$ satisfies all the constraints in Problem~\ref{prob:kcoreGlobal}.

We denote the algorithm above by {\tt Global}, as it finds the community in a global manner. By using some special optimization techniques~\cite{KDD2010,charikar2000greedy}, {\tt Global} is able to achieve linear time and space complexities, i.e., $O(n+m)$. Note that the function $f(H)$ above can be generalized to any monotone function, and the corresponding problem can also be solved by {\tt Global}~\cite{KDD2010}.

It is easy to observe that since {\tt Global} peels all the vertices with low degrees, the subgraph returned is the largest connected subgraph, in which each vertex has at least $k$ neighbors. As a result, the returned subgraph is a connected $k$-core containing $Q$, where $k$ equals to the minimum core number of vertices in $Q$.

\noindent\textbf{$\bullet$ A local search algorithm.}
According to Problem~\ref{prob:kcoreGlobal}, there may exist some subgraphs of $G_O$, which satisfy all the constraints and achieve the same value on the function $f$, but have smaller sizes. Thus, they can be considered the communities as well.

\begin{example}
\label{eg:kcoreGlobal}
Let the graph be the one in Fig. \ref{fig:kcoreEg}(a), $Q$=$\{E\}$. {\tt Global} will return the subgraph of vertices $\{A,B,C,D,E\}$ as the community, and the value of function $f$ is 2.
However, there are other three subgraphs, whose vertex sets are $\{A,B,C,E\}$, $\{A,B,D,E\}$, and $\{A,B,E\}$,
also satisfy the constraints of Problem~\ref{prob:kcoreGlobal}, and their values on $f$ are 2. Thus, they can be considered as communities.
\end{example}

In \cite{local2014}, Cui et al. proposed a local CS method, denoted by {\tt Local}, which works in a local expansion manner and finds a community that may have smaller size than that of {\tt Global}. Specifically, it assumes that there is only one query vertex $q$ (i.e., $Q$=$\{q\}$). {\tt Local} consists of three steps: First, it expands the search space from $q$. Second, it generates a candidate vertex set $C$ in the search space. Third, it finds the community from $C$.

The key step is the second step, which works in an iterative manner. In each iteration, it selects the vertex that is the local optimal and adds it into the candidate set $C$. To decide the local optimal vertex, some heuristic criteria are adopted. One typical criterion is to select the vertex that leads to the largest increment of the function $f$; another one is to select the vertex which has the largest number of connections to vertices of the candidate set.
The iterations stop when the candidate set $C$ theoretically guarantees that it contains a community satisfying the constraints of Problem~\ref{prob:kcoreGlobal}.

Let $H$ and $H'$ denote the communities returned by {\tt Global} and {\tt Local} respectively. Then, we have $f(H')=f(H)$ and $H'\subseteq H$. Besides, since in the worst case the candidate set $C$ could be the same as vertex set $V$, the time complexity of {\tt Local} is the same as that of {\tt Global}, but in practice for large graphs, the candidate set is often much smaller than the entire graph, and thus {\tt Local} achieves higher efficiency.

\noindent\textbf{$\bullet$ An index-based algorithm.}
In~\cite{barbieri2015efficient}, Barbieri et al. proposed an index structure, called {\tt ShellStruct}, which organizes all the connected $k$-cores in an offline manner. Based on {\tt ShellStruct}, Problem~\ref{prob:kcoreGlobal} can be answered in optimal time cost, i.e., $O(|H_V|)$, where $H_V$ is the set of vertices in the returned community and it is the same with that of {\tt Global}.

The index is built based on the key observation that cores are nested. That is, for any integer $0\textless k\leq k_{\max}$, the $k$-core is contained by the ($k$--1)-core, where $k_{\max}$ is the maximum core number. {\tt ShellStruct} is a tree-like structure with $k_{\max}$ levels. The root of the tree corresponds to the 1-core, and the $k$-th level keeps track of the information about the $k$-th core. In $k$-th level, each tree node, $p_k$, corresponds to a connected component $C_k$ of the $k$-core, and it keeps:
\begin{enumerate}
  \item the set of ``children" nodes, each of which corresponds to a connected component that is in the ($k$+1)-core and contained by $C_k$;
  \item the set of vertices in $C_k$ but not in ($k$+1)-core.
\end{enumerate}

It is easy to observe that in {\tt ShellStruct}, all the connected $k$-cores are well organized. The space cost is exactly $O(n)$ because each vertex appears only once. To build the index, Barbieri et al. proposed an index construction algorithm, which builds the tree level by level, starting from the root level. As a result, its time complexity is $O(n\cdot k_{\max} +m)$. We remark that a more efficient algorithm for building the same index is proposed in~\cite{Fang:VLDB:2016}, which takes $O(m\cdot\alpha(n))$ time, where $\alpha(n)$ is the inverse Ackermann function and it is less than 5 for all remotely practical values of $n$.

Based on {\tt ShellStruct}, a query algorithm is proposed. Specifically, it starts from the $l$-th level where $l$ is the maximum core number of vertices in $Q$ and checks its upper levels, until there is a connected component containing all the query vertices. By using the lowest-common-ancestor (LCA) data structure~\cite{LCA1983}, the time cost of the query algorithm can be reduced to $O(|H_V|)$.

In Problem~\ref{prob:kcoreGlobal}, the cohesiveness function is required to be maximized. However, for some applications, such as infectious disease control discussed in Section~\ref{sec:intro}, this constraint may need to be relaxed so that vertices which have less connections with the query vertices can also be involved. Motivated by this, a variant of Problem~\ref{prob:kcoreGlobal} is also studied in the literature~\cite{local2014}:

\begin{problem}
\label{prob:kcoreLocal}
Given an undirected simple graph $G(V,E)$, a query vertex $q\in V$, and a non-negative integer $k$, return a subgraph $H(V_H,E_H)$ of $G$, such that
\begin{enumerate}
  \item $V_H$ contains $q$;
  \item $H$ is connected;
  \item for each vertex $v\in H$, $deg_H(v)\geq k$.
\end{enumerate}
\end{problem}

In Fig. \ref{fig:kcoreEg}(a), let $q$=$A$ and $k$=2. Then, the subgraph of $\{A,B,C,D,E\}$ satisfies all the constraints, and thus is a community for Problem~\ref{prob:kcoreLocal}. Note that if we maximize the minimum degree as required by Problem~\ref{prob:kcoreGlobal}, we will return a smaller subgraph, i.e., $\{A,B,C,D\}$, since the minimum degree is 3.
The algorithms {\tt Global} and {\tt Local} can be easily adapted for answering the query of Problem~\ref{prob:kcoreLocal}. For details, please refer to~\cite{local2014}.

\subsubsection{Size-Bounded Community Search}
\label{sec:sizeBounded}

One drawback of Problem~\ref{prob:kcoreGlobal} is that the returned subgraph may contain a large number of vertices. Notice that although {\tt Local} may find communities which are smaller than those of {\tt Global}, it does not have any guarantee on the sizes of the returned communities, which implies that the returned communities may still have very large sizes.

For many real applications, such as holding a cocktail part, they often require the size of the output community is less than a pre-specified upper bound. Thus, it is desirable to search communities with bounded-size. By imposing the size constraint, we obtain another problem:

\begin{problem}
\label{prob:kcoreGlobalSize}
Given an undirected simple graph $G(V,E)$, a set of query vertices $Q\subseteq V$, a size constraint $k$, and a goodness function $f$, return a subgraph $H(V_H,E_H)$ of $G$, such that
\begin{enumerate}
  \item $V_H$ contains $Q$;
  \item $H$ is connected;
  \item $|V_H|\leq k$ ($H$ has at most $k$ vertices);
  \item $f(H)$ is maximized among all feasible choices for $H$.
\end{enumerate}
\end{problem}

Unfortunately, due to the size constraint, Problem~\ref{prob:kcoreGlobalSize} is NP-hard~\cite{KDD2010}.
This implies that an exact algorithm for solving Problem~\ref{prob:kcoreGlobalSize} will take exponential time cost, and thus it is impractical for large graphs. To alleviate the computational issue, some heuristic algorithms are developed~\cite{KDD2010}, and they are able to achieve reasonable efficiency, although they do not have any provable quality guarantee.

To further reduce the size of the returned community, Barbieri et al.~\cite{barbieri2015efficient} proposed the {\it minimum community search problem}, which aims to find a community that satisfies all the constraints of Problem~\ref{prob:kcoreGlobal} and has the minimum number of vertices.

\begin{problem}
\label{prob:kcoreMinSize}
Given an undirected simple graph $G(V,E)$, a set of query vertices $Q\subseteq V$, and a minimum degree based function $f$, let $H^*$ be the subgraph returned by {\tt Global}. Find a subgraph $H$ of $G$, such that
\begin{enumerate}
  \item $V_H$ contains $Q$;
  \item $H$ is connected;
  \item $f(H)$=$f(H^*)$;
  \item the size of $H$ is the smallest.
\end{enumerate}
\end{problem}

Similar to Problem~\ref{prob:kcoreGlobalSize}, Problem~\ref{prob:kcoreMinSize} is also NP-hard. It can be proved by a reduction from the \textsc{Steiner Tree} problem: given a graph $G(V,E)$ and a set of terminal vertices $T\subseteq V$, find a connected subgraph $G'$ of $G$ such that it contains all the terminal vertices and has the minimum number of edges.
Note that the most efficient algorithm~\cite{kou1981fast} of \textsc{Steiner Tree} problem achieves an approximation ratio of (2-2/$|Q|$), and takes linear time cost by the Mehlhorn's implementation~\cite{mehlhorn1988}.

To answer the query in Problem~\ref{prob:kcoreMinSize}, Barbieri et al.~\cite{barbieri2015efficient} proposed an algorithm, and it consists of two steps:
First, it reduces the size of $H^*$ as much as possible using some local greedy search. Note that after the reduction, the subgraph $H^*$ is still a qualified community of Problem~\ref{prob:kcoreGlobal}, but may have much smaller size.
Second, it finds a subgraph from $H^*$ by adopting the above approximation algorithm for the \textsc{Steiner Tree} problem.

\noindent\textbf{Remark.} Some other factors, such as distances among vertices~\cite{KDD2010} and local distance dynamics~\cite{khop2017,khop2018}, have also been considered for CS on simple graphs. Due to the space limitation, we skip the details.

\subsection{Directed Graphs}
\label{sec:kcoreDirected}

\begin{figure}[]
\centering
\begin{tabular}{c c}
  \begin{minipage}{3.7cm}
	\includegraphics[width=3.7cm]{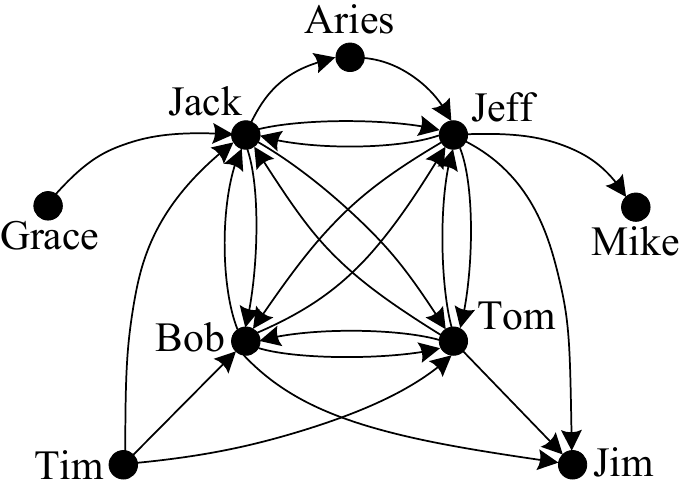}
  \end{minipage}
  &
  \begin{minipage}{3.8cm}
	\includegraphics[width=3.8cm]{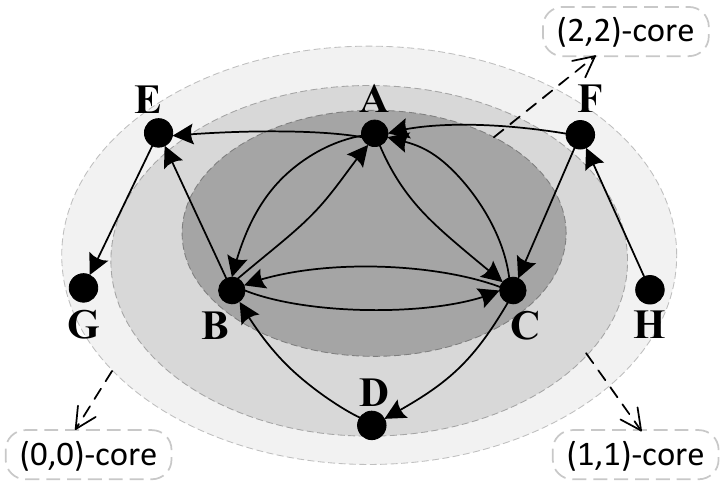}
  \end{minipage}
  \\
  (a) a directed graph
  &
  (b) illustrating D-cores
\end{tabular}
\caption{Two directed graphs~\cite{Fang:TKDE:CSD}.}
\label{fig:CSDEg}
\end{figure}

A directed graph is a graph $G(V,E)$, which contains a set of vertices $V$ and a set of directed edges $E$.
The in-degree and out-degree of a vertex $v$ in $G$, denoted by $deg_G^{in}(v)$ and  $deg_G^{out}(v)$, are the number of its in-neighbors and out-neighbors, respectively. The minimum in-degree and out-degree of the graph $G$ are denoted by $\delta_{in}(G)$ and $\delta_{out}(G)$ respectively.
Fig. \ref{fig:CSDEg}(a) depicts a directed graph with nine users.

A straightforward method of performing CS on directed graph is to ignore the directions and then use the method {\tt Global} in Section \ref{sec:sizeUnbounded} to find the community. In Fig. \ref{fig:CSDEg}(a), if we let $q$=Jack, then we will find a community with members \{{\tt Jack}, {\tt Jeff}, {\tt Bob}, {\tt Tom}, {\tt Tim}, {\tt Jim}\}. However, {\tt Tim} has no in-neighbors and {\tt Jim} has no out-neighbors in the community, which implies their interactions with other members are quite weak.

In~\cite{Fang:TKDE:CSD}, Fang et al. extended the minimum degree measure for directed graphs, and study the problem of \underline{C}ommunity \underline{S}earch on \underline{D}irected graph (or CSD problem), based on the D-core, also called ($k$, $l$)-core \cite{Dcore2014}. 

\begin{definition}[($k$, $l$)-core~\cite{Dcore2014}]
\label{def:Dcore}
Given a directed graph $G(V,E)$ and two non-negative integers $k$ and $l$, the $(k,l)$-core is the maximum subgraph $C$ of $G$ such that $\delta_{in}(C)\ge k$ and $\delta_{out}(C) \ge l$ .
\end{definition}

\begin{problem}[CSD]
\label{prob:CSD}
Given a directed graph $G(V,E)$, two positive integers $k$ and $l$, and a query vertex $q$, return a connected subgraph $G_q\subseteq G$, such that it contains $q$ and $\forall v\in G_q$, $\delta_{in}(G_q)\geq k$ and $\delta_{out}(G_q)\geq l$.
\end{problem}

Fig. \ref{fig:CSDEg}(b) shows a directed graph with its D-cores. Let $q$=$B$, $k$=2, and $l$=2. Then, the subgraph of \{$A$, $B$, $C$\} is the returned community for $B$.

Similar to {\tt Global}, a simple solution to the CSD problem is to peel vertices iteratively until each remaining vertex satisfies the in-degree and out-degree constraints. As a result, its time complexity is $O(m+n)$, which may be inefficient for large graphs.
To improve efficiency, Fang et al.~\cite{Fang:TKDE:CSD} proposed an index-based method. Specifically, it first performs D-core decomposition (i.e., computing all the ($k$, $l$)-cores), then organizes these cores in an index with a 2-dimensional table, and finally answers queries using the index.

To keep all D-cores, a simple method takes $O(n^3)$ space since $k$, $l$$\leq$$n$--1 and each D-core takes $O(n)$ space. To alleviate this issue, three methods are proposed. For ease of exposition, let $V_{i,j}$ denote the set of vertices in ($i$, $j$)-core.
The first one exploits the nested property of D-cores, i.e., for any $l\geq0$, we have ($k$, $l$+1)-core $\subseteq$ ($k$, $l$)-core, so if ($k$, $l$+1)-core has been kept, we only need to keep vertices $V_{k,l}\backslash V_{k,l-1}$ for the ($k$, $l$)-core. As a result, for any $k$, it takes $O(n)$ space to keep all ($k$, $l$)-cores (0$\leq{l}$$\leq$$n$), so the overall space cost is $O(m)$.

The second method relies on a key observation that for any $k,l\geq0$, we have both ($k$+1, $l$)-core $\subseteq$ ($k$, $l$)-core and ($k$, $l$+1)-core $\subseteq$ ($k$, $l$)-core. After keeping ($k$, $l$+1)-core and ($k$+1, $l$)-core, for ($k$, $l$)-core, if $|V_{k+1,l}|\geq|V_{k,l+1}|$, we only keep $V_{k,l}\backslash V_{k+1,l}$; otherwise, we keep $V_{k,l}\backslash V_{k,l+1}$. Thus, it takes less space than the first method. For the third method, after keeping ($k$, $l$+1)-core and ($k$+1, $l$)-core, it only keeps vertices $V_{k,l}\backslash(V_{k+1,l}\cup V_{k,l+1})$ for the ($k$, $l$)-core and takes the least space cost.

In addition, although the community $G_q$ of a CSD query is a connected subgraph, it may not be a strongly connected component (SCC)~\cite{hopcroft1983data} (i.e., each vertex of the SCC is reachable from each other vertex). To tackle this issue, a variant of the CSD problem is to find a community, which not only satisfies the minimum degree constraints, but also is an SCC. The CSD algorithms can be extended for solving this variant \cite{Fang:TKDE:CSD}.

\subsection{Keyword-Based Attributed Graphs}
\label{sec:kcoreKeyword}

\begin{figure}[]
\centering
\begin{tabular}{c c}
  \begin{minipage}{4.0cm}
	\includegraphics[width=4.0cm]{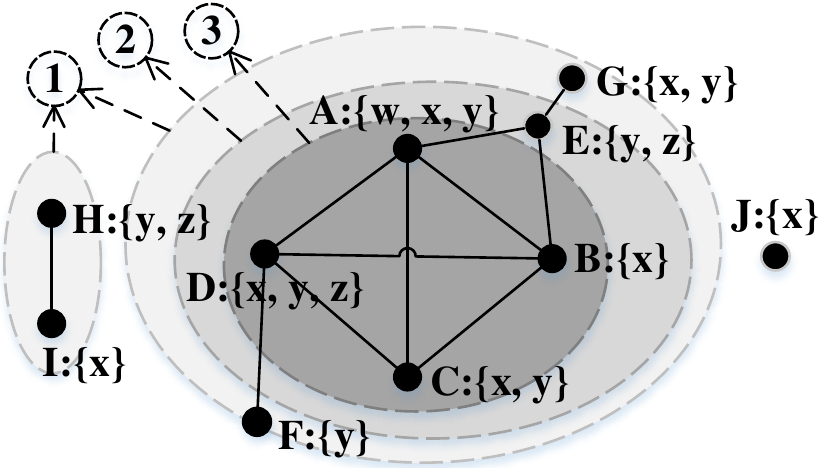}
  \end{minipage}
  &
  \begin{minipage}{3.8cm}
	\includegraphics[width=3.379cm]{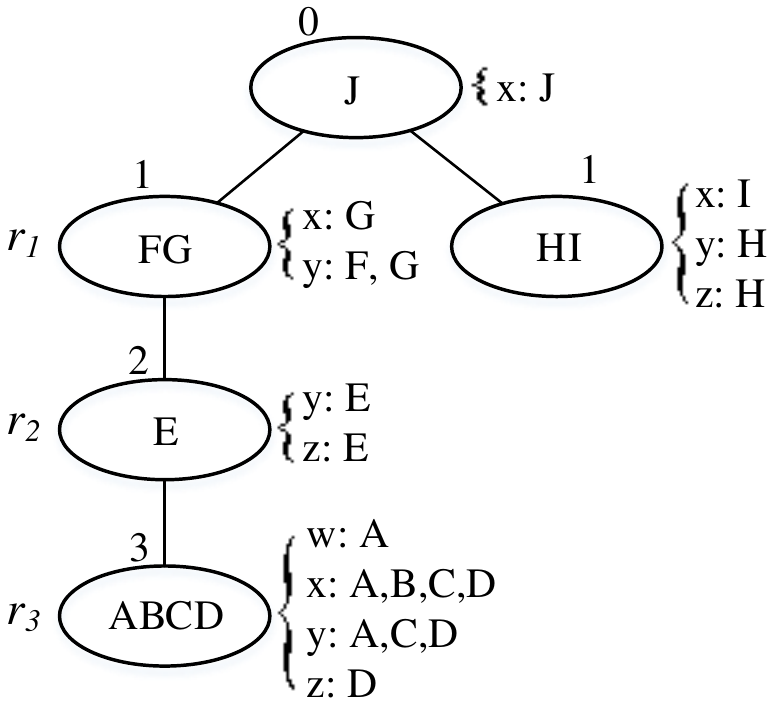}
  \end{minipage}
  \\
  (a) a graph $G$
  &
  (b) CL-tree index of $G$
\end{tabular}
\caption{An example for illustrating ACQ \cite{Fang:VLDB:2016}.}
\label{fig:kcoreACQEg}
\end{figure}

A keyword-based attributed graph is an undirected graph $G(V,E)$, with vertex set $V$ and edge set $E$. Each vertex $v \in V$ is associated with a set of keywords, $W(v)$. The keyword-based attributed graphs are prevalent in social media, bibliographical networks, and knowledge bases. In Fig. \ref{fig:kcoreACQEg}(a), a keyword-based attributed graph is depicted. For example, vertex $A$ has a set of keywords $\{w,x,y\}$.
In \cite{Fang:VLDB:2016,Fang:VLDBJ:2017,fang2017workshop,Shang2017}, CS on keyword-based attributed graphs has been studied extensively.

\begin{problem}[ACQ \cite{Fang:VLDB:2016}]
\label{prob:acq}
Given a keyword-based attributed graph $G(V,E)$, a positive integer $k$, a vertex $q \in V$ and a set of keywords $S\subseteq W(q)$, return a set $\mathcal {G}$ of subgraphs of $G$, such that $\forall G_q \in \mathcal {G}$, the following properties hold:
\begin{enumerate}
  \item \textbf{Connectivity}. $G_q$ is connected and contains $q$;
  \item \textbf{Structure cohesiveness}. $\forall v\in G_q$, $deg_{G_q}(v)\geq k$;
  \item \textbf{Keyword cohesiveness}. The size of $L(G_q, S)$ is maximal, where $L(G_q, S)=\cap_{v \in G_q}(W(v)\cap S)$ is the set of keywords shared in $S$ by all vertices of $G_q$.
\end{enumerate}
\end{problem}

For example, in Fig. \ref{fig:kcoreACQEg}(a), if $q$=$A$, $k$=2 and $S$=$\{w$, $x,y\}$, then the output of Problem~\ref{prob:acq} is the subgraph of $\{A,C,D\}$, with a shared keyword set $\{x,y\}$, meaning that these vertices share the keywords $x$ and $y$.

The subgraph $G_q$ is called an {\it attributed community} (or AC) of $q$, and $L(G_q, S)$ is the {\it AC-label} of $G_q$. In Problem~\ref{prob:acq}, the first two properties ensure the structure cohesiveness.
Property 3 enables the retrieval of communities whose vertices have common keywords in $S$.
It requires $L(G_q, S)$ to be maximal, because it aims to find the AC(s) only containing the most related vertices, in terms of the number of common keywords. In Fig. \ref{fig:kcoreACQEg}(a), if we use the same query
($q$=$A$, $k$=2, $S$= $\{w,x,y\}$),
without the ``maximal'' requirement, we can obtain communities such as $\{A,B,E\}$ (which share no keywords), $\{A,B,D\}$, or $\{A,B,C\}$ (which share 1 keyword). Note that there does not exist an AC with AC-label being exactly $\{w$, $x,y\}$.

Two outstanding features of ACQ are as follows:
(1) {\it Ease of interpretation.}
An AC contains tightly-connected vertices with similar contexts or backgrounds. Thus, an ACQ user can focus on the common keywords or features of these vertices, i.e., the AC-labels facilitate understanding of the vertices that form the AC.
(2) {\it Personalization.}  The user of an ACQ can control the semantics of the AC, by specifying a set of $S$ of keywords. Intuitively, $S$ decides the meaning of the AC based on the user's need.

The ACQ problem is challenging. A simple method to answer an ACQ runs three steps. First, all non-empty subsets of
$S$, $S_1,S_2,\cdots$, $S_{2^l-1}$ ($l$=$|S|$),
are enumerated. Then, for each subset $S_i$(1$\leq i\leq2^l-$1), it checkes whether there is a subgraph which satisfies the first two properties. Finally, it outputs the subgraphs having the most shared keywords. However, since there are exponential number of subsets, it is impractical for large graphs. To alleviate this issue, the authors observed the {\it anti-monotonicity} property, which states that given a set $S$ of keywords, if it appears in every vertex of an AC, then for every subset $S'$ of $S$, there exists an AC in which every vertex contains $S'$. Based on this property, many subsets of $S$ can be pruned, and thus faster online query algorithms can be developed.

An index, called {\tt CL-tree}, is proposed for organizing the vertex keyword data in a hierarchical structure. The CL-tree has the same tree structure as {\tt ShellStruct} (see Section~\ref{sec:sizeUnbounded}), but for each node $p$, it maintains an additional inverted list such that for each keyword $e$ that appears in the vertices of $p$, a list of IDs of vertices which contain $e$ is stored.
Since each graph vertex and each keyword appear only once, the space cost of keeping such an index is $O({\widehat l}\cdot n)$, where $\widehat l$ denotes the average size of $W(v)$ over $V$. As a result, the space cost is linear to the size of $G$. As shown in~\cite{Fang:VLDB:2016}, the {\tt CL-tree} structure can be built level by level in a bottom-up manner and it takes  linear time cost, i.e., $O(m\cdot\alpha(n))$. In addition, index maintenance algorithms for the {\tt CL-tree} are developed \cite{Fang:VLDBJ:2017}. Fig. \ref{fig:kcoreACQEg}(b) presents the {\tt CL-tree} index for the graph in Fig. \ref{fig:kcoreACQEg}(a).

Based on the {\tt CL-tree}, two incremental algorithms (from examining smaller candidate keyword sets to larger ones) and one decremental algorithm (from examining larger candidate keyword sets to smaller ones) are developed. For each candidate keyword set, they check whether there is a connected $k$-core containing $q$, and finally return the one with largest keyword set.

\subsection{Location-Based Attributed Graphs}
\label{sec:kcoreLocation}

A location-based attributed graph, also called geo-social network, is an undirected graph $G(V,E)$ with vertex set $V$ and edge set $E$. For each vertex $v\in V$, it has a location position $(v.x, v.y)$, where $v.x$ and $v.y$ denote its positions along $x$- and $y$-axis in a two-dimensional space. Geo-social networks widely exist in many location-based services, including Twitter, Facebook, and Foursquare \cite{armenatzoglou2013general,fang2014detecting,fang2016scalable}.
In Fig. \ref{fig:SACEg}(a), a geo-social network with ten vertices is depicted.

\begin{table}[h]
  \centering
  \caption {CS works on geo-social networks.}
  \label{tab:CSLocation}
  \begin{tabular}{c|l}
    \hline \textbf{CS query} & \textbf{Spatial cohesiveness}\\
    \hline\hline
        SAC search \cite{Fang:VLDB:2017,Fang:TKDE:SAC} & smallest minimum covering circle\\
    \hline
        RB-$k$-core search \cite{wangkai2018}   & radius-fixed covering circle\\
    \hline
        GSGQ \cite{zhu2017geo}           & rectangle, center-fixed circles\\
    \hline
  \end{tabular}
\end{table}

Three kinds of CS queries have been studied on geo-social networks, namely {\it spatial-aware community (SAC) search} \cite{Fang:VLDB:2017}, {\it radius-bounded $k$-core (RB-$k$-core) search} \cite{wangkai2018}, and {\it geo-social group queries with minimum acquaintance constraint (GSGQ)} \cite{zhu2017geo}.
Generally, they all require that the communities are structurally and spatially cohesive. For structure cohesiveness, they all adopt the $k$-core model, but for spatial cohesiveness, they use different constraints, as outlined in Table~\ref{tab:CSLocation}. In SAC search, the community is in the smallest minimum covering circle (MCC); in RB-$k$-core search, the community is in a circle with radius less than an input threshold; in GSGQ, the community is in a given rectangle or circle centered at the query vertex.

\subsubsection{Spatial-Aware Community (SAC) Search}
\label{sec:sac}

The MCC and SAC search are defined as follows. Note that the notion of MCC has been widely adopted to describe a set of spatially compact objects \cite{mcc-ts,mcc-sigmod}.

\begin{definition}[MCC]
\label{def:mcc}
Given a set $S$ of vertices with locations, the MCC of $S$ is the spatial circle, which contains all the vertices in $S$ with the smallest radius.
\end{definition}

\begin{problem}[SAC search]
\label{prob:sac}
Given a geo-social network $G(V,E)$, a positive integer $k$ and a vertex $q\in V$, return a subgraph $G_q\subseteq G$, and the following properties hold:
\begin{enumerate}
  \item \textbf{Connectivity}. $G_q$ is connected and contains $q$;
  \item \textbf{Structure cohesiveness}. $\forall v\in G_q$, $deg_{G_q}(v)\geq k$;
  \item \textbf{Spatial cohesiveness}. The MCC of vertices in $G_q$ satisfying Properties 1 and 2 has the smallest radius.
\end{enumerate}
\end{problem}

\begin{figure}[]
\centering
\begin{tabular}{c c}
  \begin{minipage}{3.8cm}
	\includegraphics[width=3.8cm]{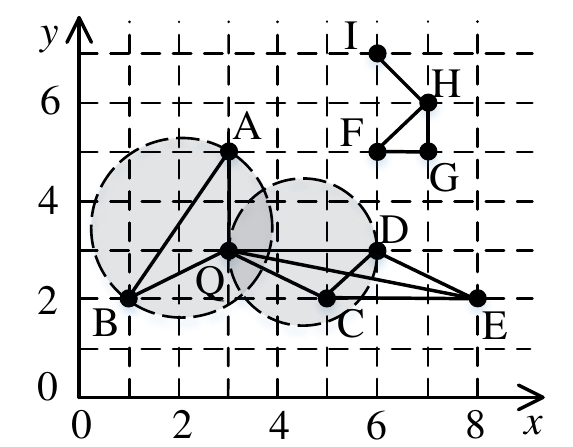}
  \end{minipage}
  &
  \begin{minipage}{3.8cm}
	\includegraphics[width=3.8cm]{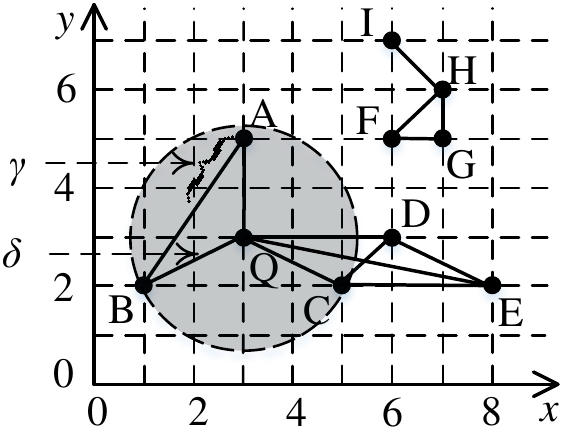}
  \end{minipage}
  \\
  (a) a graph
  &
  (b) Illustrating {\tt AppInc}
\end{tabular}
\caption{Illustrating SAC search \cite{Fang:VLDB:2017}.}
\label{fig:SACEg}
\end{figure}

A subgraph satisfying properties 1 and 2 is a~\emph{feasible} solution, and the subgraph satisfying all the three properties is the \emph{optimal} solution (denoted by $\Psi$). The radius of the MCC containing $\Psi$ is denoted by $r_{opt}$.
In Fig. \ref{fig:SACEg}(a), the two circles denote the MCCs of $C_1$=$\{Q, C, D\}$ and $C_2$=$\{Q, A, B\}$. Let $q$=$Q$ and $k$=2. Then, $\Psi$ contains vertex set $C_1$ with $r_{opt}$=1.5.

The SAC search problem is challenging. A basic exact approach takes $O(m\times{n^3})$ time, which relies on an observation that a spatial circle can be determined by three points on its boundary~\cite{mcc-ts}. This implies, we can enumerate all the three-vertex combinations, and for each combination we find a connected $k$-core in its circle, and finally get $\Psi$. This approach, however, is impractical for large graphs due to its high complexity.

To improve efficiency, the authors resorted to approximation algorithms. The first one, called {\tt AppInc}, returns the feasible solution in a circle $O(q,\delta)$ which centers at $q$ and has the smallest radius $\delta$, and it has an approximation ratio of 2. Here, the approximation ratio is defined as the ratio of the radius of MCC returned over $r_{opt}$. In Fig. \ref{fig:SACEg}(b), let $q$=$Q$ and $k$=2. Then, {\tt AppInc} returns the subgraph of $\{A,B,Q\}$.

The circle $O(q,\delta)$ can also be approximated by performing binary search on the radius $\delta$. As a result, we can get another approximation solution with ratio of (2+$\epsilon_F$), where $\epsilon_F\geq0$ is an input parameter. To achieve an approximation ratio of (1+$\epsilon_A$) where $0\textless\epsilon_A\textless1$, the authors developed another algorithm, called {\tt AppAcc}. It first locates the area containing the center of the circle of $\Psi$, then approximates the center by splitting the area into small grids, and finally finds an approximation solution by using these grids. Overall, these approximation algorithms guarantee that the radius of the MCC of $\Psi$ has an arbitrary expected approximation ratio. Based on {\tt AppAcc}, an advanced exact algorithm is developed. An interesting observation is that there is a trade-off between the quality of results and efficiency, i.e., algorithms with lower approximation ratios tend to have higher complexities. In addition, the SACs can be returned in a continuous manner, as shown in \cite{Fang:TKDE:SAC}.

\subsubsection{Radius-Bounded $k$-core Search}
\label{sec:rbkcore}

Problem~\ref{prob:rbkcore} defines the radius-bounded $k$-core search.

\begin{problem}[RB-$k$-core search]
\label{prob:rbkcore}
Given a geo-social network $G(V,E)$, a positive integer $k$, a radius $r$ and a vertex $q\in V$, return all the subgraphs $G_q\subseteq G$, and the following properties hold:
\begin{enumerate}
  \item \textbf{Connectivity}. $G_q$ is connected and contains $q$;
  \item \textbf{Structure cohesiveness}. $\forall v\in G_q$, $deg_{G_q}(v)\geq k$;
  \item \textbf{Spatial cohesiveness}. The MCC of vertices in $G_q$ has a radius $r' \leq r$;
  \item \textbf{Maximality constraint}. There exists no other subgraph $G_q'$ satisfying properties above and $G_q\subset G_q'$.
\end{enumerate}
\end{problem}

Similar to SAC search, it adopts the MCC, but imposes a constraint on its radius. To solve Problem \ref{prob:rbkcore}, Wang et al. proposed three algorithms. The first one, denoted by {\tt TriV}, is a triple-vertex-based algorithm, which is also based on the observation that a spatial circle can be determined by three points on its boundary~\cite{mcc-ts}. It proposes to generate all the candidate circles containing $q$ at first and then compute the maximum $k$-core for the subgraphs contained in the candidate circles with radius $r' \leq r$. The time complexity of {\tt TriV} is $O(mn^3)$, since there are $O(n^3)$ candidate circles in the worst case and each circle needs $O(m)$ time to verify.

To reduce the number of candidate circles, a binary-vertex-based algorithm {\tt BinV} is proposed. In {\tt BinV}, only the circles with radius $r'$=$r$ are generated and for each candidate circle, its arc passes a pair of vertices in $G$. In this manner, for each pair of vertices, at most two circles are generated. As a result, it reduces the number of candidate circles from $O(n^3)$ to $O(n^2)$.

To further improve the efficiency, a rotating-circle-based algorithm {\tt RotC} is proposed to reuse the intermediate computation results in the process of finding RB-$k$-cores. Fixing each vertex $v \in V$ as a pole, {\tt RotC} generates the candidate circles in a rotating way so that the computation cost can be shared among the adjacent circles. In addition, the authors also proposed several pruning techniques to early terminate the processing of invalid candidate circles.

\subsubsection{Geo-Social Group Queries with Minimum Acquaintance Constraint (GSGQs)}
\label{sec:gsgq}

The GSGQ is defined formally as follows:

\begin{problem}[GSGQ]
\label{prob:gsgq}
Given a geo-social network $G(V$, $E)$, a vertex $q\in V$, a positive integer $k$ and a spatial constraint $\Lambda$, return a subgraph $G_q\subseteq G$, and the following properties hold:
\begin{enumerate}
  \item \textbf{Connectivity}. $G_q$ is connected and contains $q$;
  \item \textbf{Structure cohesiveness}. $\forall v\in G_q$, $deg_{G_q}(v)\geq k$;
  \item \textbf{Spatial cohesiveness}. $G_q$ satisfies constraint $\Lambda$.
  \item \textbf{Maximality constraint}. There exists no other subgraph $G_q'$ satisfying properties above and $G_q\subset G_q'$.
\end{enumerate}
\end{problem}

In Problem~\ref{prob:gsgq}, for spatial constraint $\Lambda$, Zhu et al. \cite{zhu2017geo} considered three kinds of constraints:
\begin{enumerate}
  \item $\Lambda$ is a spatial rectangle for containing $G_q$;
  \item $\Lambda$ is a circle centered at $q$ with radius less than the distance from $q$ to its $k$-th nearest vertex in $G_q$ ($G_q$ may contain more than $k$+1 vertices);
  \item $\Lambda$ is a circle satisfying Constraint 2 and $G_q$ contains exactly $k$+1 vertices.
\end{enumerate}

By using an R-tree index \cite{guttman1984r}, a GSGQ with the first constraint can be answered in $O(n+m)$ time; when the second constraint is imposed, a GSGQ can be solved in $O(n(n+m))$ time; when the third constraint is applied, a GSGQ takes $O(C_k^{n-1}(m+n))$ time.

To improve efficiency, they proposed the social-aware R-trees (or SaR-tree) index, which incorporates both vertices' spatial locations and social relations. It is built based on the concept of core bounding rectangle (CBR), which projects the minimum degree constraint on the spatial layer. Specifically, the CBR of a vertex $v$ is a rectangle containing $v$, inside which any vertex group with $v$ does not satisfy the minimum degree constraint.

Unlink classical R-tree, each entry of an SaR-tree refers to two pieces of information, i.e., a set of CBRs and a minimum bounding rectangle (MBR). Perceptually, a CBR bounds a group of vertices from the social perspective, while an MBR bounds vertices from the spatial perspective. As such, SaR-tree gains power for both social-based and spatial-based pruning. In addition, they developed a variant of SaR-tree, called SaR*-tree, which optimizes the group of spatial objects to minimize the disk I/O cost. Based on these indexes, they developed efficient algorithms for answering GSGQs with different spatial constraints.

\subsection{Temporal Graphs}
\label{sec:kcoreTemporal}

\begin{figure}[ht]
\begin{center}
\includegraphics[width=0.9\columnwidth]{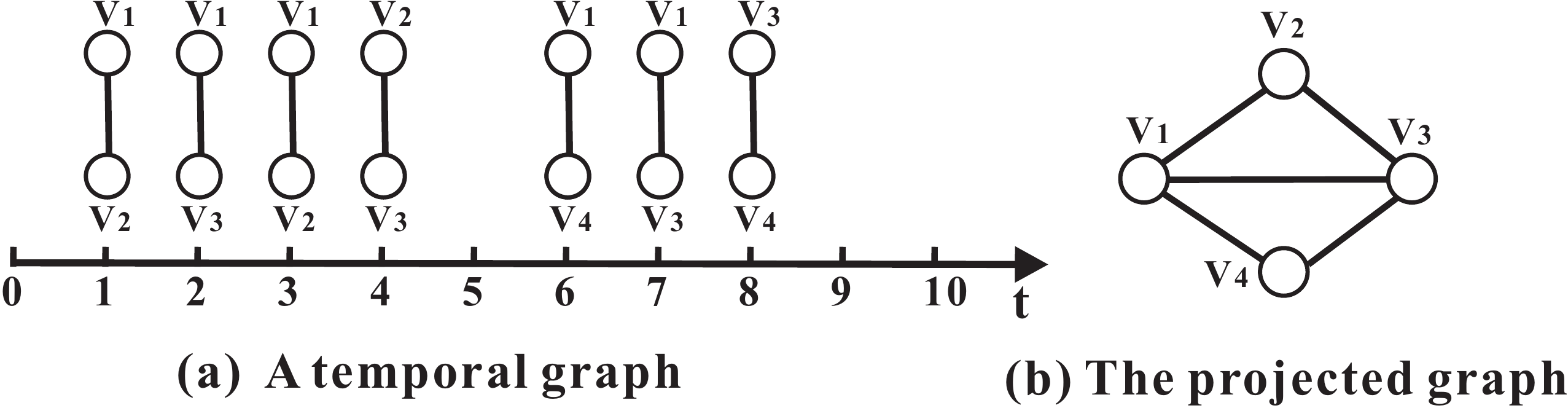}
\caption{A temporal graph and the projected graph \cite{Li:ICDE:2018}.}
\label{fig:lu-temporal}
\end{center}
\end{figure}

Li et al. \cite{Li:ICDE:2018} studied the persistent community search problem in a temporal graph.  A temporal graph is an undirected graph $G(V,E)$ with vertex set $V$ and edge set $E$. Each edge $e\in E$ is a triplet $(u,v,t)$ where $u$, $v$ are vertices in $V$, and $t$ is the interaction time between $u$ and $v$. For a temporal graph $G$, the projected graph denoted by $G_p$ over the time interval $[t_s, t_e]$ is defined as $G_p=(V, E, [t_s, t_e])$, where $V=V(G)$ and $E=\{(u,v)|(u, v, t)\in E(G), t \in [t_s, t_e]\}$. Fig.~\ref{fig:lu-temporal} (b) illustrates the  projected graph of the temporal graph in Fig.~\ref{fig:lu-temporal} (a) over the interval $[1, 8]$.

\begin{definition}
  \label{def:maximalcoreinterval} \textbf{(Maximal $(\theta, k)$-persistent-core interval)}  Given a temporal graph ${G}=({V}, {E})$ and parameters $\theta>0$ and $k>0$, an interval $[{t_s}, {t_e}]$ with ${t_e}-{t_s} \ge \theta$ is called a maximal $(\theta, k)$-persistent-core interval for $G$ if and only if the following two conditions hold. (1) For any $t \in [t_s, t_e-\theta]$, the projected graph of $G$ over the interval $[t, t+\theta]$ is a connected $k$-core subgraph. (2) There is no super-interval of $[t_s, t_e]$ such that (1) holds.
\end{definition}

\begin{definition} [Core persistence]
\label{def:corepersistence}
Let $T$ $=$ $\{[t_{s_1},$ $t_{e_1}],$ $\cdots,$ $[t_{s_r},$ $t_{e_r}]\}$ be the set of all  maximal $(\theta, k)$-persistent-core intervals of $G$. Then, the core persistence of $G$ with parameters $\theta$ and $k$, denoted by $F(\theta, $ $k,$ $G)$, is defined as
 \begin{equation*}
  \label{eq:corepersistent}
F(\theta ,k, G) = \left\{ \begin{gathered}
  \sum\limits_{i = 1}^r {(t_{{e_i}} - t_{{s_i}})}  - (r - 1)\theta , \quad if\ T \ne \emptyset  \hfill \\
  0\quad otherwise \hfill \\
\end{gathered}  \right.
\end{equation*}
\end{definition}

\begin{definition} [$(\theta, \tau)$-persistent $k$-core]
  \label{def:globalcore} Given a temporal graph $G$, parameters $\theta$, $\tau$, and $k$, a $(\theta, \tau)$-persistent $k$-core is an induced temporal subgraph $C=(V_C, E_C)$ that meets the following properties.
  \begin{enumerate}
    \item \textbf{Persistent core property.} $F(\theta, k, C) \ge \tau$;  
    \item \textbf{Maximal property.} There does not exist an induced temporal subgraph $C^\prime$ that contains $C$ and also satisfies the persistent core property.
  \end{enumerate}
\end{definition}

\begin{problem}
\label{prob:timecs}
\textbf{(The persistent community search \\ problem)} Given a temporal graph $G$, parameters $\theta$, $\tau$ and $k$, the persistent community search problem aims to find the largest $(\theta, \tau)$-persistent $k$-core in $G$.
\end{problem}

Consider the temporal graph $G$ in Fig.~\ref{fig:lu-temporal}(a). Assume that $\theta$=3 and $k$=2. We can see that there is no maximal $(3, 2)$-persistent-core interval for the entire graph $G$. There is a maximal $(3, 2)$-persistent-core interval $[1, 5]$ for the subgraph $C$ induced by vertices $\{v_1, v_2, v_3\}$. This is because $[1, 5]$ is the maximal interval such that in any $3$-length subinterval of $[1, 5]$, the vertices $\{v_1, v_2, v_3\}$ form a connected $2$-core. Let $\tau = 4$, we can see that the subgraph $C$ induced by vertices $\{v_1, v_2, v_3\}$ is a $(3, 4)$-persistent $2$-core. Because $F(3, 2, C)$=4, which is no less than $\tau$; and $C$ is the maximal subgraph that meets such a persistent core property.

As shown in \cite{Li:ICDE:2018}, the persistent community search problem is NP-hard. Therefore, a prune-and-search approach is proposed in \cite{Li:ICDE:2018}. In the pruning phase, a temporal graph reduction algorithm is designed by decomposing the whole time span of the temporal graph into several meta-intervals, each of which has some properties to prune vertices. In the search phase, a  branch and bound algorithm with several  pruning rules are proposed to find the maximum $(\theta, \tau)$-persistent $k$-core.

\subsection{Influence Value-Based Attributed Graphs}
\label{sec:kcoreInfluence}

\subsubsection{Single-dimensional Influential CS}

Li et al. \cite{Li:vldb:2015} proposed the influential CS problem. They considered an undirected graph $G(V,E)$ with vertex set $V$ and edge set $E$. Each vertex $v \in V$ is associated with a weight $w_u$ indicating the influence (or importance) of $u$. Without loss of generality, they assumed that the weight vector $W=(w_1, w_2, \cdots, w_n)$ forms a total order, i.e., for any two vertices $v_i$ and $v_j$, if $i \ne j$, then $w_i \ne w_j$.

\begin{definition} [Influence value of a subgraph]
  \label{def:influvalue} \\ Given an undirected graph $G(V, E)$ and an induced subgraph $H(V_H, E_H)$ of $G$, the influence value of $H$ denoted by $f(H)$ is defined as the minimum weight of the vertices in $H$, i.e., $f(H) = \mathop {\min }\nolimits_{u \in {V_H}} \{ {w_u}\} $.
\end{definition}

\begin{definition} [$k$-influential community]
  \label{def:influcore} Given an undirected graph $G=(V, E)$ and an integer $k$. A $k$-influential community is an induced subgraph $H^k=(V_H^k, E_H^k)$ of $G$ that meets all the following constraints.
  \begin{enumerate}
    \item \textbf{Connectivity.} $H^k$ is connected;
    \item \textbf{Cohesiveness.} Each vertex $u$ in $H^k$ has degree at least $k$;
    \item \textbf{Maximal structure.} There is no other induced subgraph ${\tilde H}$ such that (1) ${\tilde H}$ satisfies connectivity and cohesiveness constraints, (2) ${\tilde H}$ contains $H^k$,  and (3) $f({\tilde H}) = f(H^k)$.
  \end{enumerate}
\end{definition}

\begin{figure}[t]
\begin{center}
\includegraphics[width=0.7\hsize]{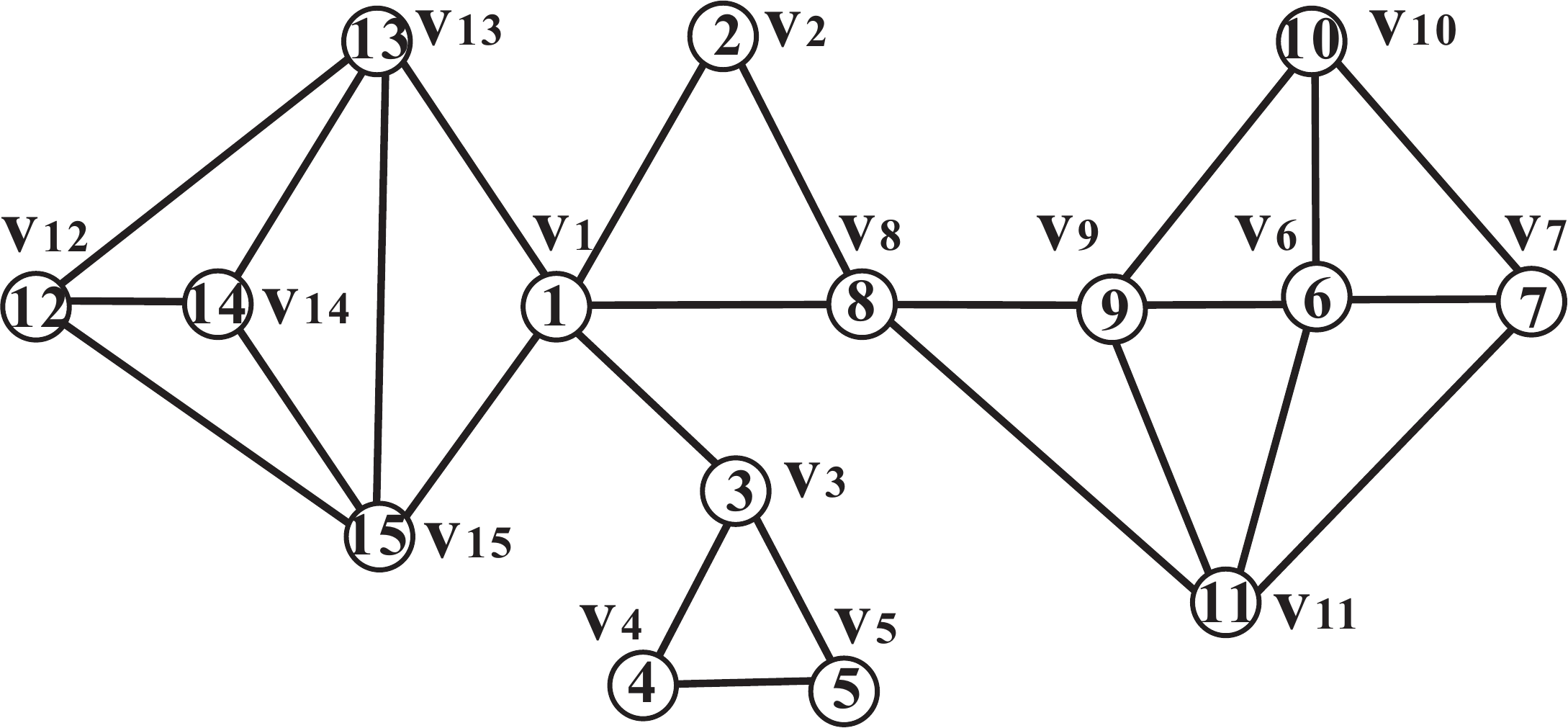}
\caption{An example of influential CS (the numbers denote the weights) \cite{Li:vldb:2015}.}
\label{fig:lu-influential}
\end{center}
\end{figure}

Consider the graph shown in Fig.~\ref{fig:lu-influential}. Suppose, for instance, that $k=2$, then by definition the subgraph induced by vertex set $\{v_{12}, v_{13}, v_{14}, v_{15}\}$ is a $2$-influential community with influence value $12$, as it meets all the constraints in Definition~\ref{def:influcore}. Note that the subgraph induced by vertex set $\{v_{12}, v_{14}, v_{15}\}$ is not a $2$-influential community. This is because it is contained in a $2$-influential community induced by vertex set $\{v_{12}, v_{13}, v_{14}, v_{15}\}$ whose influence value equals its influence value, thus fail to satisfy the maximal structure constraint.

\begin{problem} [Top-$r$ $k$-influential CS problem \\(TIC)] Given a graph $G(V, E)$ and two parameters $k$ and $r$, the problem is to find the top-$r$ $k$-influential communities with the highest influence value.
\end{problem}


\begin{definition} [Non-contained $k$-influential community]
  \label{def:noncontaincore} Given a graph $G(V, E)$ and an integer $k$. A non-contained $k$-influential community  $H^k=(V_H^k, E_H^k)$ is a $k$-influential community that meets the following constraint.
  \begin{itemize}
    \item \textbf{Non-containment.} $H^k$ cannot contain a $k$-influential community ${\bar H}^k$ such that $f({\bar H}^k) > f(H^k)$.
  \end{itemize}
\end{definition}

Consider the graph shown in Fig.~\ref{fig:lu-influential}. Assume that $k=2$. By Definition~\ref{def:noncontaincore}, we can see that the subgraphs induced by  $\{ v_{3},$ $v_{4},$ $v_{5}\}$, $\{v_8,$ $v_9,$ $v_{11}\}$ and $\{ v_{13},$ $v_{14},$ $v_{15}\}$ are non-contained $2$-influential communities. However, the subgraph induced by $\{ v_{12}, v_{13}, v_{14}, v_{15}\}$ is not a non-contained $2$-influential community, because it includes a  $2$-influential community (the subgraph induced by $\{ v_{13},$ $v_{14},$ $v_{15}\}$) with a larger influence value.

\begin{problem} [Top-$r$ non-contained $k$-influential \\ CS problem]
\label{prob:nic}
Given a graph $G(V, E)$ and parameters $k$ and $r$, find the top-$r$ non-contained $k$-influential communities with the highest influence value.
\end{problem}

\noindent\textbf{$\bullet$ Online search algorithms.}
An online search algorithm is proposed in \cite{Li:vldb:2015} to compute the top-$r$ (non-contained) $k$-influential communities given graph $G$ and parameters $r$ and $k$. The algorithm first computes the $k$-core $C$ of $G$, and then iteratively updates $C$ by removing vertices from $C$ until $C$ becomes empty. In each iteration, a vertex $u$ with the smallest influence  value is removed from $C$. After $u$ is removed, the algorithm further removes those vertices that do not belong to the $k$-core from $C$ by invoking a DFS procedure. For each iteration, the connected component that vertex $u$ belongs to forms a $k$-influential community. The $k$-influential communities obtained by the last $r$ iterations are the  top-$r$  $k$-influential communities. If after deleting a certain $u$, the vertices in the whole connected component that $u$ belongs to are removed in the DFS procedure, then the corresponding connected component is a non-contained $k$-influential community. In this way, we can obtain the top-$r$ non-contained $k$-influential communities. The algorithm runs in $O(m+n)$ time using $O(m+n)$ space.

The above algorithm needs to compute all (non-contained) $k$-influential communities before obtaining the top-$r$ (non-contained) $k$-influential communities which is costly when the graph is large and $r$ is small. Therefore, Chen et al. \cite{chen2016efficient} proposed a backward search algorithm to obtain the top-$r$ (non-contained) $k$-influential communities. The general idea is as follows. Instead of deleting the vertex with the smallest influence value each time, the backward search algorithm initializes an empty vertex set $C$ and inserts into $C$ the vertex with the largest influence value in each iteration. After a vertex $u$ with the largest influence value is inserted, if the core number of $u$ in the subgraph induced by $C$ is no smaller than $k$, the connected component containing $u$ in the subgraph induced by $C$ represents a $k$-influential community. The algorithm can terminate once $r$ $k$-influential communities are reported. The top $r$ non-contained $k$-influential communities can be computed in a similar way by checking whether each $k$-influential community is a non-contained $k$-influential community before reporting the community.

The online search algorithms in \cite{Li:vldb:2015} and \cite{chen2016efficient} need to access the whole graph to obtain the top-$r$ (non-contained) $k$-influential communities. To solve this issue, Bi et al. \cite{Bi:2018} proposed a local search algorithm. Let $G_{\geq \tau}$ be the subgraph of $G$ induced by all vertices with weights at least $\tau$, the authors proved that \textit{if the subgraph $G_{\geq \tau}$ of $G$ contains at least $r$ $k$-influential communities, then the top-$r$ $k$-influential communities in $G_{\geq \tau}$ is the query result}. The goal is to find the smallest subgraph $G_{\geq \tau^*}$ of $G$ containing at least $r$ $k$-influential communities. The general idea is as follows. The algorithm starts with a large $\tau$, and iteratively decreases the value of $\tau$  until reaching the target value. For each $\tau$, only the vertices with weights no smaller than $\tau$ need to be accessed. The authors proved that the time complexity of the algorithm is linear to the size of the smallest subgraph $G_{\geq \tau^*}$ that an online search algorithm without indexes needs to access to correctly compute the top-$r$ $k$-influential communities. Thus the algorithm is instance-optimal. Their algorithm can be easily extended to solve Problem \ref{prob:nic}.
%

\vspace*{0.1cm}\noindent\textbf{$\bullet$ An index-based algorithm.}
In \cite{Li:vldb:2015}, an index, called ICP-Index, is presented for solving Problem \ref{prob:nic}.
The index is designed based on the observation that \textit{for each $k$, the $k$-influential communities form an inclusion relationship}. Based on such an inclusion relationship, all the $k$-influential communities can be organized by a tree-shape structure. The index includes such tree structures for all possible $k$ values. In addition, instead of keeping the whole community for each tree node, a compression method is proposed to make the ICP-Index compact. Specifically, for each non-leaf node in the tree which corresponds to a $k$-influential community, the index only stores the vertices of the $k$-influential communities that are not included in their sub-$k$-influential communities. The same idea is recursively applied to all the non-leaf nodes of the tree following a top-down manner. For each leaf node which corresponds to a non-contained $k$-influential community, the index stores all the vertices of that non-contained $k$-influential community. Using the ICP-Index, the query can be answered efficiently because each node in the tree corresponds to a $k$-influential community and each leaf-node in the tree corresponds to a non-contained $k$-influential community. In \cite{Li:vldb:2015}, the authors proved that the ICP-Index can be constructed in $O(m^{1.5})$ time using $O(m+n)$ space.
%

\begin{figure}[t]
\begin{center}
\includegraphics[width=0.95\hsize]{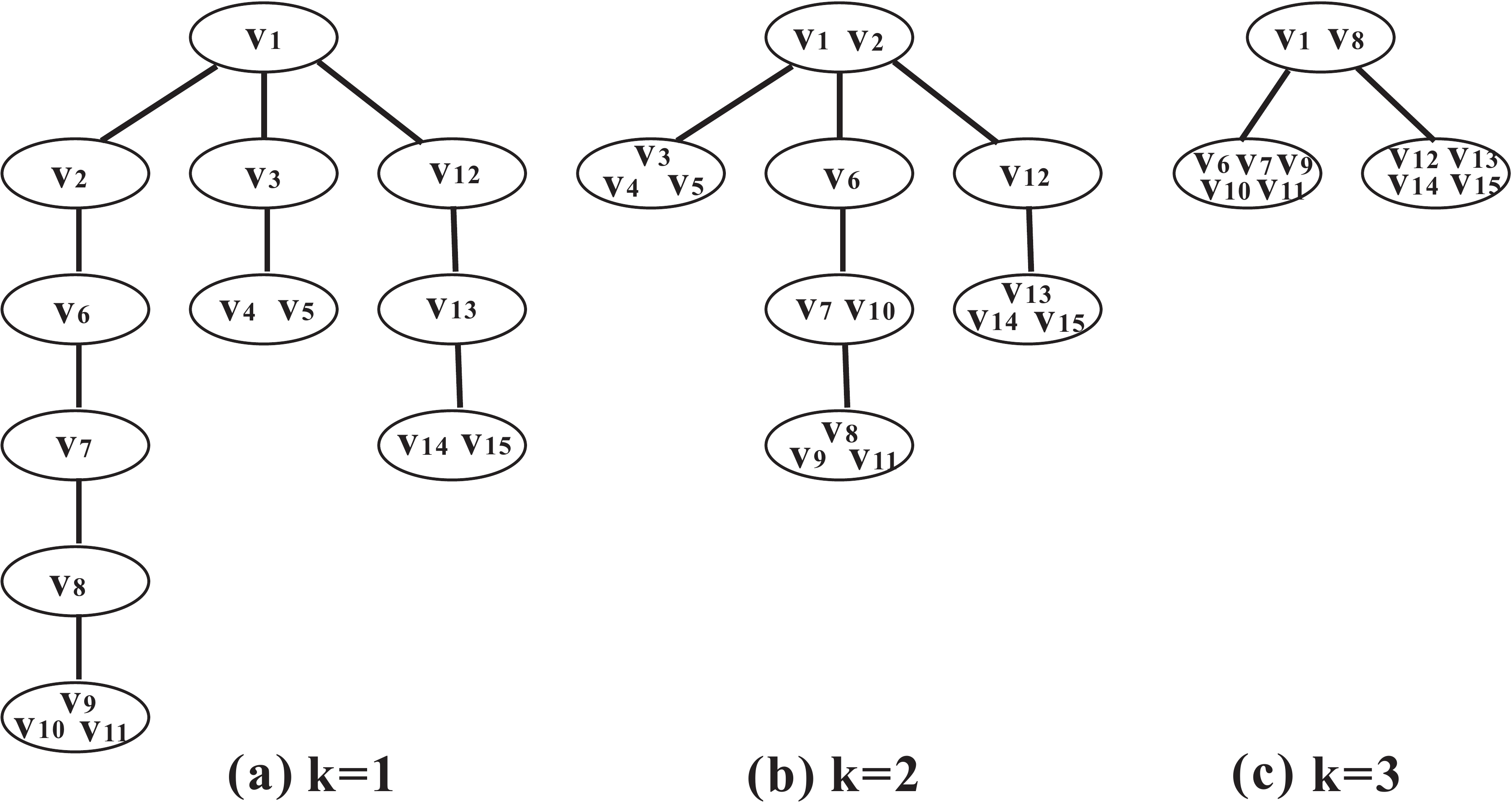}
\caption{Tree organization of all the $k$-influential communities (the ICP-Index) \cite{Li:vldb:2015}.} \label{fig:lu-icpindex}
\end{center}
\end{figure}

 Consider the graph shown in Fig.~\ref{fig:lu-influential}. Let us consider the case of $k=2$. Clearly, the entire graph is a connected $2$-core, so it is a $2$-influential community. Therefore, the root node of the tree corresponds to the entire graph. After deleting the smallest-weight vertex $v_1$, we get three $2$-influential communities which are the subgraphs induced by the vertex sets $\{v_3, v_4, v_5\}$, $\{v_6, \cdots, v_{11}\}$, and $\{v_{12}, \cdots, v_{15}\}$ respectively. Thus, we create three child nodes for the root node which correspond to the three $2$-influential communities respectively. Since $v_1$ and $v_2$ are not included in these three $2$-influential communities, we store them in the root node. The same idea is recursively applied in all the three $2$-influential communities.
 %
 %
 Fig.~\ref{fig:lu-icpindex} shows the tree organization for all $k$ for the graph shown in Fig.~\ref{fig:lu-influential}.

\noindent\textbf{$\bullet$ An I/O efficient algorithm.} An I/O efficient algorithm to compute the top-$r$ (non-contained) $k$-influential communities is presented in \cite{Li:VLDBJ:2017}. It assumes that all vertices of the graph can be stored in the main memory.
The key idea of the algorithm is that it computes the $k$-influential communities following the decreasing order of their weights, and the communities (as well as the edges in community) with large weights can be safely deleted without affecting the correctness of the algorithm to compute the tree vertices with small weights. Specifically, let $w(e)=\min\{w_u, w_v\}$ be the weight of an edge $e=(u, v)$. The algorithm first sorts the edges in a non-increasing order of their weights using the standard external-memory sort algorithm (we can use the vertex ID to break ties). Then, following this order, the algorithm loads the edges into the main memory up to the memory limit. Subsequently, the algorithm invokes an in-memory algorithm to compute the influential communities in the main memory. After that, the algorithm deletes the computed influential communities as well as the associated edges from the main memory, and then sequentially loads new edges into the main memory until reaches the memory limit. The algorithm iteratively performs this procedure until all the edges are scanned. Note that in each iteration, the algorithm only works on a partial graph, which is loaded in the main memory.
%

\begin{figure}[t]
\begin{center}
\includegraphics[width=\columnwidth]{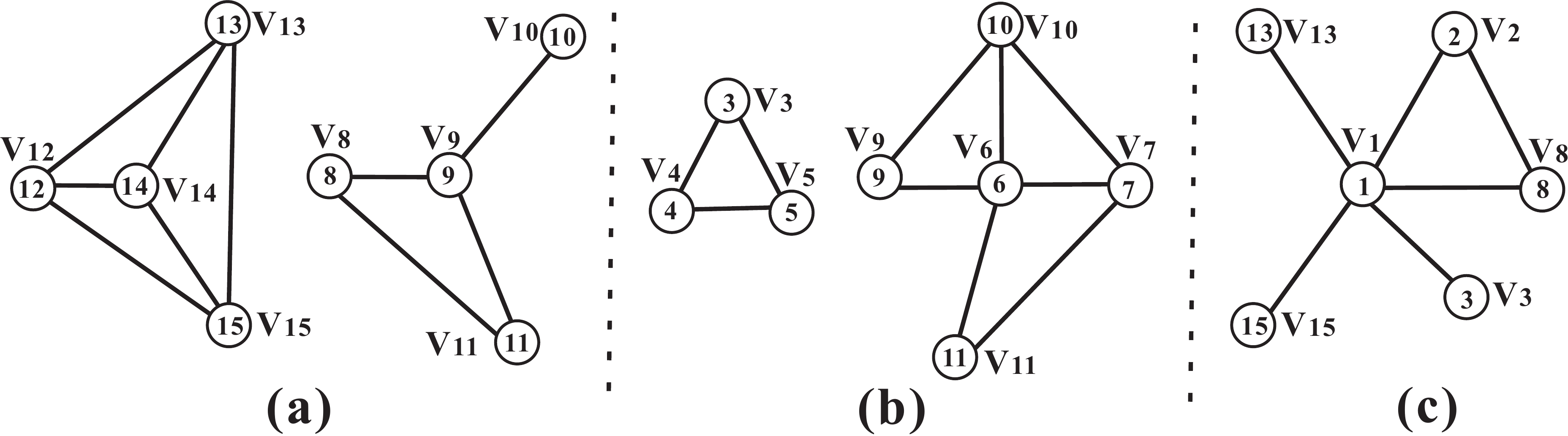}
\end{center}
\caption{Partial graphs in the memory ($k=2$, memory can hold at most 10 edges) \cite{Li:VLDBJ:2017}.}
\label{fig:lu-partialgraph}
\end{figure}

As an example, consider the graph shown in Fig.~\ref{fig:lu-influential}. Suppose $k=2$ and the memory can hold at most $10$ edges. The partial graph loaded into memory in the first three iterations for the algorithm is shown in Fig.~\ref{fig:lu-partialgraph}

\noindent\textbf{$\bullet$ Center-core CS.} Another model to capture the influence of vertices is called the centre-core community search, which is studied by Ding et al. \cite{ding2018search}. The model uses $k$-core to qualify the dense structure for the community and uses coreness to evaluate the vertex influence. Given a query vertex $q$ and an integer $k$, the center-core community is  a connected component of the maximal $k$-core containing the query vertex $q$ and the coreness of vertices in the community is no less than $q$. In addition, the community excludes those vertices with coreness equal to $q$ but cannot be reached from $q$ via vertices with the same coreness with $q$. An online search algorithm and an index based algorithm are proposed in \cite{ding2018search} to compute the center-core community.

\subsubsection{Multi-dimensional Influential CS}

In \cite{Li:SIGMOD:2018}, Li et al. studied the multi-dimensional influential CS. It deals with a multi-valued graph $G(V,E,X)$ where $V$  and $E$ denote the set of vertices and edges respectively, and $X$ ($|X|=n$) is a set of $d$-dimensional vectors. In a multi-valued graph, each vertex $v\in V$ is associated with a $d$-dimensional real-valued vector denoted by $X_v$ $=$ $(x_1^v,$ $\cdots,$ $x_d^v)$, where $X_v \in X$ and $x_i^v \in \mathbb{R}$.  Suppose without loss of generality that on the $x_i$ dimension, $x_i^v$ for all $v \in V$ form a strict total order, i.e., $x_i^v \ne x_i^u$ for any $u \ne v$. It is important to note that if this assumption does not hold, we can easily construct a strict total order by using the vertex identity to break ties for any $x_i^v = x_i^u$. The $d$-dimensional vector $X_v$ represents the values of the vertex $v$ w.r.t. $d$ different numerical attributes. The model studied in \cite{Li:SIGMOD:2018}, called the skyline community search, is based on the one-dimensional influential community model proposed in \cite{Li:vldb:2015}. The authors defined the value of $H$ on the $x_i$ dimension (for $i$=1, 2, $\cdots$, $d$) as ${f_i}(H) \triangleq \mathop {\min }\nolimits_{v \in {V(H)}} \{ x_i^v\}$.

\begin{definition}
  \label{def:communitydominate} Let $H(V_H, E_H)$ and $H^\prime(V_{H^\prime}, E_{H^\prime})$ be two subgraphs of a multi-valued graph $G$. If $f_i(H) \le f_i(H^\prime)$ for all $i=1, \cdots, d$, and there exists $f_i(H) < f_i(H^\prime)$  for a certain $i$, we call that $H^\prime$ dominates $H$, denoted by $H \prec H^\prime$.
\end{definition}

\begin{definition}
  \label{def:skylinecommunity} Given a multi-valued graph $\small G(V, E, X)$ and an integer $k$. A skyline community with a parameter $k$ is an induced subgraph $H(V_H, E_H, X_H)$ of $G$ such that it satisfies the following properties.

    \begin{enumerate}
\vspace*{-0.1cm}
    \item \textbf{Cohesive property.} $H$ is a connected $k$-core;
    \item \textbf{Skyline property.} There does not exist an induced subgraph $\small H^\prime$ of $G$ such that $\small H^\prime$ is a connected $k$-core subgraph and $\small H \prec H^\prime$;
    \item \textbf{Maximal property.} There does not exist an induced subgraph $\small H^\prime$ of $G$ such that (1) $\small H^\prime$ is a connected $k$-core subgraph, (2) $H^\prime$ contains $H$, and (3) $f_i(H^\prime)=f_i(H)$ for all $i=1,\cdots, d$.
  \end{enumerate}

\end{definition}

\begin{problem} [Skyline CS problem]
\label{prob:skyCS}
Given a multi-valued graph $G(V, E, X)$ and an integer $k$, the problem is to find all the skyline communities from $G$ with the parameter $k$. More formally, let $\cal H$ be the set of all connected $k$-core subgraphs in $G$. We aim to compute a subset $\cal R$ of $\cal H$ which is defined as:

{\scriptsize  \[{\cal R} \triangleq \{ H\in {\cal H}|\neg \exists {H^\prime, H^{\prime\prime}}\in {\cal H}:H \prec {H^\prime}, H \subset {H^{\prime\prime}} \wedge f(H) = f({H^{\prime \prime}})\},\]}

\noindent where {$H \subset  {H^{\prime\prime}}$} denotes that {$H$} is a subgraph of {${H^{\prime\prime}}$} and {$ H\ne {H^{\prime\prime}}$}, and {$f(H) = f({H^{\prime \prime}})$} means that {$f_i(H)=f_i({H^{\prime \prime}})$} for $i=1, \cdots, d$.
\end{problem}

\begin{figure}[t]
\begin{center}
\includegraphics[width=0.75\hsize]{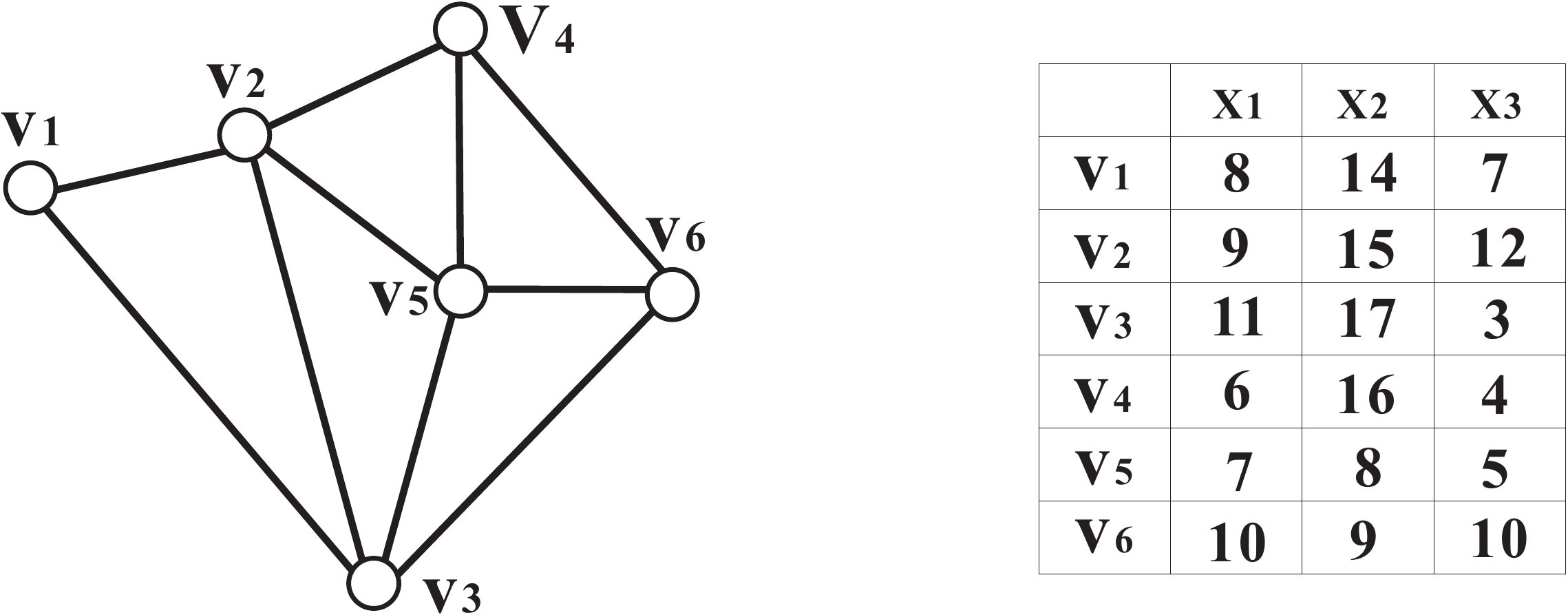}
\caption{An example of a multi-valued graph \cite{Li:SIGMOD:2018}.} \label{fig:lu-skyline}
\end{center}
\end{figure}

Consider the graph shown in Fig.~\ref{fig:lu-skyline}. The left panel is a graph with 6 vertices, and the right panel shows the values of these vertices in three different dimensions. Suppose for instance that $k=2$. Then, by Definition~\ref{def:skylinecommunity}, $H_1=\{v_1, v_2, v_3\}$ is a skyline community with values $f(H_1)=(8, 14, 3)$, because there does not exist a connected $2$-core subgraph that can dominate it, and it is also the maximal subgraph that satisfies the cohesive and skyline properties. Similarly, $H_2=\{v_2, v_4, v_5, v_6\}$ is a skyline community with $f(H_2)=(6,8,4)$. The subgraph $H_3=\{v_4, v_5, v_6\}$ is not a skyline community, because it is contained in $H_2=\{v_2, v_4, v_5, v_6\}$ which has the same $f$ values as $H_3$. The subgraph $H_4=\{v_2, v_3, v_4, v_5, v_6\}$ is not a skyline community, as $f(H_4)=(6,8,3)$ is dominated by $H_1$ and $H_2$.

In \cite{Li:SIGMOD:2018}, the authors first developed an efficient algorithm, called SkylineComm2D, to find all the skyline communities in the 2D case, i.e., $d=2$. The time complexity of SkylineComm2D is $O(s(m+n))$ where $s$ denotes the number of 2D skyline communities (i.e., the answer size), and the space complexity of SkylineComm2D is $O(m+n+s)$, which is linear w.r.t. the graph and answer size. To handle the high-dimensional case (i.e., $d \ge 3$), the authors proposed a space-partition algorithm to find the skyline communities efficiently. Two novel features of the space-partition algorithm are that (1) its worst-case time complexity is dependent mainly on the answer size, thus it is very efficient when the answer size is not very large; and (2) it is able to progressively output the skyline communities during the execution of the algorithm, and thus it is useful for applications that only require part of skyline communities.

\subsection{Profile-Based Attributed Graphs}
\label{sec:kcoreProfile}

A profiled-based attributed graph, or profiled graph, is an undirected graph $G(V,E)$ with vertex set $V$ and edge set $E$, in which each vertex is associated with {\it profile}. The profile of a vertex $v\in V$ is a set of keywords $T(v)$ that are arranged in a hierarchical manner, called a P-tree. Typical such attributes are users' affiliation, expertise, locations, etc. Profiled graphs are prevalent in knowledge bases, and social media.

\begin{figure}[]
\centering
\begin{tabular}{c}
  \begin{minipage}{6.4cm}
	\includegraphics[width=6.4cm]{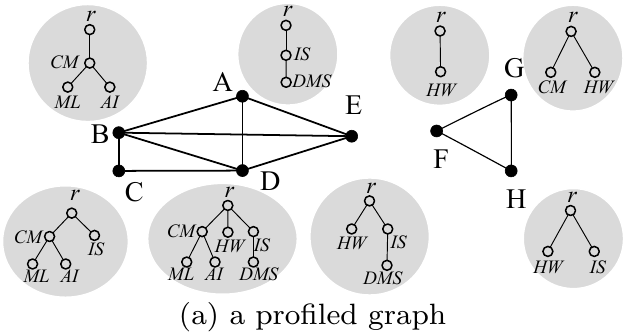}
  \end{minipage}
  \\
  \begin{minipage}{7.2cm}
	\includegraphics[width=7.2cm]{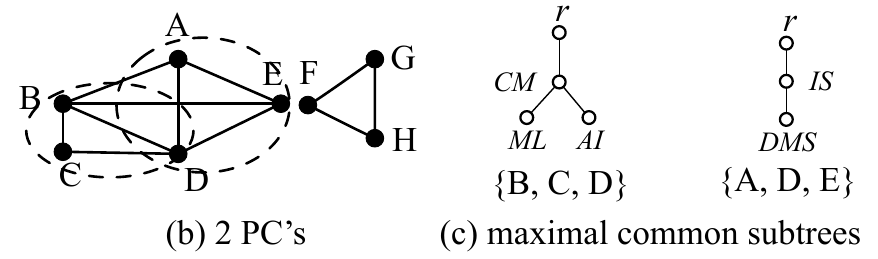}
  \end{minipage}
\end{tabular}
\caption{A profiled graph and two PC's~\cite{Yankai18}.}
\label{fig:profiledGraph}
\end{figure}

Fig. \ref{fig:profiledGraph}(a) depicts a profiled graph. For instance, vertex $D$ has a hierarchically organized profile that describes his expertise in Computer Science (e.g., abbreviation AI means Artificial Intelligence) by following the \emph{ACM Computing Classification System (CCS)}~\footnote{ACM CCS: http://www.acm.org/publications/class-2012}.

Chen et al.~\cite{Yankai18} investigated the problem of {\it profiled community search} (or PCS) on profiled graphs. To capture the profile-based cohesiveness, they introduced the concept of ``maximal common subtree'', which describes the commonality of vertices' profile.

\begin{definition}[Maximal common subtree]
\label{df:MaximalTree}
Given a profiled graph $G$, the maximal common subtree of $G$, denoted by $\mathcal M$($G$), holds the properties:
(1) $\forall v \in G$, $\mathcal M$($G$) $\subseteq T(v)$;
(2) there exists no other common subtree $\mathcal {M'}$($G$) such that $\mathcal {M}$($G$) $\subseteq \mathcal {M'}$($G$).
\end{definition}

\begin{problem}[PCS]
\label{prob:PCS}
Given a profiled graph $G(V,E)$, a positive integer $k$, a query node $q\in G$, find a set $\mathcal {G}$ of graphs, such that $\forall G_q \in \mathcal {G}$, following properties hold:
\begin{enumerate}
\item \textbf{Connectivity.} $G_q$ is connected and contains $q$;
\item \textbf{Structure cohesiveness.} $\forall v\in G_q$, $deg_{G_q}(v)\geq k$;
\item \textbf{Profile cohesiveness.} There exists no other $G'_q \subseteq G$ satisfying the above two constraints, such that $\mathcal M(G_q) \subseteq \mathcal M(G'_q)$.
\item \textbf{Maximal structure.} There exists no other subgraph $G'_q$ satisfying the above properties, such that $G_q \subset G'_q$ and $\mathcal M(G_q)$ = $\mathcal M(G'_q)$;
\end{enumerate}
\end{problem}

The subgraph $G_q$ is called a {\it profiled community} (or PC). In Problem~\ref{prob:PCS}, the first two properties guarantee the structure cohesiveness. The {\it profile cohesiveness} captures the maximal shared profile among all vertices in $G_q$. The {\it maximal structure} property aims to retrieve all qualified vertices in the community.
For instance, in Fig. \ref{fig:profiledGraph}(a), if $q$=D, $k$=2, then two PC's and their maximal common subtrees are respectively shown in Fig. \ref{fig:profiledGraph}(b) and (c).
These two common subtree sufficiently reflects the ``theme'' of the community. For example, in the PC grouped by vertices \{B, C, D\}, all the researchers involved share interest in ML (i.e., Machine Learning) and Artificial Intelligence, whereas for the other PC, the researchers are all interested in other research domains.

The PCS problem is technically challenging, because the number of subtrees of a P-tree could be exponentially large, and thus enumerating all of them is impractical. To answer the PCS query efficiently, Chen et al.~\cite{Yankai18} introduced the anti-monotonicity property, based on which the query can be performed much faster. To further improve efficiency, they developed the \emph{CP-tree} index, which systematically organizes all the graph vertices and their P-trees into a compact tree structure. The CP-tree index enables the development of two fast PC discovery algorithms.

\subsection{Discussions}
\label{sec:kcoreDiscuss}

In this section, we review CS studies that use the $k$-core model. For simple graphs, we can divide them into two groups, where the first group \cite{KDD2010,local2014,barbieri2015efficient} focuses on undirected graphs while the second group \cite{Fang:TKDE:CSD} only considers directed graphs. In particular, for the first group, the first work \cite{KDD2010} returns the maximal $k$-core containing the query vertex, while communities of the other two studies \cite{local2014,barbieri2015efficient} may not be the maximal $k$-core or with size constraints.

For attributed graphs, all the corresponding CS studies take both link relationship and attributes into consideration, because the attributes often make communities more meaningful and easy for interpretation. As a result, the solutions for different attributed graphs are different. Generally, both online and index-based algorithms are developed for CS on these graphs. The index-based algorithms run faster, but incur an offline computational cost.

In practice, the query users can select the CS solutions based on the graph models since the community models are formulated based on the graph models. For example, for keyword-based attributed graphs, ACQ can be considered.
Meanwhile, if the CS queries are executed with high frequency, the index-based algorithms should be better choices as they are faster, although they have to build the index in an offline manner.

\section{$K$-Truss-Based Community Search}
\label{sec:ktruss}

In this section, we review CS works that use the $k$-truss as structure cohesiveness metrics, including triangle-connected truss community \cite{k-truss2014,Akbas:VLDB:2017}, closest truss community \cite{huang2015approximate}, attribute-driven truss community \cite{Huang:2017:ATC}, and weighted truss community \cite{Zheng:IS:2017}. In the following, we will introduce the community models, and compare their algorithms and applications.

\subsection{Simple Graphs}
\label{sec:ktrussLocation}
In a simple and undirected graph $G(V,E)$, triangle-connected $k$-truss community model proposed by Huang et al. \cite{k-truss2014}, finds all communities containing a query vertex. We first introduce the definitions of $k$-truss and triangle connectivity, and then present the model below.

\begin{figure}[t]
\centering
\includegraphics[width=0.5\linewidth]{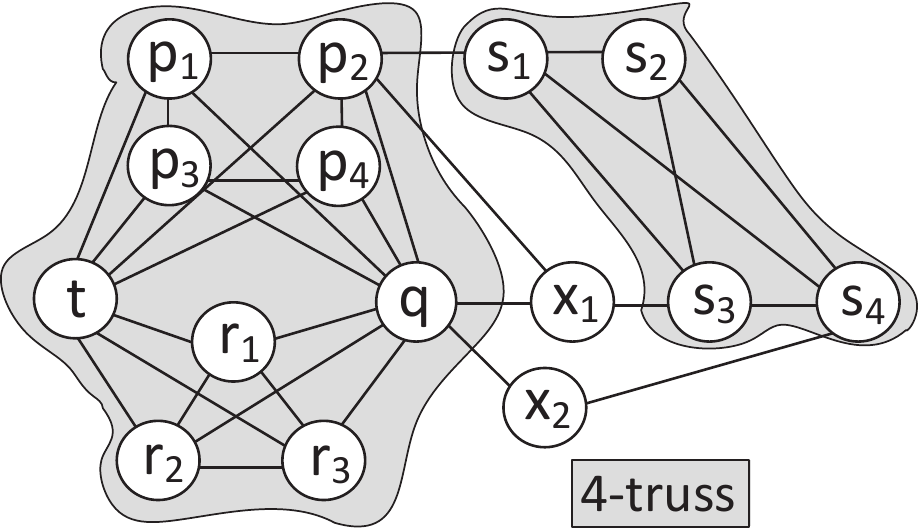}
\caption{Example of 4-truss with 2 disconnected components.}
\label{fig.4-truss}
\end{figure}

A $k$-truss is the largest subgraph $H$  of $G$ such that every edge is contained in at least $k-2$ triangles in $H$, i.e., $\forall e\in E$, its support $sup(e, H)\geq k-2$ by Definition~\ref{def:ktruss}. However, $k$-truss may be disconnected with several components in a graph, which is similar with $k$-core. Consider the graph $G$ in Fig. \ref{fig.4-truss}. There exist two components in the shaded regions to form the 4-truss of $G$, which are obviously disconnected. Disconnected subgraphs are insufficient to define a cohesive and meaningful community.

To address the disconnectivity problem of $k$-truss, \emph{triangle connectivity} is imposed on top of the $k$-truss in \cite{k-truss2014}. Given two triangles $\triangle_1$ and $\triangle_2$ in $G$, $\triangle_1$ and $\triangle_2$ are said to be adjacent if they share a common edge. Then, for two edges $e_1, e_2\in E$, $e_1$ and $e_2$ are \emph{triangle connected} if they either belong to the same triangle, or are reachable from each other through a series of adjacent triangles. In other words, $\exists \triangle_1, \triangle_2$ such that $e_1 \in \triangle_1$, $e_2\in \triangle_2$, then either $\triangle_1=\triangle_2$, or $\triangle_1$ is triangle connected with $\triangle_2$. Based on the $k$-truss and triangle connectivity, the problem of triangle-connected truss community (TTC) search is formulated as follows.

\begin{problem}
[TTC search]
\label{problem:TTC}
Given an undirected simple graph $G(V,E)$, a query vertex $q\in V$, and an integer $k \geq 2$, return all subgraphs $H\subseteq G$ satisfies the following three properties:
\begin{enumerate}
\item \textbf{Structure Cohesiveness.} $H$ contains the query vertex $q$ such that $\forall e\in E(H)$, $sup(e, H)$ $\geq (k-2)$;

\item \textbf{Triangle  Connectivity.}  $\forall e_1, e_2\in E(H)$, $e_1$ and $e_2$ are triangle connected;

\item \textbf{Maximal Subgraph.} $H$ is the maximal subgraph of $G$ satisfying Properties 1 and 2.  
\end{enumerate}
\end{problem}

\begin{figure} [t]
\vskip -0.1in
\centering \mbox{
\subfigure[Graph $G$]{\includegraphics[width=0.67\linewidth]{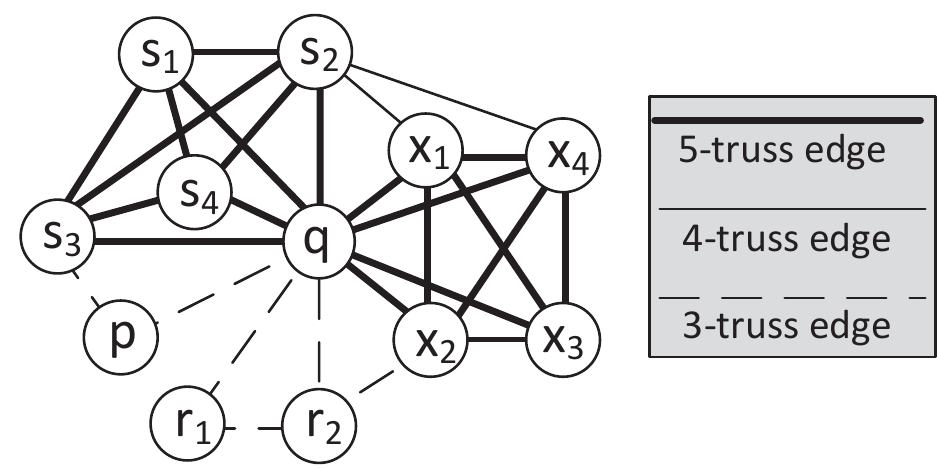}} \hskip 0.1in
\subfigure[TTCs]{\includegraphics[width=0.28\linewidth]{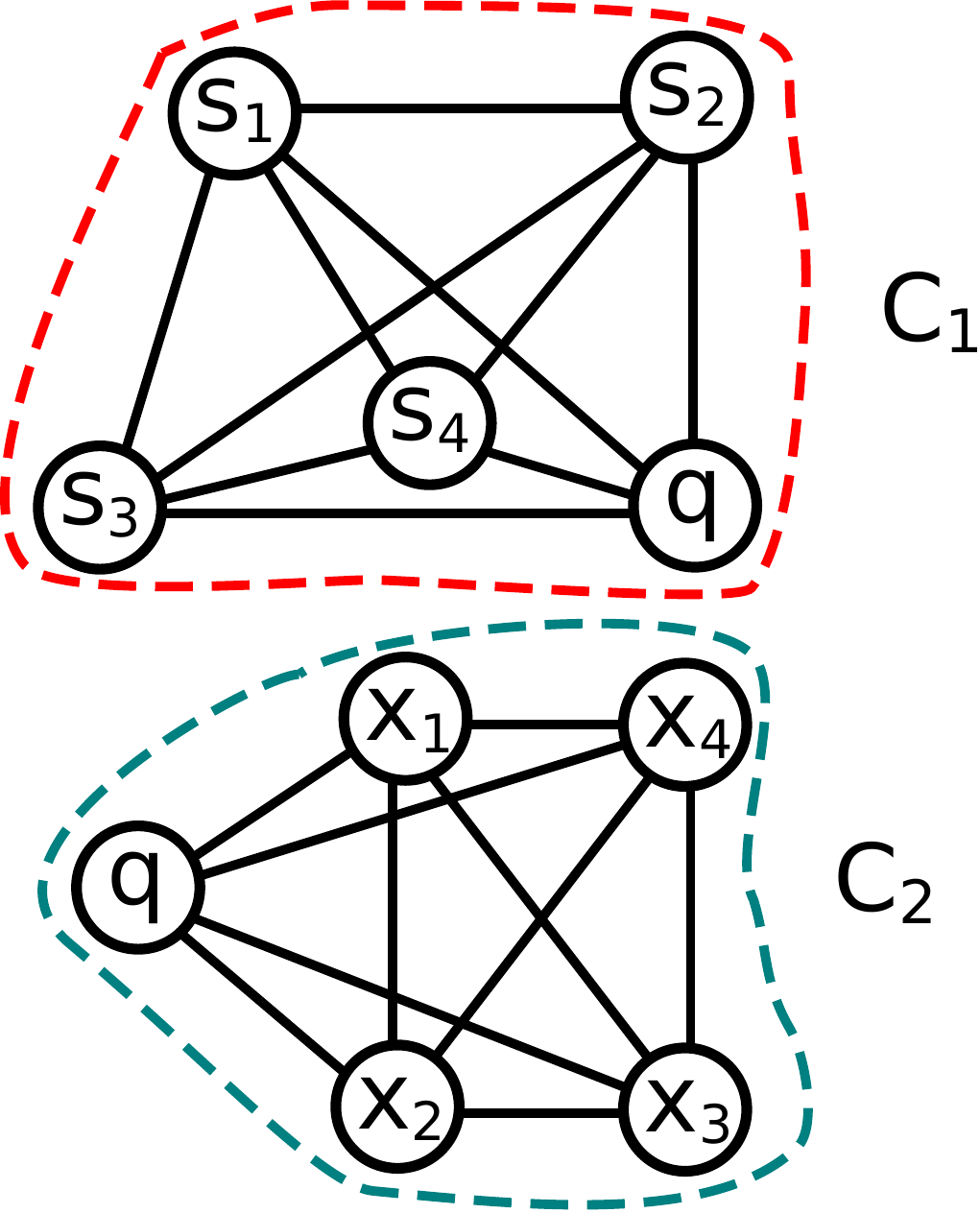} }
}
\caption{An example of TTC search. Here, $k=5$.}\vskip -0.1in
\label{fig.ttc}
\end{figure}

TTC model imposes the triangle connectivity requirement in Property 2 to ensure the discovered communities are connected.  This requirement also allows the query vertex to participate in multiple overlapping communities. For example, consider the graph $G$ in Fig. \ref{fig.ttc}(a), a query vertex $q$, and parameter $k=5$. Two triangle-connected 5-truss communities $C_1$ and $C_2$ containing vertex $q$ are shown in Fig. \ref{fig.ttc}(b). As the edges in $C_1$ cannot reach the edges in $C_2$ through adjacent triangles, $C_1$ and $C_2$ cannot merge as one large community. This is reasonable, as there are few connections between the two vertex sets $\{s_1,s_2,s_3,s_4\}$ and $\{x_1,x_2,x_3,x_4\}$.

Thanks to $k$-trusses, truss-based community model inherits several good structural properties of
$k$-trusses \cite{k-truss2014}, such as $(k-1)$-edge-connected, bounded diameter and hierarchical
structure. Specifically, the diameter of a $k$-truss community $H$ with $|V(H)|$ vertices is no larger than $\lfloor\frac{2|V(H)|-2}{k}\rfloor$ \cite{cohen2008trusses}. Small diameter has been considered as an important feature of a good community in \cite{Edachery99graphclustering}.  Second,
a $k$-truss community is ($k-1$)-edge-connected \cite{cohen2008trusses}, i.e., the community keeps connected whenever fewer than $k-1$ edges are deleted \cite{DBLP:books/fm/GareyJ79}.  Third, truss-based communities have a strong decomposability for analyzing large-scale networks at different levels of granularity.

To tackle the problem of TTC search, there exists one online search algorithm \cite{k-truss2014}, and two index-based search algorithms, which are respectively based on TCP-index \cite{k-truss2014} and EquiTruss \cite{Akbas:VLDB:2017}. In the following, we briefly introduce the key ideas of these algorithms one by one.

\noindent\textbf{$\bullet$ Online search algorithm \cite{k-truss2014}.} Huang et al. \cite{k-truss2014} proposed an online query algorithm to process a TTC query on a graph $G$. The algorithm firstly applies the truss decomposition \cite{wang2012truss} on graph $G$ to compute the trussness of all edges in $G$. By the community definition, it starts from the query vertex $q$ and checks an incident edge of $(q, v)\in E$ with trussness $\tau((q,v))\geq k$ to search triangle-connected truss communities. It explores all edges that are triangle-connected to $(q, v)$ and having trussness no less than $k$ in a BFS manner. This process iterates until all incident edges of $q$ have been processed.  Finally, a set of $k$-truss communities containing $q$ are returned.

However, this online search algorithm may incur a large number of wasteful edge accesses on checking disqualified edges, which is inefficient.

\begin{figure} [t]
\vskip -0.1in
\centering \mbox{

\includegraphics[width=0.60\linewidth]{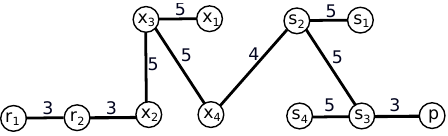}
}
\caption{TCP-index $T_{q}$ for vertex $q$ of $G$ in Figure~\ref{fig.ttc}(a).}\vskip -0.1in
\label{fig.truss-index}
\end{figure}

\noindent\textbf{$\bullet$ TCP-index based search algorithm \cite{k-truss2014}.} To avoid the computational issues mentioned above, Huang et al. \cite{k-truss2014} designed a Triangle Connectivity Preserving index (TCP-index). TCP-Index preserves the truss number and  triangle adjacency relationship in a compact tree-shape index, and supports the query of $k$-truss community in linear time with respect to the community size, which is optimal. Given a graph $G$, it needs to construct a TCP-index for each vertex in $G$, which is denoted as ${\cal T}_x$. Take a vertex $x$ as an example for TCP-index construction.
Essentially, $T_x$ is the maximum spanning forest of $G_x$, where $G_x$ is the induced subgraph of $G$ by vertex set of $x$'s neighbors as $N(x)$. For each edge $(y,z)\in E(G_x)$, a weight $w(y,z)=\min\{\tau((x,y)), \tau((x,z)),$ $ \tau((y,$ $z))\}$ is assigned to it, which indicates that $\triangle_{xyz}$ can appear only in $k$-truss communities where $k\leq w(y,z)$.
Fig. \ref{fig.truss-index} presents a TCP-index $T_{q}$ for vertex $q$ in graph $G$ shown in Fig. \ref{fig.ttc}(a). Vertices $x_1$, $x_2$, $x_3$ and $x_4$ are connected via the weighted edges of 5, indicating these vertices present in a triangle-connected 5-truss community.

Based on the TCP-index, an efficient query processing algorithm is developed for CTC search. Assume that we want to query 5-truss communities containing a query vertex $q$ in $G$ in Fig. \ref{fig.ttc}(a), we first visit an incident edge on $q$, say $(q, x_1)$, where $\tau((q,x_1))=5$. From TCP-index ${\cal T}_q$ in Fig. \ref{fig.truss-index}, we retrieve the vertex set $\{x_1,x_2,x_3,x_4\}$ belong to the same 5-truss community. Since ${\cal T}_q$ is a spanning forest, which does not keep all the edges between the vertices, the query processing algorithm then performs the reverse operations on the TCP-index for each vertex $x_1, x_2, x_3,x_4$ and gets the complete 5-truss community.

Remarkably, the TCP-index supports the $k$-truss community query in the optimal time, which accesses each edge in the answer community exactly twice \cite{k-truss2014}.
Meanwhile, the TCP-index can be constructed in $O($ $\sum_{(u,v)\in E}$ $\min\{deg_G(u), deg_G(v)\})$ time and stored in $O(m)$ space.

\noindent\textbf{$\bullet$ EquiTruss-index based search algorithm \cite{Akbas:VLDB:2017}.} To further improve efficiency, Akbas and Zhao \cite{Akbas:VLDB:2017} proposed a novel indexing technique of $k$-truss equivalence, to represent the triangle connectivity and $k$-truss cohesiveness in the triangle-connected truss communities.


We introduce the definition of $k$-truss equivalence as follows. Given two edges $e_1$, $e_2 \in E$, $e_1$ and $e_2$ are $k$-truss equivalence, if and only if (1) $\tau(e_1)=\tau(e_2)=k$, and (2) $e_1$ and $e_2$ are triangle-connected via a series of triangles in a $k$-truss.

The index of EquiTruss, a summarized graph $\mathcal{G}=(\mathcal{V}, \mathcal{E})$, is constructed based on $k$-truss equivalence. According to $k$-truss equivalence, all edges of a given graph $G$ are partitioned into a series of mutually exclusive equivalence classes. Each class represents a TTC. A super-node $E_i\in \mathcal{V}$ represents a distinct equivalence class $C_i$ where $e\in G$, and a super-edge $(E_i, E_j)\in \mathcal{E}$ , where $E_i, E_j\in \mathcal{V}$, indicates that the two equivalence classes are triangle-connected; that is, there exists two edges $e_1\in E_i$ and $e_2\in E_j$, s.t., $e_1$ and $e_2$ are $k$-truss triangle adjacent. Note that EquiTruss is a community-preserving graph summary, where all  triangle-connected $k$-truss communities are comprehensively recorded in the super-nodes, and the triangle connectivity across different communities is exactly encoded in super-edges. In this way, EquiTruss keeps records of all the information critical to community search. Moreover, each edge $e$ is recorded in exactly one super-node, which represents its $k$-truss equivalence class, $C_e$. Compared with TCP-Index, which may redundantly maintain an edge  in multiple maximum spanning forests, EquiTruss is significantly more succinct and space-efficient \cite{Akbas:VLDB:2017}.

\begin{figure}[t]
\centering
\includegraphics[width=0.95\linewidth]{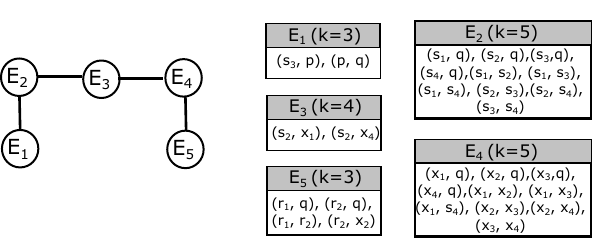}
\caption{EquiTruss index for graph $G$ in Fig. \ref{fig.ttc}(a).}
\label{fig.equitruss}
\end{figure}

For example, Fig. \ref{fig.equitruss} shows an EquiTruss index for graph $G$ in Fig. \ref{fig.ttc}(a). It has 5 super-nodes representing the $k$-truss equivalence classes for edges in $G$, as tabulated in Fig. \ref{fig.equitruss}. The super-node $E_2$ represents a 5-truss community with 10 edges: all these 10 edges are triangle connected, and belong to the 5-truss. In addition, there exist 5 super-edges in EquiTruss, which represents the triangle connectivity between super-nodes (triangle-connected $k$-truss communities).

The EquiTruss-index based community search algorithm is described as follows. Finding triangle-connected communities containing vertex $q$ can be carried out directly on EquiTruss, without the access to graph $G$. First, the algorithm finds all super-nodes containing $q$. A hash structure can help quick identification of such super-nodes. Next, starting from these super-nodes, we can traverse $\mathcal{G}$ in a BFS manner. For each unvisited neighboring super-nodes $E^*$ with $\tau(E^*)\geq k$, the edges within $E^*$ will be included into the $k$-truss community. The algorithm outputs all the discovered communities containing $q$. Consider the graph $G$ in Fig. \ref{fig.ttc}(a), $k=5$ and query vertex $q$. Based on the EquiTruss index Fig. \ref{fig.ttc}(a), we first find two super-nodes $E_2$ and $E_4$ containing $q$ with trussness no less than $5$. Super-nodes $E_2$ and $E_4$ are disconnected via any super-edges. Then, $E_2$ and $E_4$ can be respectively output as two communities.  Compared to TCP-index, EquiTruss-index based query processing only needs to access each edge exactly once, which is more efficient \cite{Akbas:VLDB:2017}.

\subsection{Closest Truss Community Search}
\begin{figure}[t]
\small
\vskip -0.1in
\centering
\includegraphics[width=1.0\linewidth]{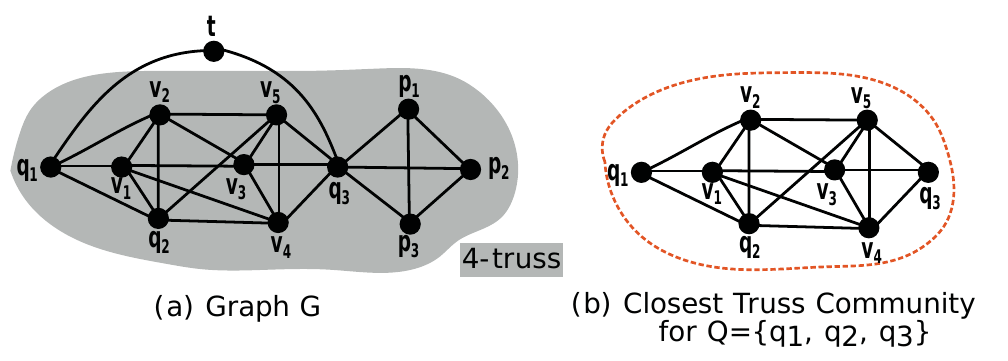}
\vskip -0.1in
\caption{Closest truss community example.}
\label{fig.ctc-community}
\end{figure}

In this section, we introduce a new truss-based community model for multiple query vertices.
Although the triangle-connected \truss community model works well to find all overlapping communities containing a single query vertex $q$, it may fail to discover any community for multiple query vertices, due to the strict requirement of triangle connectivity constraint. For example, for the graph $G$ in Fig. \ref{fig.ctc-community}(a) and query vertices $Q = \{v_4, q_3, p_1\} $, the above \truss community model cannot find a qualified community for any $k$, since the edges $(v_4, q_3)$ and $(q_3, p_1)$ are not triangle connected in any \truss. To address this limitation, Huang et al. \cite{huang2015approximate} studied the problem of closest truss community (CTC) search for multiple query vertices as follows.

\begin{problem}
\label{prob:ctc}
[CTC search] Given a graph $G$ and a set of query vertices $Q$, return a subgraph $H\subseteq G$
as a closest truss community (CTC), satisfying the following two properties:
\begin{enumerate}\label{def.ctc}
  \item Connected \truss. $H$ is containing $Q$ and a connected $k$-truss with the largest $k$, i.e., $Q\subseteq $ $H$  and $\forall e\in E(H)$, $sup(e, H) \geq k-2$;
  \item Smallest Diameter. $H$ is a subgraph of smallest diameter satisfying Property 1.
\end{enumerate}
\end{problem}

Property 1 requires that the closest community contains the query vertices $Q$ which are densely connected. In addition, to ensure every vertex included in the community is close  to query vertices and other vertices in the community, Property 2 uses graph diameter to measure the closeness of all vertices in the community. Moreover, the CTC model can avoid the free rider effect issue, that is, vertices far away from query vertices and irrelevant to them are included in the detected community \cite{huang2015approximate}.

Consider the graph $G$ in Fig. \ref{fig.ctc-community}(a), and $Q=\{q_1, q_2, q_3\}$; the subgraph in the region shaded gray is a 4-truss containing $Q$ with the largest trussness, and has a diameter of 4.   Fig. \ref{fig.ctc-community}(b) shows another 4-truss containing $Q$ but not $p_1, p_2, p_3$, and its diameter is 3. It can be verified that this is indeed the CTC, which is the 4-truss containing the query vertices $Q$ with the smallest diameter.



The problem of CTC search is very challenging. A connected $k$-truss with the largest $k$ containing query vertices can be found in polynomial time. However, finding such a $k$-truss with the minimum diameter is NP-hard \cite{huang2015approximate}. Moreover, it is even hard to approximate the \ctcp within a factor better than 2. Here, the approximation is with regard to the minimum diameter. 

To find the closest truss community, a simple but effective greedy algorithm is proposed in \cite{huang2015approximate}. The method uses a greedy strategy for finding a CTC that delivers a 2-approximation to the optimal solution, thus essentially matching the lower bound.  Here is an overview of this algorithm. First, given a graph $G$ and query vertices $Q$, we find a maximal connected \ktruss, denoted $G_0$, containing $Q$ and having the largest trussness. As $G_0$ may have a large diameter, we iteratively remove vertices far away from the query vertices, while maintaining the trussness of the remainder subgraph at $k$. Actually, this algorithm can find a connected $k$-truss with the largest $k$ containing query vertices, which achieves the smallest query distance in optimal. According to the inequality of query distance and graph diameter, this answer is a 2-approximation to CTC \cite{huang2015approximate}.

In order to improve the efficiency of CTC search, Huang et. al proposed two new techniques of bulk deletion and local exploration. One of them is based on bulk deletion of vertices far away from query vertices. This speeds up the pruning process, by deleting at least $k$ vertices in batch, to achieve quick termination while sacrificing some approximation ratio. Second, they also propose a heuristic strategy of local exploration to quickly find the closest truss community in the local neighborhood of query vertices. The key idea is as follows. It first forms a Steiner tree to connect all query vertices, and then expand the Steiner tree to a $k$-truss with the largest $k$ by involving the local neighborhood of the query vertices. Finally, to reduce the diameter, it iteratively removes the furthest vertices from this $k$-truss using the bulk deletion.










\subsection{Keyword-Based Attributed Graphs}
\label{sec:ktrussInfluence}
In this section, we introduce a \truss-based community search model on  a keyword-based attributed graph where vertices are associated with a set of keywords. Huang and Lakshmanan \cite{Huang:2017:ATC} proposed an attribute-driven truss community model, denoted by ATC, which finds the densely inter-connected communities containing query vertices with similar query attributes. ATC is equipped with two key components of \kdtruss and an attribute score function.

\begin{figure}[t]
\centering
{
\subfigure[\small{An attributed graph $G$}]{\includegraphics[width=0.54\linewidth]{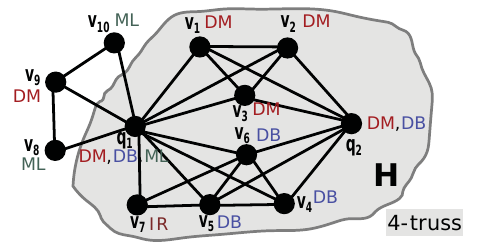} }
\subfigure[$H$]{\includegraphics[width=0.43\linewidth]{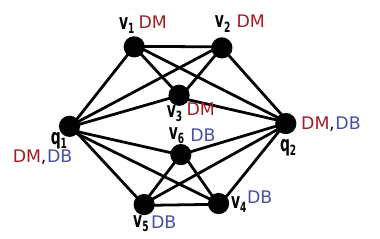} }
}
\caption{An example of attributed truss community search with query vertices $V_q=$ $\{q_1,$$ q_2\}$ and query attributes $W_q=\{$ `DB', `DM'$\}$. Here, $k=4$.}
\label{fig.subcom}
\end{figure}

To capture dense cohesiveness and low communication cost, ATC builds upon a notion of dense and tight substructure called \kdtruss. A \kdtruss requires that every edge is contained at least $(k-2)$ triangles, and the communication cost between the vertices of $H$ and the query vertices is no greater than $d$. By definition, the cohesiveness of a \kdtruss increases with $k$, and its proximity to query vertices increases with decreasing $d$. For instance, $H$ in Fig. \ref{fig.subcom}(b) for $V_q=\{q_1, q_2\}$ is a \kdtruss with $k=4$ and $d=2$.

To measure the goodness of an attributed community w.r.t. attribute coverage and correlation, an attribute score function is developed for ATC.  Let $\keyw(H, W_q)$ be the attribute score of community $H$ w.r.t. query attributes $W_q$. Then, $\keyw(H,W_q) = \sum_{w\in W_q} \frac{\score(H,w)^2}{|V(H)|}$, where $\score(H,w) = |V_w\cap V(H)|$ is the number of vertices covering query attribute $w$. The function $\keyw(H,W_q)$ satisfies three important properties as follows.  \underline{Property 1}: The more query attributes that are covered by some vertices of $H$, the higher score of $\keyw(H, W_q)$. The rationale is obvious; \underline{Property 2}: The more vertices that contain an attribute $w\in W_q$, the higher the contribution of $w$ should be toward the overall score $\keyw(H, W_q)$. The intuition is that attributes that are covered by more vertices of $H$ signify homogeneity within the community w.r.t. shared query attributes; \underline{Property 3}: The more vertices of $H$ that are irrelevant to the query, the lower the score $\keyw(H, W_q)$. The more query attributes a community has that are shared by more of its vertices, the higher its attribute score. For example, consider the query $Q=(\{q_1\}, \{$`DB', `DM'$\})$ on the running example graph of Fig. \ref{fig.subcom}(a). Intuitively, we can see that  $H$ has 5 vertices covering `DB' and `DM' each and also has the highest attribute score (i.e., $\keyw(H, W_q) = \frac{5^2}{8} + \frac{5^2}{8}=6.25$), which is the attributed truss community. On the other hand, the induced subgraph of $G$ by vertices $\{q_1, q_2, v_1, v_2, v_3\}$ and  $\{q_1, q_2, v_4, v_5, v_6\}$ are mainly focused in one area (`DB' or `DM'), achieving the score of $5.8$.

Based on the \kdtruss and $\keyw(H, W_q)$, Huang et al. \cite{Huang:2017:ATC} studied the ATC problem.

\begin{problem}
[ATC search]
\label{prob:atc}
Given a graph $G$, a query $Q = (V_q, W_q)$, and two numbers $k$ and $d$, return an attributed truss community (ATC) $H$, satisfying the following properties:
\begin{enumerate}
  \item $H$ is a \kdtruss containing $V_q$.

  \item $H$ has the maximum attribute score $\keyw(H, W_q)$ among all subgraphs satisfying property 1.
\end{enumerate}
\end{problem}



Theoretical proofs show that ATC search is NP-hard \cite{Huang:2017:ATC}, which shows the challenging for computation. To help efficiently processing of ATC queries, \cite{Huang:2017:ATC} presents a greedy algorithmic framework for finding an ATC in a top-down search manner. The general ideas of this algorithm has three steps. First, it finds the maximal \kdtruss of original graph $G$ as a candidate. Second, it iteratively removes vertices with the smallest ``attribute marginal gains'' from the candidate graph, and maintains the remaining graph as a \kdtruss, until no longer possible. The removed vertices have the smallest contribution to attribute score function $\keyw(H, W_q)$. Finally, it returns a \kdtruss with the maximum attribute score among all generated candidate graphs as the answer. If there exists more than one \kdtruss with the maximum attribute scores, the algorithms just outputs one answer.

To further improve the search efficiency while ensuring high quality, a novel index called attributed-truss index (\ati) is developed. The \ati consists of two components: structural trussness and attribute trussness, which maintain known graph structure and attribute information. \ati can quickly identify a good candidate of $(k,d)$-truss to the answer. In addition, another technique of local exploration is applied for efficiently detecting a small neighborhood subgraph around query vertices, which tends to be densely and closely connected with the query attributes.


\subsection{Weight-Based Attributed Graphs}
\label{sec:ktrussLocation}
In this section, we consider an undirected weighted graph $G=(V, E, W)$, where the weight of $e$ is denoted by $w(e) \in W$, representing the importance between vertices $u$ and $v$. Weighted graphs naturally exist in the real-world applications. For instance, in the collaboration network, the edge weights may represent the number of co-authored articles between two authors. Fig. \ref{fig.WTC} depicts an undirected weighted graph $G$, e.g., edge $(q, s_1)$  has a weight of 0.8.  Taking the edge weights into consideration, community search on weighted graphs can find communities capturing more semantics. Zheng et al. \cite{Zheng:IS:2017} proposed a model of weighted truss community (WTC):

\begin{figure}[t]
\small
\vskip -0.1in
\centering
\includegraphics[width=0.60\linewidth]{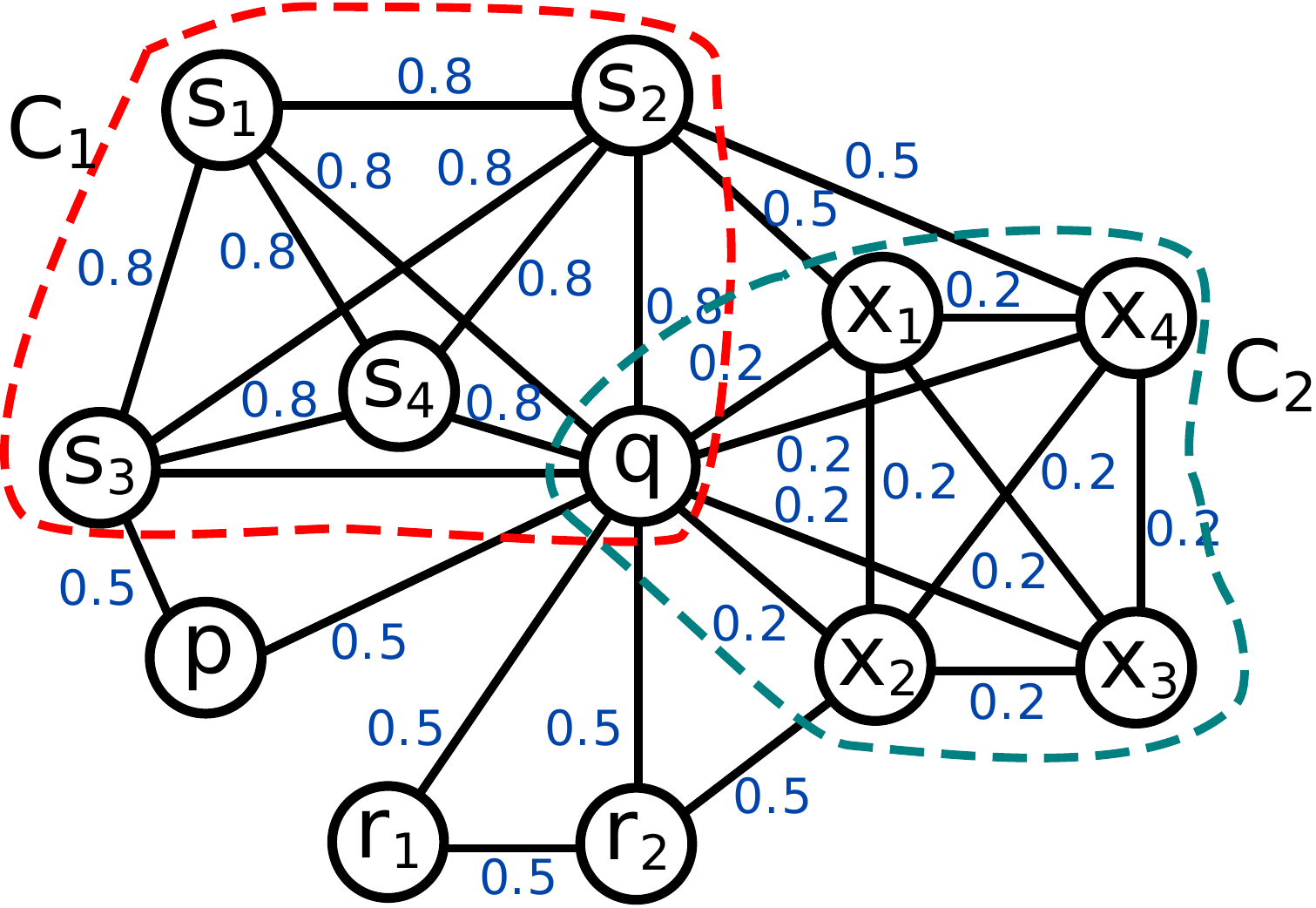}
\caption{An example of weighted truss community search.}
\label{fig.WTC}
\end{figure}


\begin{definition}[Weighted Truss Community]
\label{prob:wtc}
Given an undirected weighted graph $G$=($V$, $E$, $W$), and a positive integer $k$, a weighted $k$-truss community is an induced subgraph $H \subseteq G$, the following properties hold:
\begin{enumerate}
  \item \textbf{Connectivity}. $\forall e_1, e_2\in E(H)$, $e_1$ and $e_2$ are triangle connected in $H$;
  \item \textbf{Cohesiveness}. $\forall e\in E(H)$, $\sup_H(e) \geq k-2$;
  \item \textbf{Maximal Structure}. $H$ is a maximal induced subgraph that satisfies Properties 1 and 2.
\end{enumerate}
\end{definition}

In the weighted $k$-truss community model, Property 1 adopts the same constraint of triangle connectivity as other $k$-truss community models \cite{k-truss2014}; Property 2 requires the community to satisfy the structure of $k$-truss; Property 3 can guarantee the property of maximal structure in the weighted $k$-truss community. Given a weighted truss community $H$, the community weight of $H$ is defined as $w(H)=\min_{e\in E(H)} w(e)$. To discover the communities with large weights, Zheng et al. \cite{Zheng:IS:2017} investigated the problem of weighted truss community (WTC) search.

\begin{problem}
[WTC search]
\label{problem:WTC}
Given an undirected weighted graph $G(V,E,W)$, and parameters $k$ and $r$, find the top-$r$ weighted $k$-truss communities $H$ with the largest weights $w(H)$.
\end{problem}

Consider a weighted graph $G$ in Fig. \ref{fig.WTC}, $k=5$, and $r=1$. The community $C_1$ shown in Fig. \ref{fig.WTC} has the weight $w(C_1)=0.8$, which is larger than the weight of community $C_2$ as $w(C_2)=0.2$. Thus, $C_1$ is the answer of WTC search with the largest weight.


Straightforward to enumerate all weighted $k$-truss communities to find the $r$ communities with the largest community weights is impractical in large graphs. To  speed up the search efficiency, an index structure called KEP-Index is designed. KEP-Index is built upon the observation that all the communities can be organized into a tree-shaped structure. This is because all the weighted $k$-truss communities from a partial order relationship for each value of $k$. By indexing all the pre-computed weighted $k$-truss communities in a tree-shaped structure, WTC search can be done in the linear time w.r.t. the answer size, which is optimal.

\subsection{Discussions}
\label{sec:ktrussDiscuss}


Generally, the $k$-truss-based CS solutions on simple graphs can be divided into two groups, where the first group \cite{k-truss2014,Akbas:VLDB:2017} computes the $k$-truss community, while the second group \cite{huang2015approximate} aims to find closest communities. In the first group, Akbas et al. \cite{Akbas:VLDB:2017} improved the efficiency of \cite{k-truss2014} by developing a novel index.
For attributed graphs, there are two CS solutions, which consider keywords \cite{Huang:2017:ATC} and influence values \cite{Zheng:IS:2017} respectively. For all these studies above, both online and index-based algorithm are developed.

For practitioners, to perform CS, we would like to offer some suggestions:
(1) We should figure out the type of graph (e.g., simple graphs and attributed graphs) in the application.
(2) For simple graphs, there are two community models, i.e., triangle-connected model and closest model. Generally, the triangle-connected model \cite{k-truss2014,Akbas:VLDB:2017} is suitable for one single query vertex to discover all overlapping communities containing it, while the closest model \cite{huang2015approximate} is suitable to discover one closest community containing multiple query vertices, which is not strict to one query vertex. Moreover, triangle connectivity is weaker than the optimization metric of minimum diameter. According to our experience in the real-world applications, the discovered closest community has smaller graph size than triangle-connected truss community.
(3) For triangle-connected model \cite{k-truss2014,Akbas:VLDB:2017}, the index-based algorithm in \cite{Akbas:VLDB:2017} is faster than that in \cite{k-truss2014}.

\section{$K$-Clique-Based Community Search}
\label{sec:kclique}

In this section, we survey CS solutions that use $k$-clique or its variants to capture the structure cohesiveness.
We first briefly introduce the $k$-clique model and its variants in Section~\ref{subsec:kclique_def}.
Then, we present CS solutions using $k$-clique component and $k$-plex models in Sections~\ref{subsec:kplex} and \ref{subsec:kplex}. After that, we discuss the most influential CS using $k$-clique in Section \ref{subsec:kcliqueInfluence}. Finally, we discuss these studies in Section~\ref{subsec:kclique_discussion}.

\subsection{$K$-Clique and Its Variants}
\label{subsec:kclique_def}

Recall that by Definition \ref{def:kclique}, a $k$-clique is a complete graph with $k$ vertices where there is an edge between every pair of vertices. The $k$-clique model has been widely used for the overlapping community detection (e.g.,~\cite{article05clique,DBLP:journals/bioinformatics/AdamcsekPFDV06}).
As the condition of $k$-clique is strict, some relaxed variants
such as $\gamma$-quasi-$k$-clique~\cite{brunato2007effectively,online-sigmod2013} and
$k$-plex~\cite{seidman1978graph}, are proposed to identify cohesive subgraphs.
Below are detailed definitions.

\begin{definition}[$\gamma$-quasi-$k$-clique~\cite{brunato2007effectively,online-sigmod2013}]
\label{def:quasikclique}
A $\gamma$-quasi-$k$-clique is a graph with $k$ vertices and at least $\lfloor \gamma\frac{k(k-1)}{2} \rfloor$ edges, where $0\leq\gamma\leq1$.
\end{definition}
When $\gamma = 1$, the corresponding $\gamma$-quasi-$k$-clique is a $k$-clique.
We can tune the desired cohesiveness of the $k$ vertices by varying $\gamma$ value.

\begin{definition}[$k$-plex~\cite{seidman1978graph}]
\label{def:kplex}
A graph $G(V,E)$ is a $k$-plex, if for each vertex $v\in V$, $v$ has at least $|V|-k$ neighbors in $G$, where $1\leq k\leq |V|$.
\end{definition}

When $k$=1, the $k$-plex is exactly a $k$-clique. Clearly, by setting a smaller value of $k$, we can obtain a more cohesive $k$-plex. The problem of finding a $k$-plex from a given graph for an integer $k$ is NP-hard \cite{balasundaram2011clique}.

Another way to relax the constraint of $k$-clique is to consider the connection of two vertices.

\begin{definition}[$kr$-clique~\cite{kclique2017}]
\label{def:krclique}
Given a graph $G$ and two integers $k$ and $r$, a $kr$-clique $S$ is an induced subgraph of $G$
such that: (1) the number of vertices in $S$ is at least $k$;
and (2) any two vertices in $S$ can reach each other within $r$ hops.
\end{definition}

Clearly the problem of finding  $kr$-clique is NP-hard because $kr$-clique is a $k$-clique when $r$=1.

\subsection{$K$-Clique-Based Community Search}
\label{subsec:clique_community}
In Section~\ref{subsub:kcliquecom}, we introduce the seminar work on overlapping community detection~\cite{article05clique},
in which the $k$-clique component is proposed.
Section~\ref{subsub:kcliquecomsearch} presents the community search algorithm based on the relaxation of $k$-clique component, while Section~\ref{subsubsec:densest_clique_community} studies the densest $k$-clique community search.

\subsubsection{$K$-Clique-Based Community}
\label{subsub:kcliquecom}

In~\cite{article05clique}, Palla {\emph et al.} showed that that many real networks are characterized by well-defined
overlapping communities. For instance, a person may belong to three different communities related to school, hobby and family. For a given graph $G$, a \textit{k-clique graph} $G_k$ can be derived where each node is a $k$-clique
in $G$ and there is an edge if two nodes ($k$-cliques) are adjacent,
i.e., they share $k-1$ vertices in $G$.
Then the $k$-clique communities are the union of all adjacent $k$-cliques, which are defined as follows.

\begin{definition} [$k$-clique component]
Let $C$ denote a connected component in the $k$-clique graph, then a \textit{k-clique component} is the union of all $k$-cliques represented by vertices in $C$.
\end{definition}

One may explore the communities of the graph based on the $k$-cliques and their adjacency, and a graph vertex may belong to several communities.
Efficient $k$-clique component detection algorithm is presented in~\cite{DBLP:journals/bioinformatics/AdamcsekPFDV06}.
Particularly, considering that each $k$-clique must be contained by at least one maximal clique,
they first identify all maximal cliques of the network and then
enumerate the communities by carrying out a standard component analysis of the clique overlap matrix.


\subsubsection{$K$-Clique-Based Community Search}
\label{subsub:kcliquecomsearch}

In~\cite{online-sigmod2013},  Cui {\emph et al.} showed that there are two shortcomings in the $k$-clique community model:
(1) there are overwhelming number of $k$-cliques communities in real-life graphs;
and (2) the $k$-clique constraint and the definition of adjacent (i.e., sharing $k$-1 common vertices) are not flexible in practice.
To address these two shortcomings, they proposed an online community search (OCS) problem.
Instead of enumerating all communities, they focused on the search of the communities containing a given query vertex $q$. They relaxed the $k$-clique adjacent from $k-1$ common vertices to $\alpha$ vertices, namely $\alpha$-adjacency.
They also relaxed $k$-clique model to $\gamma$-quasi-$k$-clique model (Definition~\ref{def:quasikclique}).
By doing this, the $k$-clique components in the $k$-clique communities are relaxed to the $\gamma$-quasi-$k$-clique components. Below is the formal problem definition.

\begin{problem}
[($\alpha$, $\gamma$)-OCS]
\label{prob:kcliquesearch}
Given an undirected simple graph $G(V,E)$, a query vertex $q \in V$, and an integer $k$,
an integer $\alpha \leq k-1$, and a real value $\gamma$ with $0 \leq \gamma \leq 1$,
find all  $\gamma$-quasi-$k$-clique components containing  query vertex $q$.
\end{problem}

Clearly, a $k$-clique component search is a special case of ($\alpha$, $\gamma$)-OCS
with $\alpha=k-1$ and $\gamma = 1$. By reducing to $k$-clique decision problem, it is shown in~\cite{online-sigmod2013} that the ($\alpha$, $\gamma$)-OCS problem is $\#P$-Complete. It is shown that the density of each community in ($\alpha$, $\gamma$)-OCS is at least $2 \max\{0, \min \{ f(1), f(\alpha) \} \}$
where $f(x) = \frac{ \gamma {k \choose 2} {k-x \choose 2}}{x}$.
Both exact and approximate solutions are proposed in~\cite{online-sigmod2013}.
A naive algorithm for exact solution is to enumerate all $\gamma$-quasi-$k$-cliques containing the query vertex $q$,
and then compute the $\gamma$-quasi-$k$-clique components based on the $\alpha$-adjacency.
To avoid enumerating cliques belonging to none of the valid communities, a new computing framework is proposed
to check the adjacency when a clique is discovered. By carefully maintaining the visit status of each clique, authors further optimize the searching cost.
Authors also proposed an approximate solution.
To reduce the search space, the approximate algorithm only enumerates an unvisited clique which contains
at least one new vertex not contained by any existing community.
A heuristic is proposed to choose a vertex sequence such that the resulting clique sequence is short, leading to a good approximation solution.

\subsubsection{Densest Clique Percolation Community Search}
\label{subsubsec:densest_clique_community}

Following the $k$-clique community model in~\cite{article05clique}, Yuan et al.
studied the problem of densest clique percolation community search~\cite{kclique2018},
where a \textit{k-clique percolation community} (KCPC) is a \textit{k-clique component} in~\cite{article05clique}.
In particular, they aimed to find the $k$-clique percolation community with the maximum $k$ value
that contains a given set of query vertices.

\begin{problem}
\label{prob:densest}
Given an undirected simple graph $G(V,E)$ and a set of query vertex $Q \subseteq V$,
the problem of the \textit{\underline{d}{ensest} \underline{c}lique \underline{p}ercolation \underline{c}ommunity} (DCPC) search is to find the $k$-clique component with the maximum $k$ value that contains all the vertices in $Q$.
\end{problem}

\begin{figure}[hbt]
\centering
\includegraphics[width=\columnwidth]{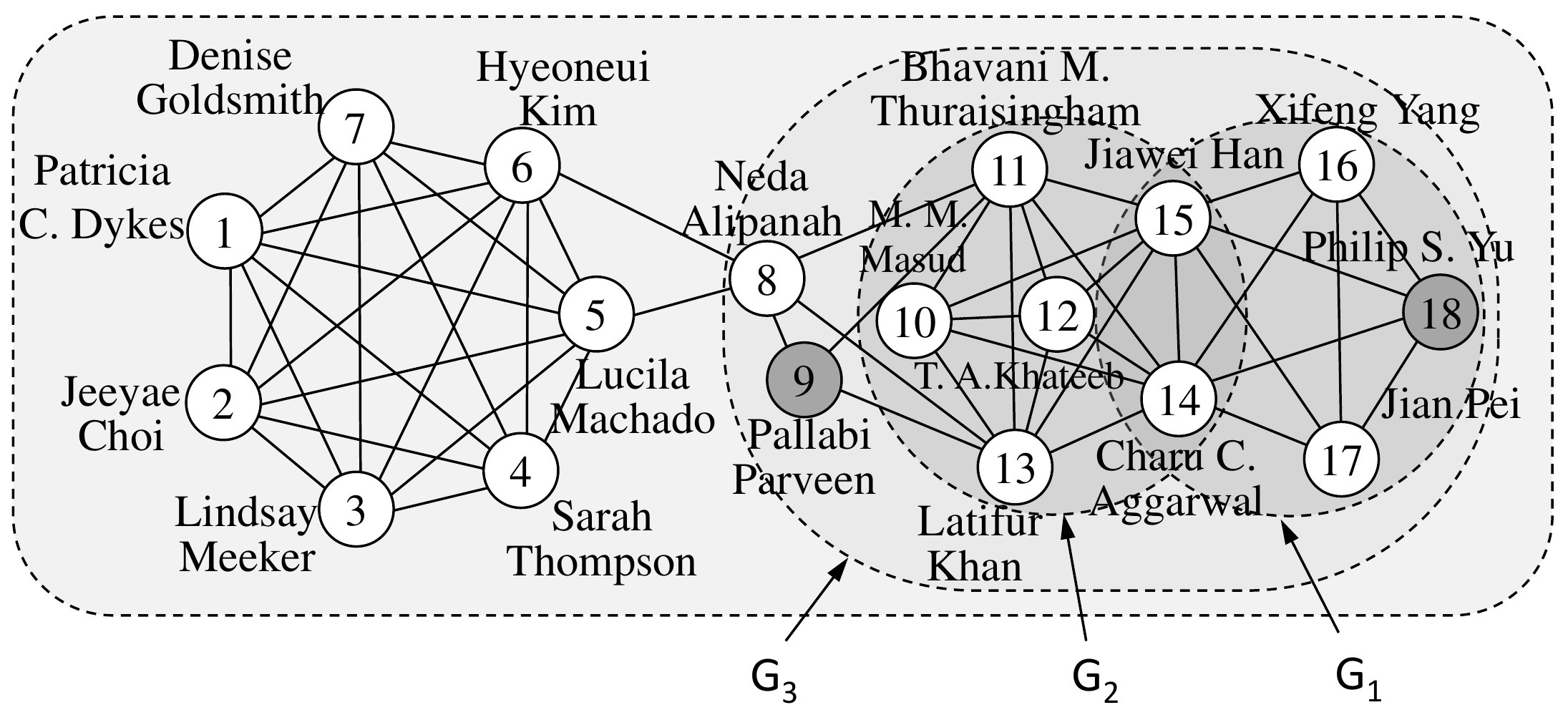}
\caption{Illustrating DCPC search \cite{kclique2018}.}
\label{fig:densest_community}
\end{figure}

Fig. \ref{fig:densest_community} in~\cite{kclique2018}
illustrates a part of the collaboration network in DBLP,
in which each vertex represents an author and each edge indicates the co-author relationship between two authors.
$G_1$ is a 4-clique percolation community as it is a maximal union of five adjacent $4$-cliques: $\{v_{14}, v_{15}, v_{16}, v_{17}\}$, $\{v_{14}, v_{15}, v_{16}, v_{18}\}$, $\{v_{14}, v_{15}, v_{17}, v_{18}\}$, $\{v_{14}, v_{16}, v_{17}, v_{18}\}$, $\{v_{15}, v_{16}, v_{17}, v_{18}\}$, and any two $4$-cliques share 3 nodes. Similarly, $G_2$ is also a 4-clique percolation community. $G_1$ overlaps $G_2$ with nodes ${v_{14}, v_{15}}$. Given a query $q = \{v_9, v_{18}\}$, the densest clique percolation community of $q$ is the  $3$-clique percolation community $G_3$ since $G_3$ is the $k$-clique percolation community with  maximum $k$ value that contains $v_9$ and $v_{18}$.

A baseline solution is to start from the maximal possible $k$ value
and check if there is a KCPC by applying the $k$-clique component detection algorithm in~\cite{article05clique}.
If there is no KCPC detected, the $k$ value will be decreased by one until a KCPC is detected.
To efficiently support online DCPC search, an index-based approach is developed in~\cite{kclique2018}.
Particularly, based on the observation that a $k$-clique component can be treated as a union of \textit{maximal} cliques, they take maximal cliques as building blocks of $k$-clique components and propose a  tree-structure named
\textit{clique adjacency tree} which can efficiently identify the $k$-clique components for a given $k$ value.
The authors further developed a new tree-structure named \textit{ordered adjacency tree}
such that only the subtrees related to the query vertices will be explored.
Together with maximal cliques and their inverted indexes, a compact index structure named DCPC-Index
is proposed to support efficient DCPC queries.

\subsection{$K$-Plex-Based Community Search}
\label{subsec:kplex}


\subsubsection{Social Group Query (SGQ)}
\label{sec:sgq}

Problem~\ref{prob:sgq} presents SGQ, which was designed for suggesting attendees in activity planning \cite{yang2011social}.

\begin{problem}[SGQ]
\label{prob:sgq}
Given a simple undirected graph $G(V,E)$, an activity initiator $q\in V$, three integers $p$, $s$, and $k$,
return a set $F$ of vertices from $G$ such that the following properties hold:
\begin{enumerate}
  \item $|F|$=$p$;
  \item The length of the minimum distance path between $v$ and $q$, $d_{v,q}$, is at most $s$;
  \item Each vertex $v\in F$ is allowed to share no edges with at most $k$ other vertices in $F$;
  \item The total social distance $\Sigma_{v\in F}d_{v,q}$ is minimized.
\end{enumerate}
\end{problem}

In Problem~\ref{prob:sgq}, Property 1 controls the expected number of attendees in the activity;
Property 2 specifies a radius constraints which requires each attendee is close to $q$ in the graph $G$;
Property 3 requires that each attendee is acquainted with other attendees by following the $k$-plex model;
Property 4 ensures that the returned group is the most compact one among all the groups satisfying all the above properties.

The SGQ problem is computationally challenging because it is NP-hard, which can be proved by a reduction from the $k$-plex problem \cite{balasundaram2011clique}. To answer SGQ, Yang et al. \cite{yang2011social} proposed an efficient solution {\tt SGSelect}. The idea is that we can first extract a subgraph $H\subseteq G$ by using the radius constraint. Then, starting from $q$, we iteratively explore vertices in $H$ to derive the optimal solution. In each iteration, we can keep track of a set of vertices that satisfy the constraint of $k$, until the set has $p$ vertices. To further speedup this process, some effective pruning criteria have been developed. For example, to choose vertices, we can give high priorities for vertices that may significantly increase the total social distance. Also, during the search process, we can prune vertices which would not lead to eventual answer by considering the acquaintance constraint $p$ and social radius constraint $s$.

In addition, Yang et al. \cite{yang2011social} studied another query, called social-temporal group query (STGQ), which generalizes SGQ by considering the available time of each candidate attendee. In specific, it finds a group of vertices satisfying:
(1) all constraints in an SGQ; and
(2) all the attendees are available in a time period $[t, t+{\delta_t}]$, where $t$ is time slot and $\delta_t$ is query parameter.
The STGQ problem is also NP-hard and some efficient solutions are developed. For details, please refer to \cite{yang2011social}.

\subsubsection{Maximum $k$-Plex Community Query (MCKPQ)}
\label{sec:mckpq}

In \cite{wang2017query}, Wang et al. proposed and studied the maximum $k$-plex community query (MCKPQ):

\begin{problem}[MCKPQ]
\label{prob:mckpq}
Given a simple undirected graph $G(V,E)$, a set of query vertices $Q\in V$, an integer $k$,
return a subgraph $G_Q(V_Q,E_Q)\subseteq G(V,E)$ such that the following properties hold:
\begin{enumerate}
  \item \textbf{Connectivity}. $G_Q$ is connected and contains $Q$;
  \item \textbf{Structure cohesiveness}. $G_Q$ is a $k$-plex;
  \item \textbf{Maximal structure}. There exists no other $G_Q'\subseteq G$ satisfying the above properties and $G_Q\subset G_Q'$.
\end{enumerate}
\end{problem}

A good property of MCKPQ is that the communities returned by an MCKPQ can avoid the free rider effect, which has been introduced and discussed in Section~\ref{sec:ktruss}. Nevertheless, the MCKPQ problem is very computationally challenging, because it is NP-complete, which can be proved by a reduction from the $k$-plex problem \cite{balasundaram2011clique}. Moreover, it is hard to approximate for MCKPQ problem in polynomial time within a factor $n^{1-\epsilon}$.

A basic solution to the MCKPQ problem is to use the generate-and-verify method, which enumerates all the $k$-plexes in the whole search space, and then returns the one with the largest size. Obviously, this method is too expensive and impractical for large graphs. To alleviate this issue, Wang et al. developed a more advanced method based on the branch-and-bound paradigm with some effective pruning criteria and a heuristic method which performs fast but has no theoretical guarantee \cite{wang2017query}. We skip the details due to space limitation.

\subsection{Most Influential Community Search}
\label{subsec:kcliqueInfluence}

In~\cite{kclique2017}, Li et al. proposed the problem of most influential community search,
which aim to find the most influential cohesive subgraph.
The concept of $kr$-clique community (Definition~\ref{def:krclique}) is proposed to capture the cohesiveness of a set of vertices. In addition to cohesiveness, authors also considered the influence of the community.
Following the popular Linear Threshold (LT) model~\cite{DBLP:journals/tkde/LeeC15},
the aggregate influence probability of a community $C$ w.r.t a vertex $v$,
denoted by $Pr(v|C)$, is defined as follows:
$$
Pr(v|C) = 1 - \prod_{u \in C} (1 - P_{u \rightarrow v})
$$
where $P_{u \rightarrow v}$ is the probability that $v$ is influenced by $u$.
Note that there is a influence probability $P_{uv}$ for each edge ($u$, $v$) in $G$,
and $P_{u \rightarrow v}$  is computed by multiplying the influence
of the edges along the maximum influence path~\cite{DBLP:journals/tkde/LeeC15}
from $u$ to $v$.
Given a probabilistic threshold $\Delta$,
the influence score of the community $C$ is the number of vertices in $G \setminus C$
with aggregate influence not less than $\Delta$, denoted by $score(C)$.
Below is the problem definition.

\begin{problem}
\label{prob:infCS}
Given a simple graph $G$ where each edge has an influence probability, the problem of the most influential community search is to find a maximal $kr$-clique community with the highest influence score.
\end{problem}

It is shown in~\cite{kclique2017} that the problem is NP-hard because of the clique computation.
A baseline solution is to access the vertices by their individual influence and
compute the maximal $kr$-clique for each vertex.
To improve efficiency, a tree structure named $C$-Tree is proposed such that any $kr$-clique community
can be generated efficiently. Four efficient search algorithms are developed to
significantly prune the search space based on the $kr$-clique constraints and the influence scores.

\subsection{Discussions}
\label{subsec:kclique_discussion}

In this section, we survey the CS solutions \cite{online-sigmod2013,kclique2018,yang2011social,wang2017query,kclique2017} using $k$-clique model. We can divide them into two groups, where the first group \cite{online-sigmod2013,kclique2018,yang2011social,wang2017query} focuses on simple graphs, while the second group \cite{kclique2017} is developed for attributed graphs.
In the first group, the first one \cite{online-sigmod2013} uses quasi-clique model, the second one \cite{kclique2018} adopts $k$-clique model, and the last two \cite{yang2011social,wang2017query} are based on $k$-plex model.
However, to our best knowledge, there is no systematic study to compare the goodness of different $k$-clique based models in real-life applications, which is crucial for researchers and practitioners to choose desirable models in practice. Moreover, there is no investigation on the trade-off between the computing time complexity and the flexibility of these models. It will be interesting to fill these two gaps in the future study.

\section{$K$-ECC-Based Community Search}
\label{sec:kecc}

In this section, we review CS studies \cite{Chang:SIGMOD:2015,hu2016querying} that use the $k$-ECC model as the community structure cohesiveness. Given a graph $G$ and a set $Q$ of vertices, their general goals are to find a subgraph $H$ of $G$, which contains $Q$ and has the maximum edge-connectivity, also called the Steiner Maximum-Connected Subgraph (SMCS). Their difference is that one maximizes the size of $H$ \cite{Chang:SIGMOD:2015}, while the other one tries to minimize the size of $H$ \cite{hu2016querying}.

\subsection{Maximum SMCS}
\label{sec:maxSMCS}

In \cite{Chang:SIGMOD:2015}, Chang et al. computed the maximum SMCS for a set of query vertices $Q$, which is defined as follows.

\begin{problem}
\label{prob:keccMax}
Given an undirected simple graph $G(V,E)$, and a set of query vertices $Q\subseteq V$, return a subgraph $H(V_H,E_H)$ of $G$, such that
\begin{enumerate}
  \item $V_H$ contains $Q$;
  \item $\lambda(H)$ is maximized;
  \item There exists no other subgraph $H'$ satisfying the above properties, such that $H\subset H'$.
\end{enumerate}
\end{problem}

\begin{figure}[ht]
\centering
\hspace*{-.3cm}
\begin{tabular}{c c}
  \begin{minipage}{5.7cm}
	\includegraphics[width=5.7cm]{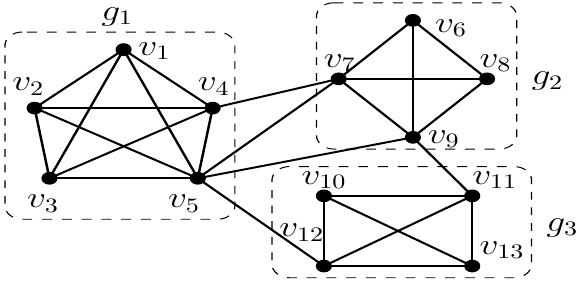}
  \end{minipage}
  &
  \begin{minipage}{2.4cm}
	\includegraphics[width=2.4cm]{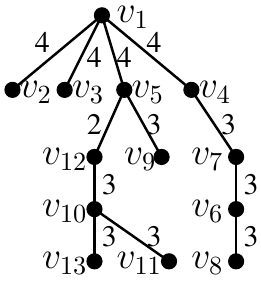}
  \end{minipage}
  \\
  (a) a graph
  &
  (b) the index
\end{tabular}
\caption{An example for illustrating maximum SMCS \cite{Chang:SIGMOD:2015}.}
\label{fig:keccMaxEg}
\end{figure}

For example, consider the graph in Fig. \ref{fig:keccMaxEg}(a). Let $Q$=$\{v_1$,$v_4\}$. Then, for this query we will return the subgraph $g_1$, and its connectivity is $\lambda(g_1)$=4.

A basic solution of Problem \ref{prob:keccMax} is to sequentially enumerate all the maximal $k$-ECCs by varying $k$ from $|V|$ to 1, and stops when the first $k$-ECC which contains $Q$ is found. Then, the first $k$-ECC is returned as the community.
In the literature, there are two efficient $k$-ECC enumeration algorithms. One is based on graph decomposition \cite{Chang:SIGMOD:2013}, while the other one is based on the random contraction \cite{akiba2013linear}. As shown in \cite{Chang:SIGMOD:2015}, the basic solution takes $O(|V|\cdot h\cdot l\cdot |E|)$ time if the first $k$-ECC enumeration algorithm is adopted, or $O(|V|\cdot t\cdot |E|)$ time if the second one is used, where $h$ and $l$ are bounded by small constants for real graphs, and $t$=$O(log^2\cdot |V|)$. Obviously, both of them are inefficient for large graphs.

To improve the query efficiency, Chang et al. proposed a novel compact index structure, which allows the query can be answered in optimal time cost, i.e., the time cost is linear to the size of $H$. The index is built based on a key observation that for any pair of vertices $u$ and $v$ in $H$, their connectivity $\lambda(u,v)$ is at least $\lambda(H)$. This implies, if the connectivity of each pair of vertices in $G$ is preserved, then the query can be answered in linear time cost, because we can first get $\lambda(H)$ by checking the connectivity of vertex pairs in $Q$, and then find $H$ by traversing the connected edges whose connectivity are at least $\lambda(H)$.

To preserve all the connectivity information of $G$, Chang et al. developed the concept of {\it connectivity graph} $G_c$ for the graph $G$, which has the same sets of vertices and edges with $G$, and for each edge $(u,v)\in G_c$, it is associated with a connectivity value denoting the edge-connectivity between vertices $u$ and $v$ in $G$. Then, the maximum spanning tree (MST) of $G_c$ is the index structure built for $G$. For example, Fig. \ref{fig:keccMaxEg}(b) presents the index structure for the graph in Fig. \ref{fig:keccMaxEg}(a). The index can be built by first constructing the connectivity graph $G_c$ and then computing the MST from $G_c$. Clearly, the space cost of the MST is $O(|V|)$ since it has $|V|$ vertices and at most $|V|$--1 edges.

Based on the index MST, Chang et al. proposed an efficient query algorithm to solve Problem \ref{prob:keccMax}. Specifically, it first computes $\lambda(H)$ by using the MST, and then finds the maximum SMCS by collecting the subtree of MST, whose edges have connectivity values being at least $\lambda(H)$. By using the technique of lowest common ancestor (LCA), the query can achieve a time cost of $O(|H_V|)$, which is optimal since outputting the vertex set of $H$ takes $O(V_H)$ time.

In addition, the authors studied a variant of Problem \ref{prob:keccMax} by imposing an additional constraint, which requires the number of vertices in $H$ is at least $L$, where $L$ is a parameter specified by the user. It can also be solved in optimal time cost with the index MST.

\subsection{Minimum and Minimal SMCS's}
\label{sec:minSMCS}

In~\cite{hu2017querying}, Hu et al. found that although the maximum SMCS has a high cohesiveness (i.e., high {\it connectivity}), the size of maximum SMCS's are often extremely large and complex. For example, on the DBLP bibliographical network that contains $803K$ vertices and $3.2M$ edges, the average number of vertices in a maximum SMCS is over $400K$. This not only hinders the analysis of the SMCS structure, but also makes it difficult to be used in real situations. To remedy this issue, Hu et al. examined the discovery of an SMCS that has a small number of vertices. Particularly, they studied the minimum SMCS and minimal SMCS problems:

\begin{problem}[Minimum SMCS]
\label{prob:keccMinimum}
Given an undirected simple graph $G(V,E)$, and a set of query vertices $Q\subseteq V$, return a subgraph $H(V_H,E_H)$ of $G$, such that
\begin{enumerate}
  \item $V_H$ contains $Q$;
  \item $\lambda(H)$ is maximized;
  \item $|H_V|$ is minimized.
\end{enumerate}
\end{problem}

\begin{problem}[Minimal SMCS]
\label{prob:keccMinimal}
Given an undirected simple graph $G(V,E)$, and a set of query vertices $Q\subseteq V$, return a subgraph $H(V_H,E_H)$ of $G$, such that
\begin{enumerate}
  \item $V_H$ contains $Q$;
  \item $\lambda(H)$ is maximized;
  \item There exists no other subgraph $H'\subset H$ satisfying the above properties.
\end{enumerate}
\end{problem}

\begin{figure}[htb]
\centering
\includegraphics[width=0.32\textwidth]{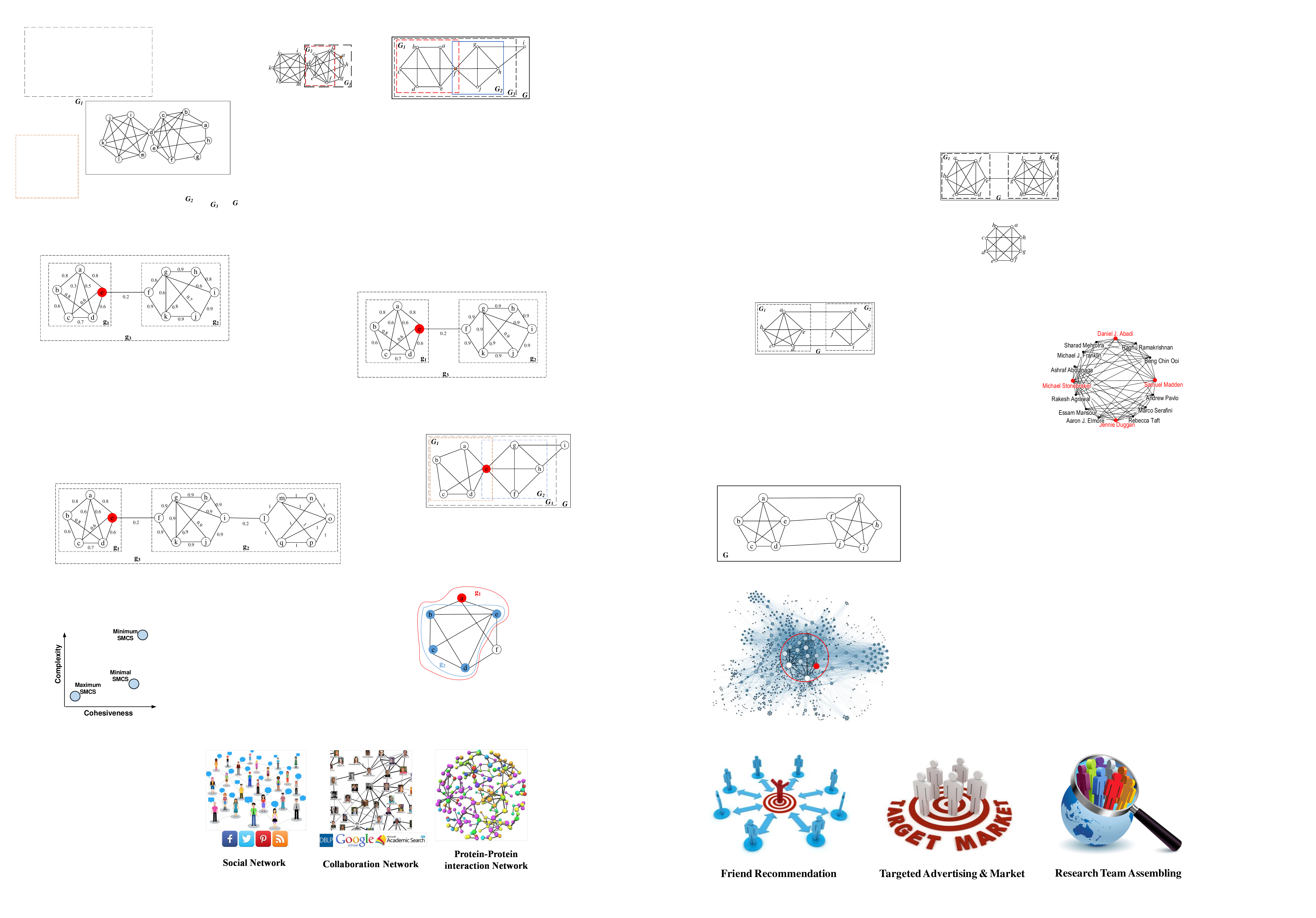}
\caption{The maximum SMCS ($G_3$), minimum SMCS ($G_2$), and minimal SMCS's ($G_1$ and $G_2$) for query Q=\{$f$\}~\cite{hu2017querying}.}
\label{fig:example_smcs}
\end{figure}

Obviously, a minimum SMCS is also a minimal SMCS, and both of them are much smaller than the maximum SMCS. For example, on the DBLP network, their average sizes are less than $0.23K$, while the average size of maximum SMCS is over $400K$. We illustrate these three kinds of SMCS in Fig. \ref{fig:example_smcs}.

In \cite{hu2017querying}, Hu et al. showed that the minimum SMCS problem is APX-hard, since it is a generalization of the \textsc{Steiner Tree} problem (see Section \ref{sec:sizeBounded}). Furthermore, unless P=NP, there does not exist any polynomial-time algorithm that approximates the minimum SMCS problem within any constant ratio. Therefore, it is not only intractable to obtain a minimum SMCS, but also hard to get its approximate version in an accurate manner. To trade off the efficiency and result quality, Hu et al. \cite{hu2017querying} focused on the minimal SMCS problem.

A naive solution for Problem \ref{prob:keccMinimal} is to first adopt the solution in~\cite{Chang:SIGMOD:2015} to compute the maximum SMCS $G'$, and then iteratively refine $G'$ to ensure its minimality. While this solution is simple, it has a high time complexity, since the cost of testing the minimality of an SMCS is high.
To achieve higher efficiency, Hu et al. proposed an {\it Expand-Refine} framework to find a minimal SMCS, which consists of three steps. First, the Steiner-connectivity of the query vertex set $Q$ (i.e., the maximum $\lambda(H)$) is computed. Then, in the \emph{Expand} step, through local expansion of vertices starting from vertices in $Q$, a subgraph $H'$ of $G$ with connectivity being $\lambda(H)$ is obtained. In the \emph{Refine} step, an algorithm is proposed to remove vertices based on the dependence of vertices on their minimal SMCS's.  As a result, the minimal SMCS problem can be solved in a polynomial time cost, i.e., $O(t\cdot\ h\cdot\ l\cdot |E|)$, where $t\textless|H_V|$, and $h$ and $l$ are usually bounded by small constants.
Besides, to further improve the efficiency, the authors relaxed the constraints from two perspectives, namely connectivity and minimality, and computed the approximate SMCS with theoretical guarantee.

In addition, for an important special case with only one query vertex (i.e., $|Q|$=1), Hu et al. developed a customized algorithm for it. The main idea is to keep the processing information related to the current query in a small cache structure, and use these information to answer the subsequent queries. As a result, it performs faster than the solution above.

\subsection{Discussions}
\label{kecc:discuss}

In this section, we review two CS studies that adopt the $k$-ECC model as the community cohesiveness metric. The first one \cite{Chang:SIGMOD:2015} aims to find the maximum SMCS, while the second one \cite{hu2016querying,hu2017querying} tries to find the minimum SMCS. In terms of efficiency, the maximum SMCS can be computed more efficiently. For example, by using the MST index \cite{Chang:SIGMOD:2015}, it can be computed in the optimal time cost. Nevertheless, the maximum SMCS may have size much larger than that of the minimum or minimal SMCS's. This also implies that for practitioners, they have to choose the specific algorithm, based on their specific requirements on community sizes and efficiency.

We remark that these two CS studies mainly focus on simple graphs. It is not clear how to adapt for them for other kinds of graphs, such as directed graphs and attributed graphs. Thus, an interesting future topic is to investigate how to perform CS on other kinds of graphs by adopting the $k$-ECC model.

\section{Other Metrics-Based Community Search}
\label{sec:other}

In this section, we review a particular kind of community search, namely local community detection, which takes an input vertex as a seed and expands the community from the seed according to a specific goodness function. The representative goodness functions are local modularity \cite{clauset2005finding,Luo2006ELC}, query biased density \cite{Wu:VLDB:2015}, personalized pagerank \cite{Kloumann2014}, and neighbor expansion \cite{Mehler2009}.

\subsection{Local Modularity-Based Community Search}
\label{sec:localModu}

Generally, studies of local modularity-based CS follow Problem~\ref{prob:kcoreGlobal} with a local modularity-based goodness function $f$. Two typical such functions are as follows.

\noindent$\bullet$ \textbf{Boundary-based local modularity \cite{clauset2005finding}.}
Assume we have a simple undirected graph $G$ and three sets of vertices, i.e., $\mathcal C$, $\mathcal U$, $\mathcal B \in G$. The known set $\mathcal C$ contains vertices in the known proportion of the community;
the unknown set $\mathcal U$ is a set of vertices that are adjacent to vertices in $\mathcal C$;
and the boundary set $\mathcal B$ is a subset of $\mathcal C$, which contains vertices having neighbors in $\mathcal U$.

By considering all the edges linked to sets $\mathcal B$ and $\mathcal C$, Clauset et al. \cite{clauset2005finding} defined the local modularity of $C$ as $f(\mathcal C)$=$I/T$, where $I$ is the number of edges with no end vertex in $\mathcal U$, and $T$ is the number of edges with at least one end vertex in $\mathcal B$. Intuitively, a good community has a sharp boundary, which means that there are few connections from its boundary set $\mathcal B$ to the unknown set $\mathcal U$, resulting in a higher value of $f(\mathcal C)$.

To uncover a community, Clauset et al. developed an algorithm that works in vertex-at-a-time manner. Let $q$ be a source (seed) vertex. Initially, it lets $\mathcal C$=$\{q\}$ and puts $q$'s neighbors into set $\mathcal U$. At each step, it adds to $\mathcal C$ the neighboring vertex that results in the largest increase of the local modularity. This process continues until it has agglomerated either a given number of vertices $k$, or it has discovered the entire enclosing component, whichever happens first. As a result, its time complexity is $O(k^2d)$, where $d$ is the mean degree and $k$ is the number of vertices to be explored.

\noindent$\bullet$ \textbf{Subgraph degree-based local modularity \cite{Luo2006ELC}.}
Given a subgraph $\mathcal C$ of a graph $G$, Luo et al \cite{Luo2006ELC} defined its indegree, $ind(\mathcal)$, as the number of edges within $\mathcal C$, and its out-degree, $outd(\mathcal C)$, as the number of edges that connect $\mathcal C$ to the remaining part of $G$. Then, they defined the subgraph modularity of $S$ as $f(\mathcal C)$=$ind(\mathcal)/outd(\mathcal)$. Clearly, its value will increase if $\mathcal C$ has more internal edges and fewer external edges.

To find a community, Luo et al. proposed an algorithm consisting of an addition step and a deletion step. Initially, $\mathcal C$ contains a seed vertex $q$ and its neighbors are in a set $\mathcal N$. In the addition step, it iteratively adds vertices from $\mathcal N$ to $\mathcal C$ that result in the greatest increase of $f(\mathcal C)$, until a certain number of neighbors have been in the subgraph. In the deletion step, it iteratively removes vertices in $\mathcal C$ that result in the increase of $f(\mathcal C)$ but not separating $\mathcal C$. The addition and deletion steps will be repeated until no vertex is added to $\mathcal C$. Note that there is no guarantee whether $q$ will be in the returned community as it may be removed during the deletion step. It has the same time complexity as the algorithm for the boundary-based local modularity.

\subsection{Query Biased Density-Based Community Search}
\label{sec:bias}

In \cite{Wu:VLDB:2015}, Wu et al. proposed the query biased density as the goodness function for CS. Before introducing the query biased density, the authors presented a vertex weighting scheme, which ensures that vertices far away from the query vertices will have large weights, resulting in high penalties to be included in the community. To assign each vertex $u$ a weight $r(u)$ w.r.t a set $Q$ of query vertices, they adopted the penalized hitting probability, which can be computed by random walk. Then, the query biased vertex weight of vertex $u$, $\pi(u)$, can be defined as the reciprocal of $r(u)$, i.e., $\pi(u)$=$1/r(u)$.

Based on the weights, the authors defined the {\it query biased density} of a graph $S$ as $\rho(S)$=$\frac{e(S)}{\pi(S)}$, where $e(S)$ is the sum of edges weights and $\pi(S)$ is the sum of query biased weights for vertices in $S$. After that, the authors proposed and studied the problem of finding the query biased densest subgraph $S$ from a graph $G$ (or QDS problem), which theoretically guarantees that QDS is a connected subgraph and contains $Q$.

Clearly, if $\pi(u)$=1, the query biased density degenerates to the classical edge-density (i.e., $\frac{e(S)}{|S|}$), and accordingly the QDS problem is reduced to the problem of densest subgraph discovery \cite{goldberg1984finding}. This also implies that after weighting $\pi(u)$, it forces the global densest subgraph shift to the neighborhood of the query vertices.

Unfortunately, the QDS problem is computationally intractable. To improve efficiency, the authors introduced two variants of the QDS problem by removing constraints that $S$ is connected and $Q$ is included in $S$, respectively. They showed that these variants can be solved in polynomial time and the results can be used to find an optimized solution for the QDS problem.

\subsection{Personalized PageRank-Based Community Search}
\label{sec:pagerank}

In \cite{Kloumann2014}, Kloumann et al. studied the use of personalized PageRank (PPR) model for identifying the community of a set of seed vertices $Q$. We first introduce the PageRank model: suppose there are an infinite number of surfers walking on a graph. If at a certain timestamp a surfer is staying at vertex $i$, at the next timestamp she goes to a random neighbor vertex $j$. As time goes on, the expected percentage of surfers at each vertex $i$ converges (under certain conditions) to a limit $r(i)$, called PageRank score of vertex $i$. Since $r(i)$ is independent of the distribution of starting vertices, it reflects the global importance of the vertex $i$.

Notice that $r(i)$ is computed with no preference for any particular vertices. However, in reality, for a particular user, some vertices, denoted by a set $Q$, may be more interesting than others, and they could be considered as the {\it preferred vertices}. To incorporate preferences of $Q$ into the model above, we can make a modification: at each step, a surfer jumps back to a vertex in $Q$ with probability $c$, and with probability ($1-c$) continues forth along a neighbor. The limit distribution of surfers in this model would favor vertices in $Q$ and vertices which are close to $Q$. The modified model is also called PPR model. Clearly, if we let $Q$ be a set of query vertices, the vertices whose limit probabilities are highest can be considered as $Q$'s community members.

Now we formally introduce the PPR model. Consider a graph $G$ and let $deg_G(i)$ denote the degree of vertex $i$ and $\mathbf A$ be the adjacent matrix of $G$, i.e., $A_{i,j}$=$\frac{1}{deg_G(i)}$ if vertex $i$ is linked to vertex $j$, where $deg_G(i)$ is the degree of vertex $i$. The {\it preference vector} $\mathbf u$ is defined over the seed vertices such that $|\mathbf u|$=1 and $u(i)$=$\frac{1}{|Q|}$ if the $i$-th vertex is in $Q$. Then, the PPR equation is
${\mathbf v}$= $(1-c){\mathbf {Av}}+c\mathbf u$,
where $c\in(0, 1]$ is the decay factor and a typical value of $c$ is 0.10 \cite{Kloumann2014}. The solution $\mathbf v$, called PPR vector, is a steady-state distribution of surfers.

\begin{problem}
\label{prob:PPR}
Given a graph $G(V,E)$, a set of query vertices $Q\subseteq V$, and an integer $k$, return a set $C$ of vertices, such that
\begin{enumerate}
  \item $Q\subseteq C$;
  \item $C$ contains $k$ vertices, whose corresponding values in the PPR vector w.r.t $Q$ are the highest;
\end{enumerate}
\end{problem}

In the literature \cite{andersen2006communities,Kloumann2014}, many efficient PPR algorithms have been developed, and thus can be applied to CS. We skip the details due to space limitation.

\subsubsection{Neighbors Expansion-Based Community Search}
\label{sec:neighbors}

In \cite{Mehler2009}, Mehler et al. presented a neighbor expansion method to discover the community from representative seeds. Specifically, given a graph $G(V,E)$ and a set $S$ of seed vertices, it repeatedly identifies the optimal ``next" vertex $v$, which is not in the community $C$ (initially $C$=$S$) but linked with vertices of $C$, based in some manner on the number or strength of $v$'s neighbors who had previously been identified as community members.
Details of vertex selection criteria and stopping rules of the expansion process are introduced as follows.

\noindent$\bullet$ \textbf{Selection criteria.} Mehler et al. proposed to assign a score to each vertex in the graph and select the highest-scoring outside vertex to join the community. The score assignment criteria are as follows:
\begin{itemize}
  \item {\it neighbor count:} the number of $v$'s neighbors in $C$;
  \item {\it juxtaposition count:} consider the weights of edges when counting the number of $v$'s neighbors in $C$;
  \item {\it neighbor ratio:} normalize vertices' degrees and count the degree-normalized neighbors in $C$;
  \item {\it juxtaposition ratio:} consider the weights of edges when computing the neighbor ratio;
  \item {\it binomial probability:} compute the binomial probability that $v$ is in $C$, given its neighbor count.
\end{itemize}

\noindent$\bullet$ \textbf{Stopping rules.} The authors proposed to reserve some fraction of seed vertices as validation members, and then monitor the frequency with which these validation members are incorporated into the community, during the expansion process. In the first phase, when community members are identified with high precision, we expect to add a new validation member with frequency equal to the fraction of community comprised by the validation set. After leaving the natural boundaries of the neighborhood, we expect to rediscover validation members according to their frequency in the entire graph.
As a result, we can find the stopping vertex as the one that best splits the validation interval (i.e., the difference between the discovery times of the $i$th and ($i-1$)-st validation members) into two groups.

\subsection{Discussions}
\label{sec:othersDiscuss}

In this section, we review CS studies that do not rely on metrics introduced in Section \ref{sec:pre}, which are often referred as local community detection. These studies mainly focus on simple undirected graphs, and uncover the communities by seed expansion using link-based metrics, such as modularity, density, pagerank, etc. Unlike CS studies introduced before, these works often rely on good seed selection algorithms \cite{moradi2014local} and assume that there are some ground truth communities. In other words, they might not aim to search communities in an online manner over big graphs, based on a query request. As a result, some of them may cost high running time for searching communities. Consequently, an interesting research direction is to develop index-based solutions for supporting efficient online CS queries using these metrics. Moreover, it would be interesting to study how to apply them for CS on attributed graphs.

\section{Community Search Systems}
\label{sec:demo}

Recently, many graph processing systems have been developed \cite{Batarfi:2015}. Generally, they can be classified into two groups. The first group (e.g., GraphX \cite{gonzalez2014graphx} and Pregel \cite{malewicz2010pregel}) aims to provide a platform for supporting general graph tasks (e.g., computing PageRank scores).
The second group is customized for specific graph tasks. For example, in \cite{fei2013expfinder}, Fan et al. developed a graph system, called Expfinder, for finding experts in social networks; in \cite{jayaram2015viiq}, a system called VIIQ is developed for interactive graph query formulation; in \cite{yi2017autog}, AutoG shows an interactive system to facilitate graph query formulation. However, none of them can be readily used for CS. To address this issue, recently some systems have been developed for searching, visualizing, and analyzing communities in large graphs. Below, we introduce two systems, namely C-Explorer \cite{Fang:demo:2017} and VizCS \cite{Huang:ICDE:2018}.

\subsection{C-Explorer}
\label{sec:cexplorer}

\begin{figure}[]
  \centering
  \includegraphics[width=3.32in]{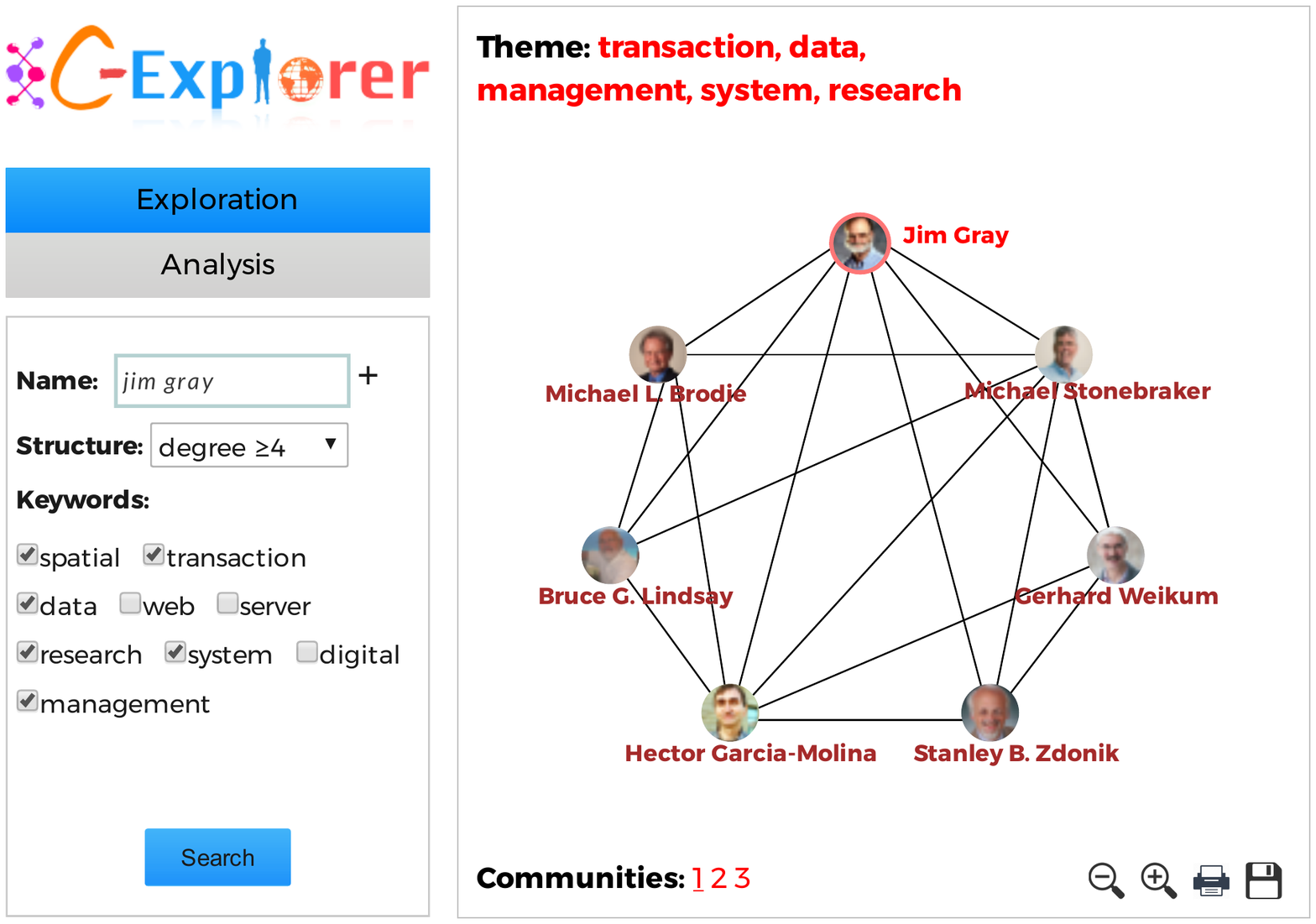}
  \caption{Interface of C-Explorer~\cite{Fang:demo:2017}.\label{fig:ce-ui}}
\end{figure}

C-Explorer is a web-based system that enables community retrieval in a simple, online, and interactive manner. The key features of C-Explorers are as follows:

First, it implements several typical CS algorithms on simple undirected graphs and keyword-based attributed graphs, including {\tt Global} and {\tt Local} (see Section~\ref{sec:kcoreSimple}), ACQ algorithm (see Section~\ref{sec:kcoreKeyword}). In addition, a CD algorithm called {\tt CODICIL}~\cite{attr-www2013} is included.

Second, it offers a user-friendly facility that enables online visualization of communities. Fig. \ref{fig:ce-ui} shows the user interface of C-Explorer configured to run on the DBLP bibliographical network. On the left panel, a user inputs the name of an author (e.g., ``jim gray") and the minimum degree of each vertex in the community she wants to have. The user can also indicate the labels or keywords related to her community. Once she clicks the ``Search" button, the right panel will display a community of Jim Gray. The user can further click on one of the vertices (e.g., Michael Stonebraker), and continue to examine its community.

Third, it allows users to compare the communities retrieved by various CS and CD algorithms, in terms of community quality and statistics.

Finally, it provides a list of API functions so that other CS and CD algorithms can be plugged in. For public users, they can easily plug their own algorithms into C-Explorer using these API functions.


\subsection{VizCS}
\label{sec:vizcs}
VizCS is an online query processing system for searching and visualizing communities in graphs \cite{Huang:ICDE:2018}. VizCS exhibits four key innovative features as follows.

First, VizCS adopts a triangle-connected truss community model for dynamic graphs where vertices/edges undergo frequently insertions/deletions \cite{k-truss2014}. It provides the feature of CS over dynamic graphs, which can be uploaded with one file of graph updates by users.

\begin{figure}[]
  \centering
  \includegraphics[width=2.5in]{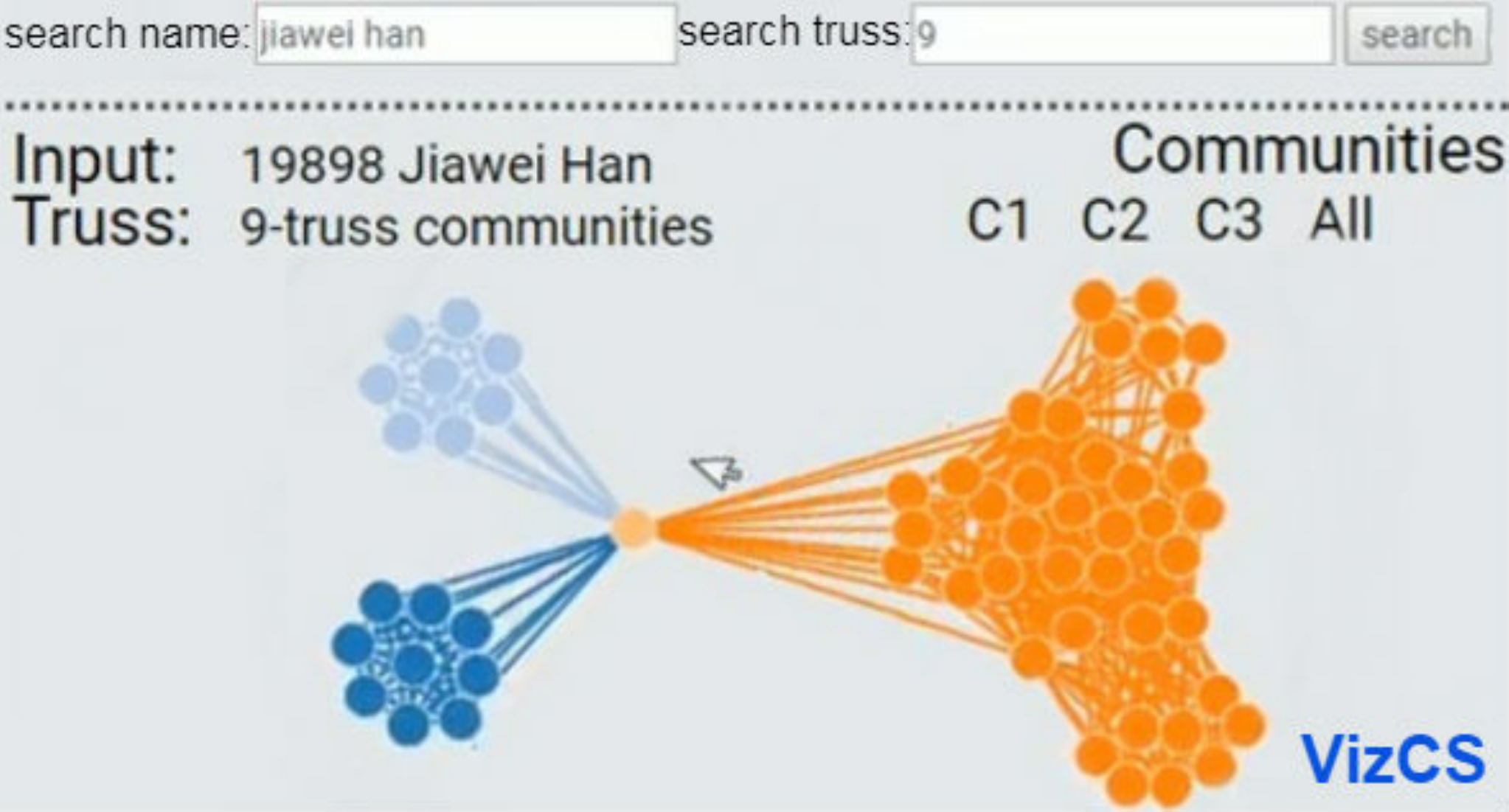}
  \caption{Interface of VizCS~\cite{Huang:ICDE:2018}.}\label{fig:vizcs}
\end{figure}

Second, VizCS offers a user-friendly visual interface to formulate queries and a real-time response query processing engine. Fig. \ref{fig:vizcs} shows an example query of author vertex $q$=``Jim Gray" and parameter $k$=8. Thanks to efficient $k$-truss CS algorithms, the query results can be quickly obtained in real-time.

Third, VizCS generates a community exploration wall by offering interactive community visualization, which facilitates users to in-depth understanding of the data. The community exploration wall uses graph visualization techniques to depict the community results and also presents informative features to users through various exploration channels, such as the profile search of community members by Google, structural statistic report, collaborator recommendation, and tag cloud.  Fig. \ref{fig:vizcs} shows the community exploration wall.

Last but not least, VizCS is a CS platform that can visualize and compare different community results by various state-of-the-art algorithms and user-uploaded approaches. It benefits users to understand different models vividly and directly.

\section{Comparison Analysis}
\label{sec:compare}

Recall that in the last subsections of Sections \ref{sec:kcore}, \ref{sec:ktruss}, \ref{sec:kclique}, and \ref{sec:kecc}, we have compared and analyzed the CS solutions using $k$-core, $k$-truss, $k$-clique, and $k$-ECC, respectively.
In this section, we would like to further compare these CS solutions across different metrics. Due to the space limitation, we are unable to compare all the surveyed 27 CS problems as well as their solutions. In the following, we mainly compare the representative CS problems and solutions on simple graphs and attributed graphs respectively, while other solutions can be considered as either their variants or less representative studies.


\begin{table*}[]
  \small
  \centering
  \caption {Comparison analysis for representative CS solutions on simple graphs.}
  \label{tab:compare}
  \begin{tabular}{c|c|c|c|c|c|c|c|c|c}
     \hline
     \multirow{2}{*}{\bf Metric}&
     \multicolumn{2}{c|}{\bf Online algorithm}&
     \multicolumn{4}{c|}{\bf Index-based algorithm}&
     \multirow{2}{*}{\tabincell{c}{\textbf{Cohesi}\\ \textbf{-veness}}}&
     \multirow{2}{*}{\tabincell{c}{\textbf{O}}}&
     \multirow{2}{*}{\tabincell{c}{\textbf{D}}}\\
     \cline{2-7} &\bf query &\bf scalab. &\bf time  &\bf space  & \bf scalab. & \bf query & & &\\
     \hline\hline
     $k$-core & $O(n)$ \cite{KDD2010} & $\bigstar\bigstar\bigstar\bigstar$ &
     $O(m)$\cite{kcore2003} & $O(n)$\cite{kcore2003} & $\bigstar\bigstar\bigstar\bigstar$ & $O(|E(C)|)$&
     $\bigstar$ & $\times$ & $\surd$\\
     \hline

     $k$-truss & $O(m^{1.5})$\cite{k-truss2014} & $\bigstar\bigstar$&
     $O(m^{1.5})$\cite{Akbas:VLDB:2017} & $O(m)$\cite{Akbas:VLDB:2017} & $\bigstar\bigstar\bigstar$ & $O(\sum_{i=1}^r |E(C_i)|)$ &
     $\bigstar\bigstar\bigstar$ & $\surd$ & $\surd$
     \\
     \hline
$k$-clique & $O(\log(n)sT)$\cite{kclique2018} & $\bigstar$ &
   $O(sLp)$ \cite{kclique2018} & $O(sL)$\cite{kclique2018} & $\bigstar$ & $O(g\log(g)Q)$\cite{kclique2018} &
     $\bigstar\bigstar\bigstar\bigstar$ & $\surd$ & $\surd$\\
     \hline

     $k$-ECC & $O(hlm)$\cite{Chang:SIGMOD:2015} & $\bigstar\bigstar\bigstar$&
     $O(\alpha(G)hlm)$\cite{Chang:SIGMOD:2015} & $O(m)$\cite{Chang:SIGMOD:2015} & $\bigstar\bigstar$ & $O(|E(C)|)$&
     $\bigstar\bigstar$ & $\times$ & $\surd$\\
     \hline
  \end{tabular}
\end{table*}

\begin{table*}[ht]
  \small
  \centering
  \caption {Empirical comparison for representative CS solutions on a real large graph.}
  \label{tab:exp}
  \begin{tabular}{c|c|c|c|c|c|c|c|c|c}
     \hline
     \multirow{2}{*}{\bf Metric}&
     {\bf Online algorithm}&
     \multicolumn{3}{c|}{\bf Index-based algorithm}&
     \multicolumn{4}{c|}{\bf Community quality}&
     \multirow{2}{*}{\tabincell{c}{\textbf{Community}\\ \textbf{number}}}\\
     \cline{2-9} &\bf query &\bf time  &\bf space  & \bf query & \bf diameter & \bf degree & \bf density & \bf CC & \\
     \hline\hline
     $k$-core  & 7.2s &   8.1s &   7.9\textbf{MB} &   2.7s  & 14.0 &   19.2 &  0.044  & 0.763 & 1  \\
     \hline
     $k$-truss & 55.1s  &  103.1s  &  179\textbf{MB}  & 0.2s  &   4.1 &  13.9  &  0.476 & 0.868 &  1.31\\
     \hline
     $k$-clique & 1872s  &  61.6s & 108\textbf{MB} & 4.3s  & 10.6
        & 9.2 & 0.424 & 0.709  & 1.05 \\
    \hline
     $k$-ECC   &   39.9s  & 38.3s   & 68\textbf{MB} &   0.15s & 10.5  & 18.4 & 0.152  & 0.774  &  1\\
     \hline
  \end{tabular}
\end{table*}

%

\subsection{Simple Graphs}
\label{sec:simpleG}

In this section, we compare representative CS problems for cohesiveness metrics studied on simple graphs, which are Problem \ref{prob:kcoreGlobal} for $k$-core, Problem \ref{problem:TTC} for $k$-truss, Problem \ref{prob:densest} for $k$-clique, and Problem \ref{prob:keccMax} for $k$-ECC.
In the following, we first compare these solutions in terms of the complexities and scalability of the state-of-the-art online algorithms, index construction complexities, index-based query algorithms, community cohesiveness, and support for overlapped CS as well as dynamic graphs. After that, we perform an experiment on real large graphs by using these CS algorithms, and compare their empirical performance.

To make a fair comparison, we consider a simple undirected graph $G(V,E)$, where $n$=$|V|$, $m$=$|E|$, and its arboricity is denoted by $\alpha(G)$ ($\alpha(G)$ is often much smaller than $\sqrt{m}$). We use $h$ and $l$ to denote small values that can be bounded by small constants \cite{Chang:SIGMOD:2015}.
In Table \ref{tab:compare}, we compare these representative CS solutions on $G$. Note that to measure the strength of algorithm scalability and community cohesiveness, we use notation $\bigstar$; that is, an algorithm with more $\bigstar$ means that it has better scalability or cohesiveness. Meanwhile, if a CS solution returns only one community $C$, we denote its community edge number by $|E(C)|$. If multiple communities are returned, we use $C_i$ to denote the $i$-th (1$\leq$$i$$\leq$$r$) community, where $r$ is the total number of returned communities.
We use ``O" and ``D" to denote whether the solutions support overlapped CS and dynamic graphs respectively.

In addition, for the complexities of the $k$-clique-based algorithm, we adopt the notations in \cite{kclique2018}, where $s$ is the average size of maximal cliques, $T$ is the time to enumerate all maximal cliques, $L$ is the number of maximal cliques, $p$ is the average number of maximal cliques a vertex is contained in, $Q$ is the number of maximal cliques containing at least one query vertex, and $g$ is the height of the index tree.

From Table \ref{tab:compare}, we can make the observations:
\begin{itemize}
  \item For online query algorithms, in terms of query time complexity, we can rank them as: $k$-core $\preceq$ $k$-ECC $\preceq$ $k$-truss $\preceq$ $k$-clique, which is consistent with the efficiency ranking relationship of these metrics in Section \ref{sec:analysis}. As a result, the $k$-core-based algorithm achieves the highest scalability while the $k$-clique-based algorithm has the lowest scalability.
  \item For index construction algorithms, the ranking relationship above still holds. For index-based query algorithms, most of them except $k$-clique have the optimal time complexity, which is linear to the community edge number (i.e., $|E(C)|$).
  \item The community structure cohesiveness is in line with the cohesiveness of these four metrics.
  \item The $k$-core and $k$-ECC-based solutions can only return one community for each query, while the other two solutions may return multiple overlapped communities containing the query vertex.
  \item All algorithms support dynamic graphs where vertices and edges are inserted or deleted dynamically.
\end{itemize}

Next, we empirically evaluate the performance of algorithms in Table \ref{tab:compare}. The input of these algorithms except the $k$-truss-based one is a query vertex, and they aim to find communities containing the query vertex which will maximize the value of $k$. For the $k$-truss-based one (Problem \ref{problem:TTC}), its input is a set of query vertices and an integer $k$. To make a fair comparison, we adapt its algorithm such that its input is a query vertex and the algorithm will maximize the value of $k$. To measure the quality of returned communities (subgraphs), we introduce four metrics, i.e., diameter, degree, density (i.e., the number of edges over the maximum number of possible edges in a graph), and clustering coefficient (CC). Generally, a lower value of diameter and higher values of degree, density, and CC mean the higher quality of the community.

To conduct the experiments, we use a real-world graph Google \footnote{\scriptsize{Available at \url{http://snap.stanford.edu/data/index.html}}}, which contains 875,713 vertices and 5,105,039 edges. We randomly select 100 vertices from the graph as query vertices, perform CS queries using these vertices, compute the average running time and community quality, and report experimental results in Table \ref{tab:exp}. Generally, the efficiency results in Table \ref{tab:exp} are consistent with the complexity analysis in Table \ref{tab:compare}. More specifically, we have:

\begin{itemize}
  \item For online query algorithms, the $k$-core-based algorithm is the fastest. The $k$-truss and $k$-ECC-based algorithms have similar time cost. The $k$-clique-based algorithm takes the highest time cost.
  \item To build indexes, the $k$-core-based algorithm is the fastest and the $k$-truss-based algorithm is slower than others.
  \item For index space cost, the $k$-core-based index takes the least space, while the space cost of others is around or over an order of magnitude larger than that of $k$-core-based algorithm.
  \item For index-based query algorithms, the $k$-core-based algorithm is slower than the $k$-truss-based algorithm (which also takes optimal query time cost), because its returned communities are larger than those of other algorithms. The $k$-clique-based algorithm is the slowest, as its complexity is higher than others.
  \item In terms of community quality, the $k$-truss-based solution achieves the smallest diameter, highest density, and highest clustering coefficient, due to small and tight triangle-based community structure. The $k$-core-based algorithm achieves the highest degree, against other methods. The $k$-clique-based method achieves the smallest degree.
  \item In line with Table \ref{tab:compare}, the $k$-core and $k$-ECC-based solutions return one community, while $k$-truss-based and $k$-clique-based solutions respectively return 1.31 and 1.05 communities.
\end{itemize}

\subsection{Attributed Graphs}
\label{sec:attributedG}

As shown in Table \ref{tab:methods}, for attributed graphs, five kinds of attributes have been considered for CS, which are keywords, locations, temporal information, profile, and influence values. However, the semantics of these attributed communities are different. Moreover, the problem definitions are also different. Therefore, it may not make sense to compare them under the same metrics.

For location, temporal information, and profile-based attributed graphs, only the $k$-core model has been studied on these graphs, which have been discussed and compared extensively in Section \ref{sec:kcoreDiscuss}.
For influence value-based graphs, the meanings of influences are very different.
In $k$-core-based CS solutions \cite{Li:vldb:2015,Li:VLDBJ:2017,chen2016efficient,Bi:2018,Li:SIGMOD:2018,ding2018search}, the influence values are associated to graph vertices, denoting their influence or importance.
In $k$-truss-based CS solutions \cite{Zheng:IS:2017}, the influence values are associated to graph edges, representing the influence or importance of edges.
In $k$-clique-based CS solutions \cite{kclique2017}, the influence values are also associated to graph edges, but they are probability values, meaning how likely a vertex is influenced by another vertex.
Meanwhile, none of these influence value-based graphs has been investigated with at least two different cohesiveness metrics, so we do not compare solutions for influence value-based graphs in this paper.
In the following, we mainly focus on comparing and analyzing CS solutions on keyword-based attributed graphs.

For keyword-based attributed graphs, there are two representative studies, 
namely ACQ \cite{Fang:VLDB:2016,Fang:VLDBJ:2017} and ATC \cite{Huang:2017:ATC}. Generally, both of them seek to find a densely connected community containing query vertex(es) with similar query keywords, but ACQ adopts the $k$-core model, while ATC uses the $k$-truss model.
From the discussions in Section~\ref{sec:analysis}, we infer that
the community of ATC 
is more structurally cohesive, but may take higher computational cost.
Besides, in terms of keyword cohesiveness, ACQ model in Section~\ref{sec:kcoreKeyword} imposes a strict homogeneity constraint, requiring that each vertex shares same query attributes in the community; ATC model in Section~\ref{sec:ktrussInfluence} uses an attribute score function to quantify the query keyword coverage and allows missing some query keywords in the community.

In \cite{Huang:2017:ATC}, Huang et al. empirically compared the community quality and efficiency of ACQ and ATC. 
They used 13 real graphs with ground-truth communities. For each graph, they ran 200 CS queries.
Specifically, for each query, they randomly selected a ground-truth community, and then randomly selected a vertex from the community as the query vertex. After that, they ran ACQ and ATC with the same parameters, i.e., $k$=4 and two query keywords which are selected from the community.
The results are consistent with the discussions above. Specifically, ATC achieves higher average $F_1$ score values than ACQ on all the datasets, which means that it is more accurate to search communities. On the other hand, in terms of efficiency, ACQ consistently outperforms ATC on all the datasets, and is up to two orders of magnitude faster than ATC.

\section{Related Work}
\label{sec:related}

In this section, we review related studies, including community detection, cohesive subgraph discovery, graph keyword search, and graph pattern matching.

\subsection{Community Detection}
\label{sec:CD}

Below, we review representative CD studies on undirected graphs, directed graphs, and attributed graphs.

\subsubsection{Undirected Graphs}
\label{sec:CDSimple}

A large number of studies aim to detect communities from simple graphs, and we can classify these studies based on the techniques they use. Some representative classes are as follows, to name a few:
\begin{enumerate}
  \item community quality optimization-based methods (e.g., modularity \cite{newman2004fast});
  \item clustering methods (e.g., $k$-means \cite{tang2009scalable}, spectral clustering \cite{von2007tutorial});
  \item graph partitioning methods (e.g., Metis \cite{karypis1995metis});
  \item embedding-based methods (e.g., DeepWalk \cite{perozzi2014deepwalk}, \cite{li2018community});
  \item random walk-based methods (e.g., \cite{pons2005computing});
  \item label propagation-based methods (e.g., \cite{gregory2010finding});
  \item information diffusion-based methods (e.g., \cite{hajibagheri2012community});
  \item statistic inference-based models (e.g., \cite{hastings2006community});
  \item deep learning-based methods (e.g., \cite{Yang:2016});
  \item centrality-based methods (e.g., \cite{community-phy2004});
  \item locality sensitive hashing-based methods (e.g., \cite{macropol2010scalable});
  \item physics-based methods (e.g., Potts low \cite{wu1982potts});
  \item local metric-based methods (e.g., $k$-plex \cite{conte2018d2k});
  \item multi-commodity flow-based methods (e.g., \cite{leighton1988approximate});
  \item hybrid-based methods (e.g., \cite{henderson2010hcdf}).
\end{enumerate}

For a detailed survey of CD, please refer to the following survey and empirical evaluation papers:
\cite{CD:Survey:2009,Yang2010,community-phy2010,papadopoulos2012community,danon2005comparing,gulbahce2008art,CD:Survey:2011,CD:Survey:2011a,CD:Survey:2013,CD:Survey:2017,CD:Survey:2014-overlap,CD:Survey:2015-multilayer,CD:Survey:2018-dynamic,CD:Survey:2010-compare,CD:Survey:2014-evaluate,CD:Survey:2015-truth}.
Although these CD solutions are able to discover communities from networks, they may not well satisfy the desirable factors of CS on big graphs as we discuss in Section \ref{sec:intro}, because most of them often use a global predefined criterion for generating communities and cannot find communities in an online manner.

\subsubsection{Directed Graphs}
\label{sec:CDDirected}

In recent years, a number of studies have investigated CD on directed graphs. Here are some representative studies, to name a few.
In~\cite{prl2008}, Leicht et al. extended the concept of modularity maximization~\cite{newman2004fast}, which was originally designed for undirected graphs, for detecting community structure in directed networks that makes explicit use of information contained in edge directions. In \cite{flake2000}, Flake et al. identified communities from websites network, which can be considered as directed graphs.
In~\cite{CDBenchmark}, Lancichinetti et al. introduced new benchmark graphs to test CD methods on directed
networks.
In~\cite{pre2010}, Kim et al. also proposed a new modularity metric for CD on directed networks.
In~\cite{sdm2010}, Yang et al. developed a new stochastic block model for CD on directed networks.
In~\cite{yang2014detecting}, Yang et al. presented algorithms for detecting communities from both directed and undirected networks.
Ning et al. \cite{ning2016local} studied local community extraction in directed networks.
A recent survey can be found in~\cite{CDSurvey}.

\subsubsection{Keyword-Based Attributed Graphs}
\label{sec:CDKeyword}

To identify communities from keyword-based attributed graphs, recent works~\cite{attr-vldb2009,SDM2013,WSDM2013,cheng2012clustering,attr-www2013,huang2016attributed} often use clustering techniques. Zhou et al. \cite{attr-vldb2009} computed vertices' pairwise similarities using both links and keywords, and then clustered the graph.
Subbian et al. \cite{SDM2013} explored noisy labeled information of graph vertices for finding communities.
Qi et al. \cite{WSDM2013} dynamically maintained communities of moving objects using their trajectories.
Ruan et al.~\cite{attr-www2013} developed a method {\tt CODICIL}, which augments the original graph by creating new edges based on content similarity, and then performs clustering on the new graph.

Another common approach is based on topic models. In~\cite{attr-topic-kdd2008,attr-topic-icml2009}, the {\tt Link-PLSA-LDA} and {\tt Topic-Link LDA} models jointly model vertices' content and links based on the {\tt LDA} model. In~\cite{attr-topic-sigmod2012}, the attributed graph is clustered based on probabilistic inference. In~\cite{attr-topic-www2012}, the topics, interaction types, and the social connections are considered for discovering communities. {\tt CESNA}~\cite{yang2013community} detects overlapping communities by assuming communities ``generate'' both the link and content. A discriminative approach~\cite{attr-kdd2009} has also been considered for community detection.
However, computing pairwise similarity among vertices is very costly, and thus they are questionable for performing online CS queries.

\subsubsection{Location-Based Attributed Graphs}
\label{sec:CDLocation}

The problem of CD on location-based attributed graphs (or geo-social networks) \cite{barthelemy2011} has been extensively studied \cite{girvan2002,guo2008,pnas2011,KDD2013,IJGIS2015}. In \cite{girvan2002}, Girvan et al. introduced the geo-community, which is a graph of intensely connected vertices being loosely connected with others, but it is more compact in space.
Guo et al.~\cite{guo2008} proposed the average linkage (ALK) measure for clustering objects in spatially constrained graphs.
In~\cite{pnas2011}, Expert et al. uncovered communities from spatial graphs based on modularity maximization.
In~\cite{KDD2013}, Shakarian et al. used a variant of Newman-Girvan modularity to mine the geographically dispersed communities.
In~\cite{IJGIS2015}, Chen et al. proposed a method using modularity maximization for detecting communities from geo-social networks.

\subsubsection{Temporal Graphs}
\label{sec:CDTemporal}

Many recent studies aim to detect communities from temporal graphs. In \cite{zhou2007discovering}, Zhou et al. studied CD over a temporal heterogeneous social network consisting of authors, document content, and the venues.
In \cite{liu2014persistent}, Liu et al. studied persistent community detection for identifying communities that exhibit persistent behavior over time.
In \cite{angadi2015overlapping}, Angadi et al. detected communities from dynamic networks where data arrives as a
stream to find the overlapping vertices in communities.
In \cite{bazzi2016community}, Bazzi et al. investigated the detection of communities in temporal multi-layer networks.
In \cite{ditursi2017local}, DiTursi et al. proposed a filter-and-verify framework for community detection in dynamic networks.
In \cite{kuncheva2017multi}, Kuncheva et al. presented a method by using spectral graph wavelets to detect communities in temporal graphs.
For more related studies, please refer to survey papers \cite{CD:Survey:2018-dynamic,tamimi2015literature}.

\subsection{Cohesive SubGraph Discovery}
\label{sec:cohesive}

In this section, we review studies on cohesive subgraph discovery. Notice that CD is one kind of cohesive subgraph discovery, but the latter one is more general.

\subsubsection{Simple Graphs}
\label{sec:cohSimple}

For simple graphs, typical cohesive subgraph models are
$k$-core \cite{md1983,kcore2003},
$k$-truss \cite{saito2008extracting,cohen2008trusses,zhang2012extracting},
$k$-clique \cite{kclique,article05clique},
and $k$-ECC \cite{gibbons1985algorithmic,hu2016querying}, as discussed in Section \ref{sec:pre}.
To compute these subgraphs, there are many efficient in-memory algorithms (e.g., $k$-core \cite{kcore2003}, $k$-truss \cite{wang2012truss}, $k$-clique \cite{mauro2018}, and $k$-ECC \cite{zhou2012finding,Chang:SIGMOD:2013,akiba2013linear}).
For graphs that are too large to be kept in memory, there are also some disk-based and parallel algorithms.
For example, in \cite{cheng2011efficient,wen2018efficient}, \cite{wang2012truss,khaouid2015k}, and \cite{cheng2012fast}, disk-based algorithms for computing $k$-core, $k$-truss, and $k$-clique are developed, respectively; in \cite{montresor2013distributed} and \cite{chen2014distributed}, parallel algorithms for computing $k$-core and $k$-truss are proposed, respectively. In addition, to maintain $k$-core and $k$-truss for dynamic graphs, some efficient algorithms are developed in \cite{li2014efficient,sariyuce2016incremental,zhang2017fast} and \cite{zhou2014efficient}, respectively.

Besides, there are many other cohesive subgraph models and the representatives are as follows.
In \cite{seidman1978graph}, Seidman proposed the $k$-plex model (which is introduced in Section~\ref{sec:kclique}).
In \cite{matsuda1999classifying}, Matsuda et al. introduced the concept of quasi-clique model.
In \cite{zhang2018discovering}, Zhang et al. proposed the ($k$, $s$)-core, which considers both user engagement and tie strength.
In \cite{sariyuce2016fast}, the authors proposed the concept of nucleus, which is a generalization of $k$-core and $k$-truss.
In \cite{zhao2012large}, Zhao et al. introduced the mutual-friend subgraph.
In \cite{wang2010triangulation}, Wang et al. proposed the DN-Graphs by considering vertices' common neighbors.
In \cite{Chang:SIGMOD:2013}, Chang et al. studied the problem of enumerating $k$-ECCs in a graph for a given $k$.
In \cite{zhu2018diversified}, Zhu et al. introduced the notion of coherent cores on multi-layer graphs.
In addition, Goldberg et al. \cite{goldberg1984finding} and Fang et al. \cite{fang2019efficient} discovered the densest subgraph, Galbrun et al. \cite{Galbrun:2016} studied the top-$k$ densest subgraphs, Tsourakais et al. \cite{tsourakakis2013denser} computed the quasi-clique-based dense subgraphs, and Qin et al. \cite{Qin:2015} studied the problem of finding top-$k$ locally densest subgraphs.

\subsubsection{Attributed Graphs}
\label{sec:cohAttributed}

For attributed graphs, in addition to CD methods, there are also many studies of finding cohesive subgraphs.
In \cite{denian:2012}, Yang et al. studied the socio-spatial group query which finds a group of users that are cohesively linked and close to the rally point in a geo-social network.
In \cite{Zhang:VLDB:2017}, Zhang et al. studied the problem of finding ($k$, $r$)-cores on attributed graph and for a specific ($k$, $r$)-core, each vertex has at least $k$ neighbors, and the attribute similarity of each pair of vertices is at least $r$.
In \cite{Chen:VLDB:2018}, Chen et al. studied the problem of ($k$, $d$)-MCC (maximum co-located community) search on geo-social network, where a ($k$, $d$)-MCC is a connected $k$-truss and for any two vertices, their distance is at most $d$.
In addition, Wu et al. \cite{wu2015finding} studied the problem of finding the densest connected subgraph from the dual network, which can be considered as an attributed graph.

\subsection{Graph Keyword Search}
\label{sec:GKS}

Generally, graph keyword search \cite{wang2010survey,keyword-yu-2009,yuan2017keyword,yuan2013efficient} aims to find a tree or a subgraph, which contains a set of query keywords, from a large graph $G$. Earlier studies often output a tree structure.
In~\cite{keyword-icde2002}, Bhalotia et al. developed a backward algorithm for finding Steiner trees.
In~\cite{keyword-icde2007}, Ding et al. proposed a dynamic programming algorithm finding Steiner trees.
In~\cite{golenberg2008keyword}, Golenberg et al. presented a novel algorithm which produces Steiner trees with polynomial delay.
In~\cite{keyword-vldb2005}, Kacholia et al. proposed a bidirectional search algorithm, and He et al.~\cite{he2007blinks} improved its efficiency by introducing a new index structure.

Recently, some solutions have output subgraphs.
In \cite{keyword-sigmod2008}, Li et al. proposed to find $r$-radius Steiner graphs that contain query keywords.
Qin et al. \cite{keyword-icde2009} proposed to find multi-centered subgraphs that contain query keywords within a given distance. Kargar et al. \cite{keyword-vldb2011} studied the $r$-clique which is a set of vertices that cover query keywords and satisfy the distance constraint.

However, these works are substantially different from CS queries on keyword-based attributed graphs. First, they do not specify query vertices as required by CS queries. Second, the tree or subgraph produced do not guarantee structure cohesiveness. Third, their solutions do not ensure strong keyword cohesiveness.

\subsection{Graph Pattern Matching (GPM)}
\label{sec:GPM}
For simple graphs, the problem of GPM is NP-complete \cite{cook1971complexity} and it has been studied extensively under different settings:
(1) in main memory \cite{ullmann1976algorithm,chiba1985arboricity}. For example, Ullmann \cite{ullmann1976algorithm} proposed a backtracking algorithms.
(2) in external memory, Chu et al. \cite{chu2011triangle} and Hu et al. \cite{hu2014efficient} studied triangle counting; in \cite{qiao2017subgraph}, a novel GPM solution based on graph compression is presented.
(3) in distributed platforms, both DFS-style approaches \cite{afrati2013enumerating,park2016pte}and BFS-style approaches \cite{lai2015scalable,lai2016scalable} are developed. The DFS-style approaches avoid intermediate results by using one-round computation, while BFS-style approaches shuffle a large number of intermediate results.

For attributed graphs, there are also many studies. Tong et al. \cite{GPM-KDD2007} studied the use of lines, loops and stars for finding the matched subgraphs; Zou et al. \cite{zou2009distance} developed a novel GPM solution based on distance join;
Fan et al. \cite{GPM-VLDB2010} studied GPM by using bounded simulation; in \cite{GPM-PVLDB2015}, GPM has been studied for finding graph association rules; in \cite{GPM-ICDE2012}, Cheng et al. studied the problem of top-$k$ GPM. Recently, Fang et al. have studied a variant of the GPM problem on spatial databases \cite{Fang:ICDE:SPM,Fang:ICDE:DEMO}, and it aims to find spatial objects that are matched with a given pattern.
However, GPM is different with CS since (1) it often focuses on small patterns, so it cannot generate large communities; and (2) the subgraphs of GPM solutions often do not guarantee strong structure cohesiveness. Other related topics include subgraph search \cite{yua,yuan2012efficient}.

\section{Future Work}
\label{sec:future}

Recall that in Table~\ref{tab:methods}, the cohesiveness metrics are orthogonal to graph types,
so if a metric has not been studied for a particular type of graphs,
then it is a future research direction to study CS by applying the metric on this type of graphs.
Apart from this, we present a number of promising future directions as follows.

\subsection{Optimization for Query Parameters}
\label{sec:queryPara}

Most existing CS queries require users to input some parameters, in addition to the query vertex. A typical parameter is the integer $k$ \cite{KDD2010,local2014,barbieri2015efficient}, which controls the structure cohesiveness of returned communities. For attributed graphs, existing works also require users to input some parameters related to attributes. For example, in ACQ~\cite{Fang:VLDB:2016} and ATC~\cite{Huang:2017:ATC}, a set of query keywords are required.
Although these parameters provide strong flexibility and personalization for the query, it may not be easy for users to set proper values for these parameters. For example, if the integer $k$ is too large, a false query may incur, i.e., the query returns empty result. On the other hand, if $k$ is too small (e.g., $k$=1 or 2), the returned community may contain too many vertices, which may make the community meaningless.

Unfortunately, most existing CS works assume that users can input proper values for these parameters. This assumption, however, is too strong, especially when users do not know much about the underlying network. To suggest query parameters, a possible research direction is to exploit historical query logs and suggest some values of parameters automatically \cite{baeza2004query,marcel2011survey}. Another direction is to study how to use crowdsourcing platforms (e.g., AMT~\cite{amt}) to facilitate query suggestions.

\subsection{More Cohesiveness Metrics}
\label{sec:futureMetric}

As aforementioned, in CS solutions, a community is required to satisfy certain cohesiveness metrics. Essentially, the cohesiveness metrics formally define the communities, so they  play crucial roles in CS.

For structure cohesiveness, there are many other cohesiveness models (see Section~\ref{sec:cohesive}) which have not been used for CS. Thus, it would be interesting to study CS using these models. For example, in \cite{sariyuce2016fast,sariyuce2015finding}, the authors have proposed the concept of nucleus, which is a generalization of $k$-core and $k$-truss.

For attribute-based cohesiveness, as discussed in Section~\ref{sec:cohesive}, there are some studies finding cohesive subgraphs from attributed graphs. Thus, it is of interest to extend them for CS on attributed graphs.
Besides, each existing CS solution only focuses on one particular type of attribute (e.g., keyword). This, however, may be problematic for many real applications because a real graph often involves multiple types of attributes.
Thus, it is desirable to study how to perform CS by considering multiple types of attributes.

\subsection{Other Types of Graphs}
\label{sec:futureGraph}

In recent years, many novel network models have been developed and the representative ones are as follows:
\begin{itemize}
  \item {\it Public-private network~\cite{Archer2017,huang2018pp,chierichetti2015efficient}}. In a public-private network (e.g., Facebook), there is a public graph $G$, containing a set of vertices and a set of edges that are visible to all users of the network. In particular, each vertex $u$ is associated with a private graph $G_u$, where vertices of $G_u$ are vertices from the public graph $G$, and $G_u$ is only known to $u$.
  \item {\it Uncertain graph~\cite{hu2017embedding,li2018edbt,huang2016truss}}. In many real applications (e.g., biology), the graph data are often noisy, inexact, and inaccurate, and they can be modeled as uncertain graphs, where each edge is associated with a value denoting its existence probability.
  \item {\it Signed graph~\cite{yang2007community}}. A signed graph is a graph whose edges carry signs. For example, in social networks, the relationship of two users is either positive (e.g., friendship) or negative (e.g., hostility). Thus, users' relationship can be modeled as a signed graph.
  \item {\it Multi-dimensional graphs~\cite{fang2014detecting}}. In many scenarios, a graph often contains various types of edges, which represent various types of relationships between entities. Such graphs are often called multi-dimensional graph, or multi-layer graphs or multi-view graphs.
  \item {\it Heterogeneous information network (HIN)~\cite{shi2017survey,Hu:HIN:2019}}. HINs are networks with multiple typed objects and multiple typed links denoting different relations.
\end{itemize}

To our best knowledge, there is no prior research about CS on these graphs. Thus, it is still an open problem of how to perform CS on these graphs.

\subsection{Real Big Graphs}
\label{sec:futureEff}

Most existing CS studies assume that the graphs can be kept in the memory of a single machine. The graphs used for experimental evaluation are often million-scale, and only a few of them \cite{Fang:TKDE:CSD,Li:vldb:2015} are able to process billion-scale graphs. However, in many real applications (e.g., Facebook), the graphs may involve billions of vertices and edges \cite{li2015walking}. As a result, existing CS solutions may fail to process such real big graphs within reasonable time cost. Hence, how to efficiently perform online CS on such big graphs is a challenging task.

For big graphs that cannot be kept by a single machine, some possible research directions are as follows.
First, we can consider developing query algorithms based on distributed computation platforms (e.g., GraphX \cite{gonzalez2014graphx}), which are able to process big graphs in a cluster.
Second, to save memory space, we may keep the graph data on disk and design I/O-efficient query algorithms.

\subsection{An Online Repository for Codes and Datasets}
\label{sec:opensource}

For most of surveyed CS studies, their codes of algorithms and datasets are not publicly available. Thus, it is desirable to build an online repository to keep these codes and datasets. The major benefits of doing this are two-fold: First, for researchers, the codes and datasets can serve as a benchmark for comparison studies. Second, practitioners can easily plug these CS solutions into their applications without re-implementation.

\section{Conclusion}
\label{sec:conclude}

In this paper, we conduct an extensive survey on the topic of community search over large graphs. We systematically review over 30 research articles, which focus on the topic of community search, published between 2010 and 2019.
We first analyze and compare different community cohesiveness metrics. Then, we classify studies about CS according to these metrics, and for each class of works, we review and discuss the representative studies on different types of graphs. Furthermore, two systems that are customized for the purpose of community search are discussed. Finally, we point out a list of future research topics as well as challenges. In summary, our survey provides an overview of the start-of-the-art research achievements on the topic of community search, and it will give researchers a thorough understanding of community search.

\section*{Acknowledgments}
We would like to thank Jiafeng Hu and Kai Wang for their helpful discussions, Dan Yin for the proof-reading, and Jinbin Huang for conducting experimental comparisons.
Xin Huang is supported by the NSFC Project No. 61702435, and Hong Kong General Research Fund (GRF) Project No. HKBU 12200917.
Lu Qin is supported by DP160101513.
Ying Zhang is supported by FT170100128 and DP180103096.
Wenjie Zhang is supported by DP180103096.
Reynold Cheng is supported by the Research Grants Council of Hong Kong (RGC Projects HKU 17229116 and 17205115) and HKU (Projects 102009508 and 104004129).
Xuemin Lin is supported by 2019DH0ZX01, 2018YFB1003504, NSFC61232006, DP180103096 and DP170101628.
\blfootnote{*For lack of space, we use abbreviations for the names of major conferences and
journals in database and data mining areas (e.g., we use ``PVLDB" to mean ``Proceedings of the VLDB Endowment").
For other venues, we use full names.}

\bibliographystyle{abbrv}
\bibliography{CS,CD,demo,GS,misc,future,ICDE17-truss,truss,others}

\end{document}